%% file: main.tex
\documentclass[a4paper]{article}

\usepackage[utf8]{inputenc}
\usepackage[T1]{fontenc}
\usepackage{geometry}
\usepackage{amsmath,amssymb}
\usepackage{graphicx}
\usepackage{booktabs}
\usepackage[numbers,sort&compress]{natbib}
\usepackage{hyperref}
\usepackage{csquotes}
\usepackage{siunitx}
\usepackage{threeparttable}
\usepackage{float}
\usepackage{placeins}
\usepackage{adjustbox}
\usepackage{fancyvrb}
\usepackage{lscape}
\usepackage{soul}
\usepackage{rotating}
\usepackage{authblk}

\providecommand{\tightlist}{\setlength{\itemsep}{0pt}\setlength{\parskip}{0pt}}

\newcounter{mainhyp}

\AtBeginDocument{\setcounter{mainhyp}{1}}
\newcommand{\thehyp}{$\boldsymbol{H_{\arabic{mainhyp}}}$}
\newcommand{\newseries}{\stepcounter{mainhyp}}

\newenvironment{hyp}{%
  \par\vspace{6pt}\noindent\thehyp:\quad\ignorespaces
}{%
  \par\vspace{6pt}%
}

\title{All Public Voices Are Equal, But Are Some More Equal Than Others to LLMs?}

\author[1]{Sola Kim\thanks{Corresponding author}}
\author[1]{Marco A. Janssen}
\author[2]{Jieshu Wang}
\author[1]{Ame Min-Venditti}
\author[3]{Neha Karanjia}
\author[1,4]{John M. Anderies}

\affil[1]{School of Sustainability, Arizona State University, Tempe, AZ 85281, USA}
\affil[2]{Department of Technology and Society, Stony Brook University, Stony Brook, NY 11794, USA}
\affil[3]{W. P. Carey School of Business, Arizona State University, Tempe, AZ 85281, USA}
\affil[4]{School of Human Evolution and Social Change, Arizona State University, Tempe, AZ 85281, USA}

\date{}

\begin{document}

\maketitle

\begin{abstract}
Federal agencies are increasingly deploying large language models (LLMs) to process public comments submitted during notice-and-comment rulemaking, the primary mechanism through which citizens influence federal regulation. Whether these systems treat all public input equally remains largely untested. Using a counterfactual design, we held comment content constant and varied only the commenter's demographic attribution---race, gender, and socioeconomic status---to test whether eight LLMs available for federal use produce differential summaries of identical comments. We processed 182 public comments across 32 identity conditions, generating over 106,000 summaries. Occupation was the only identity signal to produce consistent differential treatment: the same comment attributed to a street vendor, compared to a financial analyst, received a summary that preserved less of the original meaning, used simpler language, and shifted emotional tone. This pattern held across all names, prompts, models, and regulatory contexts tested. Race effects were inconsistent and appeared driven by specific name tokens rather than racial categories; gender effects were absent. Writing quality predicted summarization outcomes through argument substance rather than surface mechanics; experimentally injected spelling and grammar errors had negligible effects. The magnitude of occupation-based differential treatment varied by model provider, meaning that selecting a model implicitly selects a level of fairness---a dimension that current procurement frameworks such as FedRAMP do not evaluate. These findings suggest that socioeconomic signals warrant attention in AI fairness assessments for government information systems, and that fairness benchmarks could be incorporated into existing federal IT procurement processes.
\end{abstract}


\section{Introduction}

Democratic governance rests on the principle that citizens' voices carry equal weight regardless of who speaks. Federal regulatory agencies issue more than 4,000 new rules each year \citep{febrizioFederalAgenciesAre2021, xie2024RegulatoryYear25}, many through notice-and-comment rulemaking, whereby citizens submit written input on proposed policies before they take effect \citep{purpuraActiveLearningERulemaking2008, usc_5_553_2022}. What matters is the quality of arguments and the diversity of perspectives, not submission volume \citep{jacksonPublicTrustAdministrative2023, coglianeseTransparencyPublicParticipation2009}; when commenters more closely reflect the demographic composition of the affected public, agencies may be more likely to receive input that captures the range of public preferences \citep{bullMakingAdministrativeState2013, jacksonPublicTrustAdministrative2023}. Following the E-Government Act of 2002, over 40 federal agencies accept electronic submissions through \texttt{Regulations.gov} \citep{rinfretBotsFakeComments2022}, broadening access but also amplifying volume: organized campaigns can generate tens or hundreds of thousands of similar messages that strain agency resources \citep{coglianeseTransparencyPublicParticipation2009, defigueiredoERulemakingBringingData2006, rinfretBotsFakeComments2022}. Yet agencies cannot simply ignore submissions, because failure to acknowledge and respond to significant comments can lead courts to invalidate rules as procedurally deficient or arbitrary and capricious \citep{jacksonPublicTrustAdministrative2023, staszewskiPoliticalReasonsDeliberative2011}.

In response, federal agencies have deployed automated tools for processing public comments. Early systems used natural language processing for tasks such as comment classification and organization; the 2024 Federal AI Use Case Inventory documents several such tools, including the Department of the Interior's Public Comment Analysis Tool (PCAT) and the Centers for Disease Control's comment processing system \citep{officeofmanagementandbudget2024FederalAI2024}. More recently, agencies have adopted AI for comment processing: the Department of Justice's Docketscope uses AI to detect duplicate and computer-generated submissions, cluster comments by topic, and identify substantive arguments that the Administrative Procedure Act requires agencies to consider \citep{DocketScope2023ImpactAIComments}. The Department of Energy's PermitAI initiative, developed by Pacific Northwest National Laboratory, uses AI to extract insights from public comments and expedite environmental review across interagency partners \citep{Nally2025NEPA_AI_Workshop}. Federal agencies typically access LLMs through cloud-based APIs certified under the Federal Risk and Authorization Management Program (FedRAMP), which addresses data security and system reliability without assessing fairness, accuracy, or consistency \citep{officeofmanagementandbudgetModernizingFederalRisk2024}.

This gap raises a question worth investigating: Do large language models treat all public voices equally? Agencies value comments that substantively engage with proposed rules and offer concrete alternatives \citep{rinfretBotsFakeComments2022}, and must consider all substantive arguments while responding to significant policy concerns \citep{staszewskiPoliticalReasonsDeliberative2011, jacksonPublicTrustAdministrative2023, purpuraActiveLearningERulemaking2008}. If LLM tools handle comments differently depending on perceived commenter characteristics, this could mean some voices are represented less faithfully, limiting the range of viewpoints agencies encounter. To test for equal treatment, we held comment content constant and varied only the commenter's identity. Any systematic change in how the summary was produced constitutes differential treatment. This approach draws on counterfactual fairness, which defines a decision as fair if it would not change were the individual to belong to a different demographic group \citep{chiappaPathSpecificCounterfactualFairness2019, kusnerCounterfactualFairness2017, russellWhenWorldsCollide2017, gargCounterfactualFairnessText2019}.

Unequal treatment in comment processing could pose concerns for democratic governance in at least two respects. Deliberative democratic theory holds that administrative decision-making should prioritize the substance of arguments over the identity of those who make them \citep{gutmannWhyDeliberativeDemocracy2016, staszewskiPoliticalReasonsDeliberative2011}; if perceived commenter identity influences how LLMs summarize content or tone, the summary may reflect more than just the argument itself, which would sit uneasily with deliberative principles. Beyond deliberative concerns, notice-and-comment rulemaking is designed to give all members of the public an equal opportunity to participate \citep{bullMakingAdministrativeState2013, jacksonPublicTrustAdministrative2023, rossiParticipationRunAmok1997}. If algorithmic processing treats some voices differently, this could raise questions about whether that equal footing is maintained in practice \citep{deanDeliberatingStateLocating2024, gonzalezPolicingDemocraticParticipation2023}.

These concerns motivate the choice of outcome measures in our study. Among them, we treat semantic similarity---the degree to which a summary preserves the meaning of the original text---as most consequential. Research suggests that up to 30\% of machine-generated abstractive summaries may contain factual inconsistencies \citep{kryscinskiEvaluatingFactualConsistency2020, kryscinskiNeuralTextSummarization2019a}. If a summary drops or misrepresents an argument, officials reviewing public input are less likely to encounter it \citep{gonzalezPolicingDemocraticParticipation2023}. Tone shifts may also matter: if a summary makes a comment sound more positive or more negative than the original, it could misrepresent the commenter's stance \citep{purpuraActiveLearningERulemaking2008}. Compression ratio---the length of a summary relative to its source---is a standard summarization evaluation metric \citep{fabbriSummEvalReevaluatingSummarization2021, liuReferencefreeSummarizationEvaluation2022}; differential compression may indicate whose contributions receive fuller or more truncated representation. Readability, most commonly operationalized through Flesch--Kincaid Grade Level in summarization research \citep{cacholaEvaluatingEvaluatorsAre2025}, captures whether models simplify language for some groups more than others.

This study addresses the following research question: \textbf{Do base LLMs available for federal use produce different summaries of identical public comments when commenter characteristics vary?} We focus on base LLMs---foundation models as provided by cloud service vendors before any agency-specific customization---because proprietary systems such as PCAT and Docketscope are not available for external research, and base models represent the most accessible starting point for systematic testing. While deployed systems may exhibit different patterns due to customizations or retrieval-augmented generation architectures, testing foundation models provides a common baseline: if base models show no differential treatment, deployed systems may be less likely to introduce such patterns; if base models do show differential treatment, this would suggest that further investigation of deployed systems is warranted. A pilot study (77,975 paired observations across eight LLMs) provided preliminary evidence of differential treatment that varied by model, demographic group, and text features (Appendix~\ref{app:pilot}).


\section{Hypotheses}

\subsection{Racial Identity Signals}

Names reliably trigger differential treatment in experimental research: job applicants with African American-associated names receive fewer callbacks despite identical qualifications \citep{bertrandAreEmilyGreg2004}, with similar patterns in housing \citep{carpusorRentalDiscriminationEthnicity2006} and academic mentorship \citep{milkmanWhatHappensField2015}. In LLM research, bias direction varies by model and task, with some studies finding pro-Black and others anti-Black patterns depending on the context \citep{karvonenRobustlyImprovingLLM2025, fangBiasAIgeneratedContent2023, anMeasuringGenderRacial2025}. Given this heterogeneity, we tested for differential treatment without predicting direction.

\vspace{11pt}
\begin{hyp}
$\Delta$ DV scores (change from baseline when identity is added) will differ across racial groups (White, Black, Hispanic, Asian).
\end{hyp}


\subsection{Gender Signals}

Gender bias manifests across diverse LLM contexts, including how LLMs describe politicians \citep{stanczakQuantifyingGenderBias2023}, generate product descriptions \citep{kellyUnderstandingGenderBias2025}, and respond to adversarial prompts \citep{kumarDecodingBiasesAnalysis2025}. Findings on direction are mixed: \citet{anMeasuringGenderRacial2025} found consistent pro-female bias, while others find anti-female patterns depending on task \citep{fangBiasAIgeneratedContent2023}. We tested for gender effects without directional predictions.

\vspace{11pt}
\newseries
\begin{hyp}
$\Delta$ DV scores will differ between male and female conditions.
\end{hyp}

\subsection{Socioeconomic Status Signals}

Socioeconomic status has received less research attention than race and gender in the LLM bias literature. Directional patterns vary: clinical settings showed bias favoring higher-income patients \citep{arzaghiUnderstandingIntrinsicSocioeconomic2025, omarSociodemographicBiasesMedical2025}, while simulated college admissions favored low-SES applicants \citep{nghiemRichDadPoor2025}. Other studies find that LLMs assign high-income professions more often to names from affluent contexts \citep{singhBornSilverSpoonInvestigating2025} and that LLMs tend to associate disability with lower income \citep{hariWhosAskingInvestigating2025}. In public comment summarization, occupation signals may trigger associations about commenter credibility, potentially resulting in differential attention to arguments.

\vspace{11pt}
\newseries
\begin{hyp}
$\Delta$ DV scores will differ between high-SES and low-SES conditions (operationalized as occupation).
\end{hyp}
\vspace{11pt}

Intersectionality theory holds that overlapping identities create experiences distinct from the sum of their parts \citep{keeanga-yamahttataylorHowWeGet2017, crenshawDemarginalizingIntersectionRace1989}. LLM research supports this: resume screening found simultaneous pro-female and anti-Black male bias \citep{anMeasuringGenderRacial2025}, resume retrieval disadvantaged Black men specifically \citep{wilsonGenderRaceIntersectional2025}, and LLM bias is intertwined across race, gender, and SES \citep{singhBornSilverSpoonInvestigating2025}. These findings suggest that examining main effects alone may obscure intersectional patterns.

\vspace{11pt}
\newseries
\begin{hyp}
The magnitude or pattern of racial differences in $\Delta$ DV scores will differ between male and female conditions.
\end{hyp}

\subsection{Writing Quality Signals} 

Writing quality varies across commenters for reasons including language background, educational access, and writing experience \citep{zabihiRoleCognitiveAffective2018, derakhshanInvestigationIranianEFL2020, johnsonTwentyYearFollowUpChildren2010}. If writing quality predicts summarization outcomes, this could produce unequal effects across demographic groups — not because LLMs consider demographics directly, but because writing quality itself is shaped by factors that vary demographically. Research suggests LLMs degrade on noisy inputs \citep{alahmariLargeLanguageModels2025, wangResilienceLargeLanguage2024} but show relative resilience to grammatical errors \citep{wangResilienceLargeLanguage2024}; direct evidence on broader quality dimensions in democratic deliberation contexts remains limited. We addressed this through two complementary analyses: an observational examination of whether writing quality predicts summarization outcomes ($\boldsymbol{H_5}$), and an experimental manipulation adding surface-level errors to lower-quality comments ($\boldsymbol{H_6}$). $\boldsymbol{H_5}$ results may partly reflect appropriate differential treatment of substantively weaker arguments; $\boldsymbol{H_6}$ isolates surface characteristics by holding content constant.

\vspace{11pt}
\newseries
\begin{hyp}
Raw DV scores will be associated with writing quality levels across identity conditions.
\end{hyp}

\vspace{11pt}
\newseries
\begin{hyp}
Raw DV scores will differ between error-added and original conditions for low and mid-low quality comments.
\end{hyp}

\section{Methods}

\subsection{Procedural Overview}

This study followed a pre-registered protocol \citep{kimAllPublicVoices2026}. Data collection proceeded in three phases: comment retrieval and preparation, experimental manipulation, and data generation.

In the retrieval phase, we collected all public comments from two dockets via the regulations.gov API: Department of the Interior docket DOI-2025-0004 ("National Environmental Policy Act Implementing Regulations") as our primary sample, chosen because agencies are already using LLMs to process NEPA-related public comments \citep{Nally2025NEPA_AI_Workshop}, and EPA docket EPA-HQ-OAR-2025-0124 ("Repeal of Greenhouse Gas Emissions Standards") as a validation sample. We extracted text from all formats using text extraction or optical character recognition (OCR) as needed, removed exact duplicates and near-duplicates exceeding 95\% cosine similarity, and applied initial filters for English language and minimum length of 50 words.

In the preparation phase, we classified comments by length and sophistication using LLM-assisted coding with human verification, sampling until stratification targets were met. We detected and redacted existing identity signals using regular expressions, named entity recognition, and LLM-based verification. Although we initially planned to validate a 10--15\% sample, automated redaction required extensive correction, so all comments were manually reviewed.

In the generation phase, we created 65 versions of each comment by inserting identity signals, then submitted each version to eight LLMs via API with randomized condition order. Dependent variables were computed by comparing attributed summaries to baseline summaries and original comments. Statistical analyses proceeded as specified in Section~\ref{analysis} and Appendix~\ref{app:statistical-models}.

The study crossed 182 comments with 65 conditions (32 identity conditions, 32 error-injection conditions, and 1 baseline) and 8 LLMs from four providers (OpenAI, Google, Anthropic, Meta), yielding over 106,000 summaries; design rationale, condition structure, and model selection are detailed in Appendix~\ref{sec:design-changes}. Power simulations indicated minimum detectable effects of $d = 0.039$ at 80\% power and $d = 0.049$ at 95\% power for identity main effects (Appendix~\ref{app:pilot:power}).

\paragraph{Identity Conditions ($\boldsymbol{H_1}$--$\boldsymbol{H_5}$)} \label{identity-condition} The primary experimental conditions manipulated identity signals while holding comment text constant. We crossed four levels of Race (White, Black, Hispanic, Asian) with two levels of Gender (Male, Female) and two levels of Socioeconomic Status (high and low occupational prestige), using two names per Race $\times$ Gender cell (16 names total). Each name appeared in both SES conditions, producing 32 identity conditions plus one baseline containing no identity signals (33 total). These conditions tested $\boldsymbol{H_1}$--$\boldsymbol{H_4}$ (identity effects) and provided observations for $\boldsymbol{H_5}$ (writing quality association).

\paragraph{Error-Injection Conditions ($\boldsymbol{H_6}$)}\label{error-injection-condition} To test whether surface-level writing errors affect summarization, we applied rule-based error injection to comments in the Low and Mid-Low quality quartiles. Each of these comments appeared in both original and error-added versions across all 32 identity conditions, creating matched pairs for within-comment comparison.

\noindent For all conditions, identity signals were inserted using a standardized format: \verb|"My name is [Name],| \verb|and I am a [Occupation],"| at the beginning, followed by the original comment text, concluded with \verb|"Sincerely, [Name]."|

\subsection{Experimental Manipulation}

\subsubsection{Name Selection} We selected two names per Race $\times$ Gender cell (Table~\ref{tab:names}), drawing first names from \citet{tzioumisDemographicAspectsFirst2018} and surnames from U.S. Census data \citep{u.s.censusbureauFrequentlyOccurringSurnames2010}, requiring greater than 80\% probability of correct racial classification. Full selection criteria and limitations of the two-name design in Appendix~\ref{app:design:name-selection}. Two names per cell enabled modeling name as a random effect to generalize beyond specific name choices; sensitivity analyses with name as a fixed effect are reported in Appendix~\ref{app:name-sensitivity}.

\begin{table}[!tb]
\centering
\caption{Name Stimuli by Demographic Category}
\begin{adjustbox}{max width=\textwidth}
\begin{tabular}{llll}
\toprule
\textbf{Race} & \textbf{Gender} & \textbf{Name 1} & \textbf{Name 2} \\
\midrule
White & Female & Mary Miller & Susan Murphy \\
White & Male & Michael Miller & John Murphy \\
Black & Female & Latoya Washington & Tamika Jefferson \\
Black & Male & Darnell Washington & Jermaine Jefferson \\
Hispanic & Female & Luz Garcia & Blanca Rodriguez \\
Hispanic & Male & Jose Garcia & Juan Rodriguez \\
Asian & Female & Phuong Nguyen & Thuy Kim \\
Asian & Male & Hung Nguyen & Minh Kim \\
\bottomrule
\end{tabular}
\end{adjustbox}
\par\medskip\noindent\small 
\textit{Note.} All names exceed 80\% racial classification probability except Jefferson (74.2\%), the second-highest Black surname after Washington (87.5\%). Asian names reflect Vietnamese/Korean origins (94--97\% classification probability). Full selection criteria and limitations of the two-name design in Appendix~\ref{app:design:name-selection}.
\end{table}

\subsubsection{Socioeconomic Status Operationalization} Socioeconomic status was operationalized through occupation rather than education to avoid confounding with policy expertise.  We used "financial analyst" (occupational prestige score: 67.31) as the high-SES occupation and "street vendor" (score: 22.04) as the low-SES occupation \citep{hughesOccupationalPrestigeStatus2024, hughesOccupationalPrestige2019}. This provided 45.3 points of separation across the scale's effective range of 14.5--88.91. We avoided occupations with greater prestige, such as physicians, lawyers, or professors, because these signal domain-specific expertise that could confound SES with perceived argument quality. Neither occupation has an obvious connection to the environmental and energy regulation tested in our study. Both occupations are high-frequency, widely recognized, and approximately gender-balanced (financial analyst: 42.7\% female; street vendor: 45.4\% female) \citep{u.s.bureauoflaborstatisticsEmployedPersonsDetailed2025}.

\subsubsection{Writing Quality Transformations}\label{sec:error-injection} To test whether surface-level writing errors affect summarization ($H_6$), we applied rule-based error injection to comments in the low and mid-low quality quartiles, creating matched pairs of original and error-added versions across all 32 identity conditions. Errors included misspellings, capitalization errors, article omissions, preposition substitutions, noun number disagreement, and verb form errors, applied at a rate of approximately 2--2.5 errors per 100 words—roughly half the rate documented among native English speakers \citep{wilcoxNatureErrorAdolescent2014}. This manipulation tested LLM sensitivity to surface-level mechanical errors specifically; it does not capture broader quality dimensions such as content depth or organizational clarity. Full error type definitions with citations, implementation details, and the rationale for applying identical injection across identity conditions appear in Appendix~\ref{app:design:error-injection-detail}.

\subsubsection{Quality Control and Blinding}
Quality control procedures addressed five areas: OCR accuracy validation, multi-layered identity redaction with human verification of a 10-15\% stratified sample, programmatic input verification confirming correct identity signal insertion, API call logging with retry protocols for failed calls, and output stability assessment via a 20-comment reliability subset processed five times each (Section~\ref{sec:reliability}). Condition order was randomized independently for each comment-model combination using a documented seed, and blinding was maintained throughout automated data collection and analysis. Full procedures are described in Appendix~\ref{app:design:quality-control}.

\subsection{Sampling}

We employed a hybrid sequential sampling approach: after automated filtering, comments were randomly shuffled and processed sequentially through text extraction, LLM-assisted classification, inclusion criteria check, and stratum assignment until all four strata (length $\times$ sophistication) reached their targets for each agency. Comments that became incoherent following redaction were excluded with reasons documented. The final sample and observation counts are reported in Appendix~\ref{app:descriptive-statistics}. Full sampling procedures, targets, and feasibility assessment are in Appendix~\ref{app:sampling}.

\subsubsection{Sampling and Stratification}

Comments were stratified by length (short vs.\ long, median split) and sophistication (substantive vs.\ non-substantive), yielding four strata. Classification used LLM-assisted coding with majority voting across three models (GPT-4o Mini, Claude Haiku 4.5, and Gemini 2.0 Flash Lite, all excluded from the summarization test set), validated by human verification (Cohen's $\kappa$ target $\geq 0.70$). Target sample sizes were 30 per stratum for DOI and 20 for EPA. We included English-language comments of at least 50 words that addressed substantive policy content. Near-duplicate submissions exceeding 95\% cosine similarity were reduced to a single representative instance. The pre-registered protocol specified excluding organizational comments; this exclusion was not applied during data collection because reliably distinguishing organizational from individual submissions proved impractical in an automated pipeline. Organizational identifiers were instead removed during redaction, substantially mitigating the original concern (See Section~\ref{sec:discussion:limitation}). Feasibility assessment procedures, full inclusion and exclusion criteria, deduplication details, and technical failure protocols are in Appendix~\ref{app:sampling}.

\subsubsection{Writing Quality Classification}\label{sec:writingquality}

We classified all sampled comments on writing quality across five dimensions drawn from linguistics and education research: Content, Organization, Language Use, Vocabulary, and Mechanics (Table~\ref{tab:writingquality}). Two LLMs (GPT-4o Mini and Claude Haiku 4.5, both excluded from the summarization test set) rated each comment on a 5-point scale with anchored descriptors (Appendix~\ref{app:writing-quality-descriptors}); comments with inter-rater disagreement $\geq 2$ on any dimension were adjudicated by two human raters, whose scores were averaged to produce the final rating. We computed an aggregate score (mean of five dimensions) and classified comments into quartiles (Low, Mid-Low, Mid-High, High). The continuous aggregate served as the primary measure for $\boldsymbol{H_5}$; the Low and Mid-Low quartiles identified comments eligible for error injection ($\boldsymbol{H_6}$). Full classification procedures, ICC targets, and adjudication protocols are in Appendix~\ref{app:sampling::writingquality}.

\begin{table}[ht]
\centering
\caption{Writing Quality Evaluation Dimensions}
\label{tab:writingquality}
\begin{adjustbox}{max width=\textwidth}
\begin{tabular}{p{2cm} p{3cm} p{4cm}}
\toprule
\textbf{Dimension} & \textbf{Description} & \textbf{Supporting Literature} \\
\midrule
Content & Clarity of position, relevance, depth of supporting points & \citet{kyleAutomaticallyAssessingLexical2015, kyleMeasuringSyntacticComplexity2018, kimEvaluatingDimensionalityFirst2014, engberRelationshipLexicalProficiency1995, uccelliMasteringAcademicLanguage2013, zhangRevisitingPredictivePower2022} \\
\midrule
Organization & Logical structure, coherence, progression, transitions & \citet{yangDifferentTopicsDifferent2015, kyleMeasuringSyntacticComplexity2018, uccelliMasteringAcademicLanguage2013, crossleyLinguisticFeaturesWriting2020, huangDoesProcessGenreApproach2020, zhangRevisitingPredictivePower2022, yangLinguisticPredictorsL22025, lauferVocabularySizeUse1995} \\
\midrule
Language Use & Grammatical accuracy, sentence variety and complexity & \citet{huangDoesProcessGenreApproach2020, barrotComplexityAccuracyFluency2021, yoonLinguisticDevelopmentStudents2017, yangDifferentTopicsDifferent2015, biberShouldWeUse2011, bulteConceptualizingMeasuringShortterm2014} \\
\midrule
Vocabulary & Lexical diversity, sophistication, range, appropriateness & \citet{kimEvaluatingDimensionalityFirst2014, crossleyLinguisticFeaturesWriting2020, barrotComplexityAccuracyFluency2021, xuNavigatingComplexityPlain2023, yangLinguisticPredictorsL22025, lauferVocabularySizeUse1995} \\
\midrule
Mechanics & Spelling, punctuation, capitalization & \citet{kimEvaluatingDimensionalityFirst2014, tsaiUsingGoogleTranslate2019, luCorpusBasedEvaluationSyntactic2011, napolesEnablingRobustGrammatical2019, glomo-narzolesWorkplaceEnglishLanguage2021, derakhshanInvestigationIranianEFL2020} \\
\bottomrule
\end{tabular}
\end{adjustbox}
\end{table}

\subsubsection{Identity Redaction}
Before experimental manipulation, we removed existing identity signals from original comments using layered detection: keyword-based regular expressions and named entity recognition identified candidate signals, GPT-4o Mini redacted or rephrased detected signals, and Claude Haiku 4.5 verified that no identity signals remained. Both models were excluded from the summarization test set. Targeted categories included personal names, occupational references, gender and family role terms, racial and ethnic identifiers, and geographic locations. Signature lines were deleted; identity references woven into argumentative substance were rephrased to preserve meaning. Although we initially planned to validate a 10--15\% sample, automated redaction required extensive correction, so all comments were manually reviewed and corrected as needed. Full redaction procedures, replacement strategies, and the exploratory redaction intensity analysis are in Appendix~\ref{app:sampling:identity-redaction}.

\subsubsection{Large Language Models and Prompts}\label{sampling:llm}

We tested eight LLMs from four providers, all available through FedRAMP-authorized cloud services via Ask Sage \citep{federalriskandauthorizationmanagementprogramfedrampFedRAMPMarketplace2026, asksageAPIEndpoints2026}: GPT-4 and GPT-4.1 (OpenAI), Claude 4 Sonnet and Claude 4 Opus (Anthropic), Gemini 2.5 Flash and Gemini 3.0 Flash (Google), and Llama 3 70B and Llama 3.3 70B (Meta, via Ollama). All models were run at temperature zero with a maximum output length of 1,024 tokens, well within every model's output capacity, to ensure summaries were not truncated. We used a minimal primary prompt ("Summarize this public comment") to isolate intrinsic differential treatment, with two alternative prompts (systematic and length-pressure) tested on a 25-comment robustness subset for all the hypotheses (Appendix~\ref{app:alternative-prompts}). We accessed commercially available model versions; government-specific deployments in AWS GovCloud or Azure Government environments may have different configurations, which represents a limitation (see Limitations). Model parameters, prompt specifications, robustness interpretation framework, and computational load details are in Appendix~\ref{app:sampling:llm} to Appendix~\ref{app:sampling:prompts}.

\subsection{Analysis}\label{analysis}

\subsubsection{Inference Framework}\label{sec:inference}

All confirmatory analyses employed both frequentist and Bayesian inference. Frequentist tests used $\alpha = .05$ (two-tailed) with Holm--Bonferroni correction. Bayes factors were approximated from BIC values of nested models fit via maximum likelihood \citep{wagenmakersPracticalSolutionPervasive2007}. We adopted three interpretation thresholds \citep{robertHaroldJeffreyssTheory2009}: $\mathrm{BF}_{10} \geq 3$ as evidence supporting the effect, $\mathrm{BF}_{10} \leq 1/3$ as evidence that failed to support the effect, and intermediate values as inconclusive. This dual approach distinguished between absence of evidence and evidence of absence. The formula and technical details are in Appendix~\ref{app:inference-detail}.

\subsubsection{Dependent Variables}\label{intro-analysis-dv}

We assessed four dependent variables computed as difference scores: for each comment and model, we subtracted the baseline summary measure (no identity signal) from the attributed summary measure, so that each gap directly captures how adding identity signals changed summarization.

\paragraph{Semantic Similarity ($\Delta_{\text{Sim}}$)} Cosine similarity between each summary and its original comment, computed using sentence embeddings from \texttt{sentence-transformers/all-mpnet-base-v2} \citep{reimersSentenceBERTSentenceEmbeddings2019}. $\Delta_{\text{Sim}}$ equals attributed minus baseline similarity to the original. Negative values indicate that adding identity reduces content preservation. This measure captures topical overlap but does not directly assess entailment or factual consistency; the measure whether summaries discuss the same content as originals rather than whether they are logically faithful to source claims.

\paragraph{Sentiment ($\Delta_{\text{Sent}}$)} Sentiment scores computed using \texttt{siebert/sentiment-roberta-large-english} \citep{hartmannMoreFeelingAccuracy2023}. $\Delta_{\text{Sent}}$ equals attributed minus baseline sentiment. Positive values indicate that adding identity makes summaries more positive.

\paragraph{Length Ratio ($\Delta_{\text{Len}}$)} Summary word count divided by original comment word count. $\Delta_{\text{Len}}$ equals attributed minus baseline length ratio. Negative values indicate greater compression when identity is added.

\paragraph{Readability ($\Delta_{\text{Read}}$)} Flesch-Kincaid Grade Level scores \citep{kincaidDerivationNewReadability1975}. $\Delta_{\text{Read}}$ equals attributed minus baseline grade level. Positive values indicate summaries written at a higher reading level; negative values indicate simpler language.

\noindent We analyzed each DV separately to preserve directional information and characterize how differential treatment manifests (Table~\ref{tab:dv-interpretation}). As a summary measure for the omnibus gate, we computed the Composite Index ($CI$) as the mean of absolute $z$-scored $\Delta$ values across all four DVs, with $z$-scores computed within each comment-model combination across conditions.

\begin{table}[htbp]
\centering
\caption{Normative Interpretation of Dependent Variables}\label{tab:dv-interpretation}
\begin{adjustbox}{width=\textwidth, center}
\begin{tabular}{llll}
\toprule
\textbf{DV} & \textbf{Positive $\Delta$ means...} & \textbf{Less desirable direction} & \textbf{Rationale} \\
\midrule
$\Delta_{\text{Sim}}$ & More content preserved & Negative & Less faithful to original \\
$\Delta_{\text{Sent}}$ & More positive tone & Context-dependent & Could favor or sanitize \\
$\Delta_{\text{Len}}$ & Longer summary & Context-dependent & Could be fuller or padded \\
$\Delta_{\text{Read}}$ & Higher reading level & Context-dependent & {Higher (low accessibility); lower (simplification)}\\
\bottomrule
\end{tabular}
\end{adjustbox}
\end{table}

\subsubsection{Analysis Sequence}

\paragraph{Identity Analysis ($\boldsymbol{H_1}$--$\boldsymbol{H_4}$)}\label{analysis:identity} Using the 32 identity conditions, we tested whether identity signals differentially affected summarization. The baseline condition was used to compute difference scores but excluded from the analysis sample. To control family-wise error rate, we employed hierarchical testing.

\subparagraph{Level 1: Omnibus Gate} We compared a full model (Race, Gender, SES, Race $\times$ Gender, and random intercepts for Comment, Model, and Name) against an intercept-only model on the Composite Index via likelihood ratio test. If $p \geq .05$, we concluded insufficient evidence for identity-based differential treatment and did not proceed to Level 2. Full model equations are in Appendix~\ref{app:statistical-models}.

\subparagraph{Level 2: Hypothesis Testing} For each hypothesis, we computed the $F$-test $p$-value on each DV and identified the minimum $p$-value across measures. We applied Holm-Bonferroni correction across the four hypothesis-level minimum $p$-values. For each hypothesis where corrected $p < .05$, the remaining three DVs were tested with Holm-Bonferroni correction within that hypothesis. Pairwise comparisons for significant effects were reported descriptively. Bayes factors were computed at each level regardless of $p$-values.

\paragraph{Writing Quality Analysis ($\boldsymbol{H_5}$--$\boldsymbol{H_6}$)}\label{analysis:writingquality}

$\boldsymbol{H_5}$ and $\boldsymbol{H_6}$ used raw scores rather than difference scores as dependent variables, reflecting a conceptual distinction: $\boldsymbol{H_1}$--$\boldsymbol{H_4}$ test whether identity signals affect summarization relative to baseline, whereas $\boldsymbol{H_5}$ and $\boldsymbol{H_6}$ ask whether writing quality predicts summarization across identity conditions.

\subparagraph{Writing Quality Association ($\boldsymbol{H_5}$)} We fit linear mixed-effects models for each raw DV with Race, Gender, SES, and WritingQuality (continuous aggregate) as fixed effects, testing WritingQuality via $F$-test with Holm-Bonferroni correction across the four DVs. As secondary analysis, we examined each dimension separately with Holm-Bonferroni correction across the five dimensions; quartile-based categorical analyses were reported for robustness.

\subparagraph{Error Injection Effect ($\boldsymbol{H_6}$)} Using comments in the Low and Mid-Low quality quartiles, we compared original and error-added versions via $F$-test for the ErrorAdded term with Holm-Bonferroni correction across the four DVs.

\subparagraph{Exploratory Interactions} We tested Race $\times$ WritingQuality, Gender $\times$ WritingQuality, and SES $\times$ WritingQuality interactions for $\boldsymbol{H_5}$, and parallel interactions with ErrorAdded for $\boldsymbol{H_6}$, reported descriptively.

\subsubsection{Statistical Model}

We used linear mixed-effects models implemented in R (\texttt{lme4} and \texttt{lmerTest}). For the identity analysis ($\boldsymbol{H_1}$--$\boldsymbol{H_4}$), models predicted each $\Delta$ DV from Race, Gender, SES, and Race $\times$ Gender as fixed effects, with random intercepts for Comment, Model, and Name nested within Race $\times$ Gender. All categorical predictors used dummy coding with White, Male, and High-SES as reference categories. Degrees of freedom were approximated using the Satterthwaite method \citep{lukeEvaluatingSignificanceLinear2017}. When models failed to converge, we simplified the random effects structure following \citet{barrRandomEffectsStructure2013}, documenting all attempted specifications. Full model equations, convergence protocol, effect size reporting conventions, assumption check procedures, and distributional contingencies are in Appendix~\ref{app:statistical-models}.

\subsubsection{Sensitivity and Exploratory Analyses}

To assess whether specific names drove observed effects, we re-ran primary models with name as a fixed rather than random effect and compared demographic coefficients. Exploratory analyses examined higher-order interactions (Race $\times$ SES, Gender $\times$ SES, Race $\times$ Gender $\times$ SES), comment characteristic moderators (stance, length, sophistication), model family effects, and prompt sensitivity across three prompt conditions. For the 20-comment reliability subset (5 repetitions each), we computed intraclass correlations for each DV, adopting ICC $\geq 0.70$ as the threshold for adequate reliability \citep{lebretonAnswers20Questions2008}. These analyses are reported in Appendices~\ref{app:sensitivity-robustness}--\ref{app:exploratory} and~\ref{app:reliability-analysis}.

\section{Results}

\subsection{Sample and Measurement Overview}

We retrieved 9,438 comments from two federal rulemaking dockets. After applying a multi-stage exclusion pipeline that included text extraction, deduplication to remove near-identical submissions, language filtering, and stratified sampling, 182 comments met all inclusion criteria. Of these, 102 originated from Department of the Interior proceedings governed by the National Environmental Policy Act (NEPA), and 80 from Environmental Protection Agency air quality regulation. Appendix~\ref{app:fig:consort_flow} details the full exclusion flow. Near-duplicate comments accounted for the largest share of exclusions (6,923), followed by comments falling below the 50-word minimum (498). The final sample fell short of our pre-registered target of 200 (120 DOI, 80 EPA) because the DOI docket contained an unexpectedly high duplication rate; the EPA docket met its target in full.

We assessed measurement reliability using intraclass correlation coefficients (ICCs) from five independent repetitions on a 20-comment subset. Three of four dependent variables showed adequate reliability: similarity ($\Delta_{\text{Sim}}$; ICC $= 0.882$), length ($\Delta_{\text{Len}}$; ICC $= 0.833$), and readability ($\Delta_{\text{Read}}$; ICC $= 0.890$). Sentiment ($\Delta_{\text{Sent}}$) fell below our pre-registered threshold of 0.70 (ICC $= 0.695$), and we interpret it with appropriate caution throughout. This lower reliability likely reflects the near-bimodal distribution of the underlying sentiment scores rather than true measurement instability (Appendix~\ref{app:dv-distributions} and~\ref{app:reliability-analysis}).

Assumption checks revealed no multicollinearity concerns (maximum generalized VIF $= 2.0$); residual diagnostics were consistent with the mathematical properties of the Composite Index measure (Appendix~\ref{app:assumption-checks}). The omnibus likelihood ratio test on the Composite Index was statistically significant ($\chi^{2}(8) = 61.17$, $p < .001$), permitting dependent-variable-specific testing per the pre-registered hierarchical procedure (Appendix~\ref{app:primary-identity}).

\subsection{Street Vendor Comments Were Summarized Less Faithfully, and Only Occupation Showed This Consistently}

Of the three demographic dimensions we tested (race, gender, and occupation), only occupation altered how LLMs summarized identical public comments. A comment attributed to a street vendor produced a measurably different summary than the same comment attributed to a financial analyst \citep{hughesOccupationalPrestigeStatus2024, hughesOccupationalPrestige2019}. This pattern held across all names, models, prompts, and both regulatory agencies.

\begin{figure}[htbp][!tb]
    \centering
    \includegraphics[width=\textwidth]{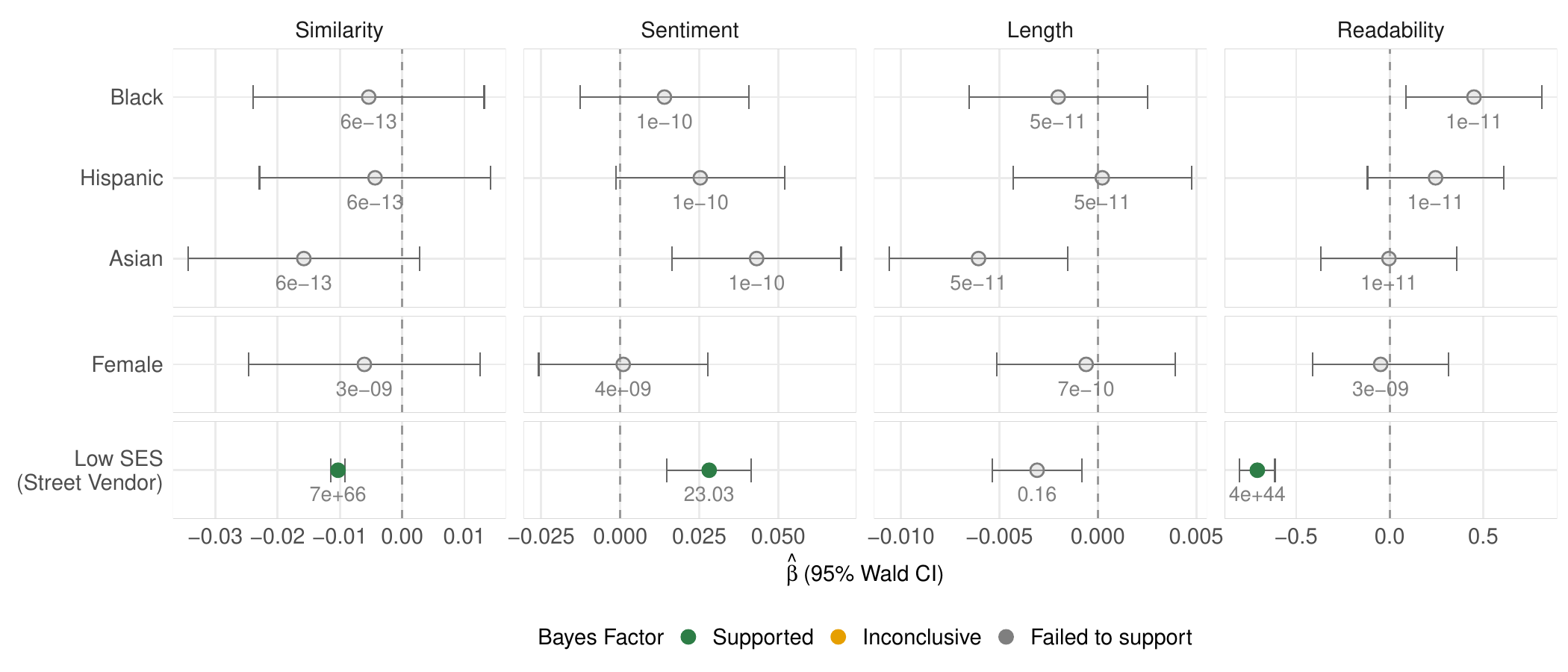}
    \caption{\footnotesize \textbf{Only Occupation Produced Supported Effects on Summarization Outcomes.} Each panel shows the estimated coefficient ($\hat{\beta}$) and 95\% Wald confidence interval for one identity signal (rows) on one summarization outcome (columns). The dashed line marks zero (no effect). Green points indicate effects that were both statistically significant and Bayesian-supported ($\text{BF}_{10} \geq 3$); amber points were statistically significant but Bayesian-inconclusive; gray points failed to support an effect ($\text{BF}_{10} \leq 1/3$). Race (Black, Hispanic, and Asian, each relative to White) and gender (Female relative to Male) produced no supported effects on any outcome, with all Bayes factors strongly favoring no effect. Only the Low SES row (street vendor relative to financial analyst) shows green points: street vendor-attributed comments preserved less of the original meaning (Similarity), shifted tone in a more positive direction (Sentiment), and were written at a lower reading level (Readability). The Length effect was statistically significant but Bayesian-inconclusive ($\text{BF}_{10} = 0.16$). Full results in Appendices~\ref{app:primary-level2} and~\ref{app:primary-holm-bonferroni-correction}.}
    \label{fig:fixed_effects_bf}
\end{figure}

After Holm-Bonferroni correction across the four hypothesis-level tests ($H_1$–$H_4$), occupation reached statistical significance on all four dependent variables (Appendix~\ref{app:primary-holm-bonferroni-correction}). Street vendor-attributed comments preserved less of the original meaning ($\Delta_{\text{Sim}}$: $p < .001$, $BF_{10} = 6.6 \times 10^{66}$), shifted sentiment in a more positive direction ($\Delta_{\text{Sent}}$: $p < .001$, $BF_{10} = 23.0$), compressed the text more ($\Delta_{\text{Len}}$: $p = .008$, $BF_{10} = 0.16$), and used simpler language ($\Delta_{\text{Read}}$: $p < .001$, $BF_{10} = 3.9 \times 10^{44}$). Negative values on similarity and length indicate less faithful and shorter summaries, while negative readability values indicate the model wrote at a lower reading level, meaning it used simpler words and sentences. This pattern is consistent with the LLM treating the street vendor's comment as warranting less careful reproduction. For length, the effect was statistically significant but too small for the Bayesian analysis to distinguish from the null (Figure~\ref{fig:fixed_effects_bf}; Appendix~\ref{app:primary-fixed-effects-effect-sizes-summary}).

In absolute terms, these effects were small: approximately 1\% less semantic similarity, 3\% sentiment shift, and 0.7 grade levels lower reading level. Because the models received only a single-sentence summarization instruction with no role assignment or behavioral constraints, these estimates reflect a narrow testing condition. Deployed systems with additional processing layers may produce different patterns.

\begin{figure}[htbp][!tb]
    \centering
    \includegraphics[width=\textwidth]{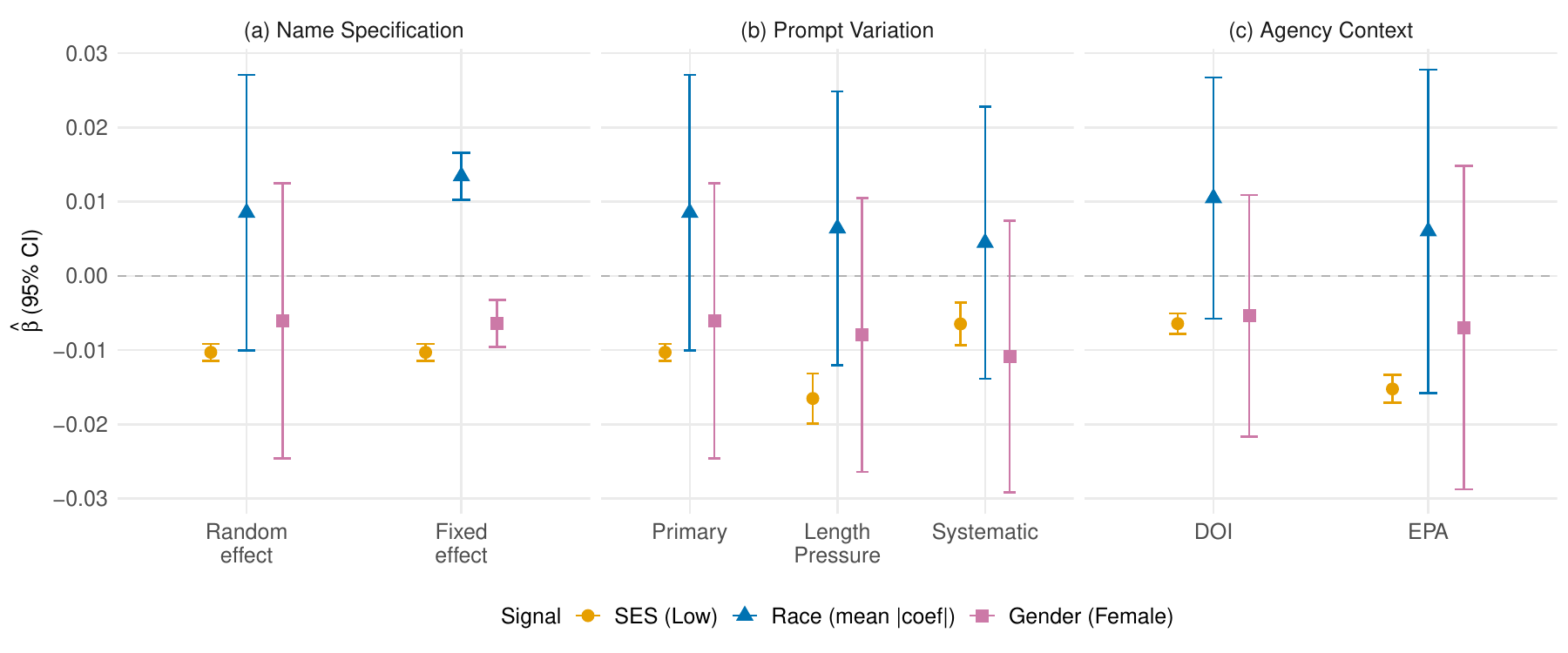}
    \caption{\footnotesize \textbf{The Occupation Effect Held Up Under Every Test; Race and Gender Did Not.} Each point shows the estimated coefficient for a identity signal on similarity ($\Delta_{\text{Sim}}$), re-estimated under different conditions. The vertical axis measures how much the signal changed the summary relative to baseline; the dashed line marks zero (no change). Three panels test robustness: (a) whether the effect depends on which specific names were used, (b) whether it holds under three different prompt instructions, and (c) whether it replicates across two regulatory agencies (DOI and EPA). Gold circles represent the occupation (SES) effect, where street vendor is measured relative to financial analyst. Blue triangles represent race, calculated as the mean absolute effect across three comparisons against the White reference group (Black, Hispanic, and Asian). Because we used absolute values, this measure is always positive. Pink squares represent gender (Female relative to Male). The occupation effect tells a stable story: gold circles cluster tightly around the same negative value in every condition, meaning that street vendor-attributed comments consistently received summaries that preserved less of the original meaning, regardless of which name carried the signal, which prompt was used, or which agency's comments were tested. Race and gender tell a different story: blue triangles shift position and cross zero across conditions, and pink squares hover near zero throughout. For example, the race coefficient nearly doubles when moving from panel (a) Random to Fixed, meaning the apparent race effect changed substantially depending on a single modeling choice. Full coefficient tables in Appendix~\ref{app:primary-identity}.}    
    \label{fig:stability_demographic_signal_coefficients}
    \end{figure}
\FloatBarrier

Four checks confirmed the robustness of this finding. First, the occupation effect was completely independent of which name carried it. Whether a comment was signed "Mary Miller, street vendor" or "Hung Nguyen, street vendor," the effect was identical; coefficients changed by 0\% across all four DVs when we tested each name individually (Figure~\ref{fig:stability_demographic_signal_coefficients}(a); Appendix~\ref{app:name-sensitivity}). Race and gender coefficients, by contrast, shifted by 40--3,000\% under the same test, a pattern we address below. Second, the occupation effect on similarity and readability replicated under both alternative prompt formulations ($p < .001$ for both); effects on sentiment and length were prompt-sensitive, meaning results varied depending on how we worded the summarization instruction (Figure~\ref{fig:stability_demographic_signal_coefficients}(b); Appendix~\ref{app:alternative-prompts}). Third, the occupation effect replicated across both regulatory contexts, DOI and EPA, on three of four DVs (Figure~\ref{fig:stability_demographic_signal_coefficients}(c); Appendix~\ref{app:cross-agency-comparison}). Fourth, an exploratory Race $\times$ SES interaction reached statistical significance on similarity ($F(3, 46320) = 10.43$, $p < .001$), suggesting the magnitude of occupational differential treatment may vary across racial groups, though this finding requires confirmatory replication (Appendix~\ref{app:higher-order-interactions}).

\subsection{Lower Writing Quality Led to Less Faithful Summaries, and Occupation Widened the Gap}

How well a comment was written shaped how it was summarized, but which aspects of writing mattered was not what one might expect. The strength of the argument (content), the clarity of its structure (organization), and the sophistication of its vocabulary (language use) all influenced how closely the summary tracked the original (similarity). Spelling and grammar (mechanics) did not. And when occupation labels accompanied lower-quality comments, exploratory evidence suggested the differences in treatment may have grown larger.

Writing quality, assessed as a composite index across five dimensions, was statistically significantly associated with three of four summarization outcomes ($p < .001$ for similarity, length, and readability; all $\text{BF}_{10} > 292$; Appendix~\ref{app:H5-writing-quality}). Higher-quality comments received summaries that preserved more of the original meaning, were more compressed relative to the original, and used more complex language. Sentiment was the exception: emotional tone did not vary with writing quality ($\text{BF}_{10} = 0.005$, strong evidence for no effect).

\begin{figure}[htbp][!tb]
    \centering
    \includegraphics[width=0.9\textwidth]{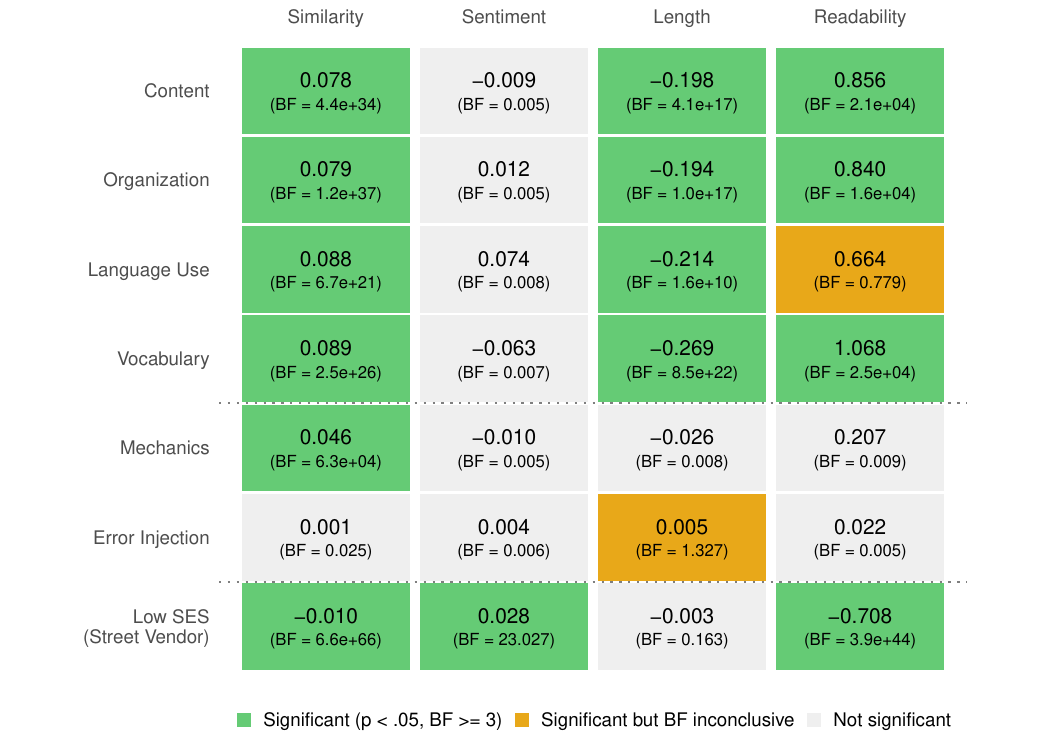}
    \caption{\footnotesize \textbf{LLMs Responded to Argument Substance, Not Surface Errors.} Rows represent text signals: the top four capture argument substance (what was argued and how), Mechanics and Error Injection capture surface quality (spelling, grammar), and the bottom row shows the occupation (SES) effect from Figure~\ref{fig:fixed_effects_bf} for comparison. Columns represent four summarization outcomes. Green indicates a supported effect; amber indicates inconclusive evidence; gray indicates no effect. The top four rows are predominantly green: stronger arguments received summaries closer to the original, more compressed, and at a higher reading level. The Mechanics and Error Injection rows are predominantly gray: spelling and grammar barely changed how summaries were produced. The low SES (street vendor) row at the bottom mirrors the substance pattern, showing green on the same three outcomes. This suggests that both argument quality and occupation influenced the same aspects of summarization, while neither affected sentiment. The Sentiment column is entirely gray except for occupation, the only signal that changed a summary's sentiment. Full results in Appendix~\ref{app:H5-writing-quality} and~\ref{app:h6-error-injection}.}    
    \label{fig:text_signal_effects_summarization_outcomes}
\end{figure}
\FloatBarrier

When we examined each dimension of writing quality separately, a clear pattern emerged (Figure~\ref{fig:text_signal_effects_summarization_outcomes}). The four dimensions related to argument substance, specifically what the commenter argued (content), how they structured it (organization), how they expressed it (language use), and what words they chose (vocabulary), each predicted similarity, length, and readability ($p < .001$ for all; all $\text{BF}_{10} > 10^{4}$; Appendix~\ref{app:h5-secondary-dimension-specific}). Mechanics, which captures spelling, punctuation, and capitalization, predicted only similarity and had no detectable effect on how models compressed or simplified summaries ($\text{BF}_{10} < 0.01$ for length and readability). LLMs appeared to respond to the quality of the argument, not the polish of the typing.

An experimental test pointed in the same direction. We introduced surface-level errors into lower-quality comments, including misspellings, capitalization changes, and grammatical article omissions, then measured whether these changes altered summarization outcomes. The effects were negligible. Only length reached statistical significance ($p = .003$), and even there the Bayesian evidence was inconclusive ($\text{BF}_{10} = 1.33$). The other three outcomes showed evidence for no effect (all $\text{BF}_{10} < 0.03$; Appendix~\ref{app:h6-primary-error-effect}). No demographic group was differentially affected by error injection (Appendix~\ref{app:h6-exploratory-interactions}). Both analyses pointed to the same conclusion: LLMs responded to argument substance, not surface-level errors.

\begin{figure}[htbp][!tb]
    \centering
    \includegraphics[width=\textwidth]{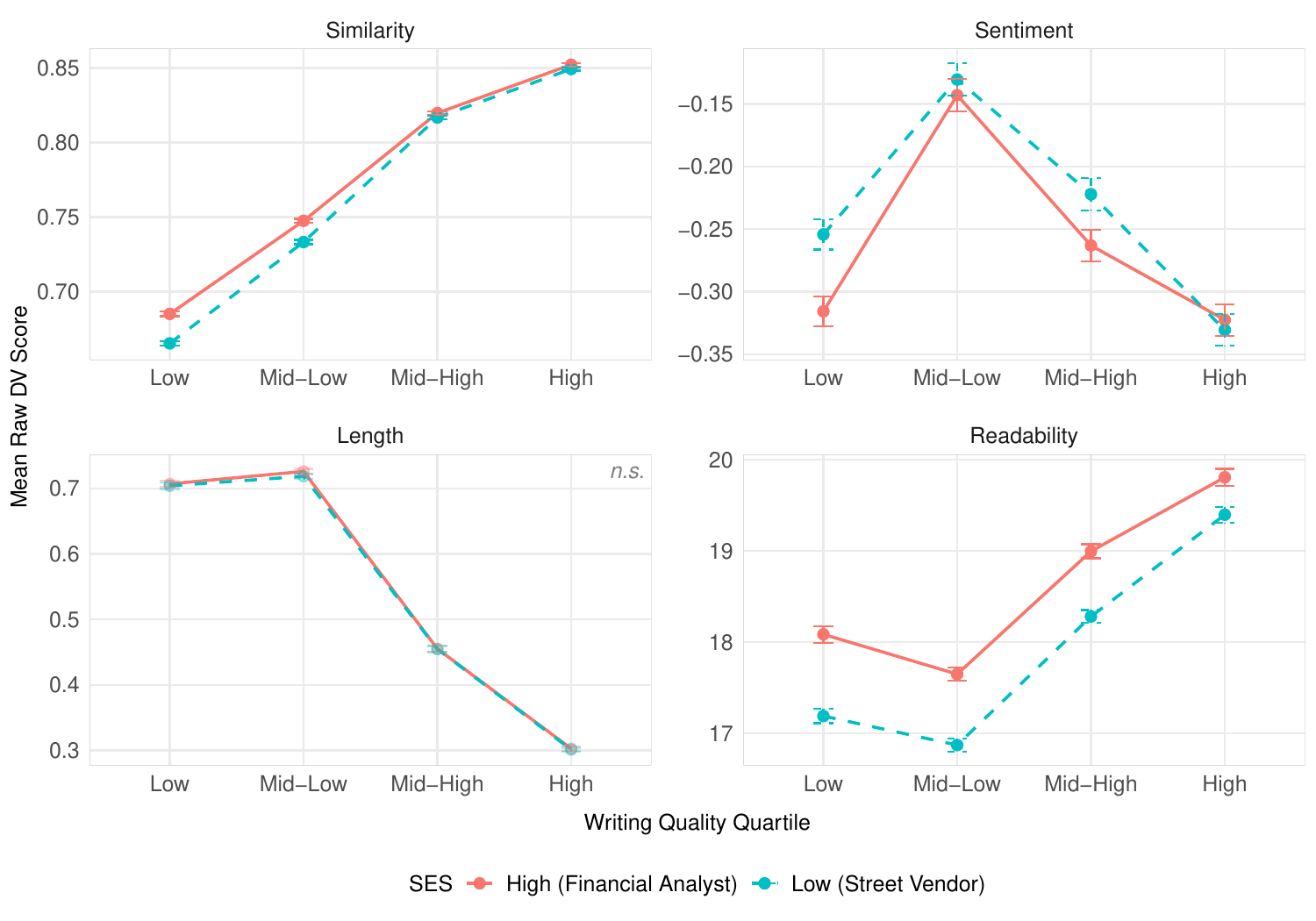}
    \caption{\footnotesize \textbf{The Occupation Gap May Widen for Lower-Quality Comments (Exploratory).} Each panel shows one summarization outcome across four writing quality quartiles (horizontal axis). The vertical axis shows the mean raw score for that outcome. Two lines compare comments attributed to a financial analyst (solid blue) and a street vendor (dashed red). The vertical gap between the lines at each quartile represents the occupation-based differential treatment for comments at that quality level. For Similarity, at the lowest writing quality quartile the street vendor line falls noticeably below the financial analyst line, meaning models preserved less of the original meaning for street vendor-attributed comments; this gap narrows at higher quality levels, where both lines converge. The Sentiment pattern shifted direction across quartiles rather than following a consistent trend. Length Ratio showed no interaction between occupation and writing quality, so the panel is dimmed to indicate the result was not statistically significant. Readability shows a similar pattern to Similarity: street vendor-attributed comments were consistently summarized in simpler language, but the gap was largest for lower-quality comments. All results are exploratory and require confirmatory replication. Occupation reflects experimentally attributed labels, not commenters' actual socioeconomic status. Error bars = $\pm 1$ SE. Full output in Appendix~\ref{app:h5-exploratory-interactions}.}    
    \label{fig:ses_writing_quality_interaction_summarization_outcomes}
\end{figure}
\FloatBarrier

Exploratory analysis raised a further possibility: occupation signals and writing quality may interact (Figure~\ref{fig:ses_writing_quality_interaction_summarization_outcomes}). For comments scoring higher on writing quality, the gap between financial analyst and street vendor attributions tended to be smaller. For lower-quality comments, this gap appeared to widen: summaries attributed to a street vendor were somewhat less faithful to the original and written at a somewhat simpler reading level than summaries of the same comment attributed to a financial analyst ($p < .001$ for similarity and readability; Appendix~\ref{app:h5-exploratory-interactions}). The sentiment interaction, while statistically significant ($p = .001$), shifted direction across quality levels rather than following a consistent trend. Length showed no interaction ($p = .318$). These patterns are exploratory and would need to be confirmed in a dedicated study before drawing firm conclusions. An important caveat applies: occupation in our design reflects the experimentally attributed label ("financial analyst" versus "street vendor"), not commenters' actual socioeconomic status. The interaction describes how models appeared to respond to the combination of an occupational signal and textual properties, not a claim about how different populations write.

Qualitative inspection illustrated the range of these differences. At the point where occupation-related differential treatment reached its median value (similarity; $\Delta_{\text{Sim}} \approx -0.033$), differences were subtle, involving minor rephrasing and slight softening of hedging words (Appendix~\ref{case-sim-median}). In extreme cases, models appeared to dismiss substantive regulatory content when it was attributed to a street vendor, characterizing identical text as lacking substantive feedback that the same models summarized faithfully without any occupation label. For example, one model summarized an EPA comment in detail under the baseline condition but, when the same text was attributed to a street vendor, reduced it to: "No further substantive feedback was provided" (Appendix~\ref{case-sim-extreme}).

\subsection{Race and Gender Effects Were Inconsistent, but Model Choice Mattered}

Race and gender, the dimensions most commonly examined in AI fairness research, did not produce consistent differential treatment in our study. Race effects appeared in some tests but dissolved when examined more closely. Gender effects were absent throughout. What did vary consistently was which model was used: the magnitude of occupation-based differential treatment differed substantially across providers.

Race reached statistical significance on two of four outcomes, specifically sentiment and length (both $p \leq .001$), but not on similarity or readability (Appendix~\ref{app:primary-holm-bonferroni-correction}). Bayesian analysis failed to support a race effect on any outcome (all $\text{BF}_{10} < 10^{-10}$; Figure~\ref{fig:fixed_effects_bf}). In practical terms, the detected shifts in sentiment and length were so small that the Bayesian framework could not distinguish them from no effect at all.

\begin{figure}[htbp][!tb]
    \centering
    \includegraphics[width=0.9\textwidth]{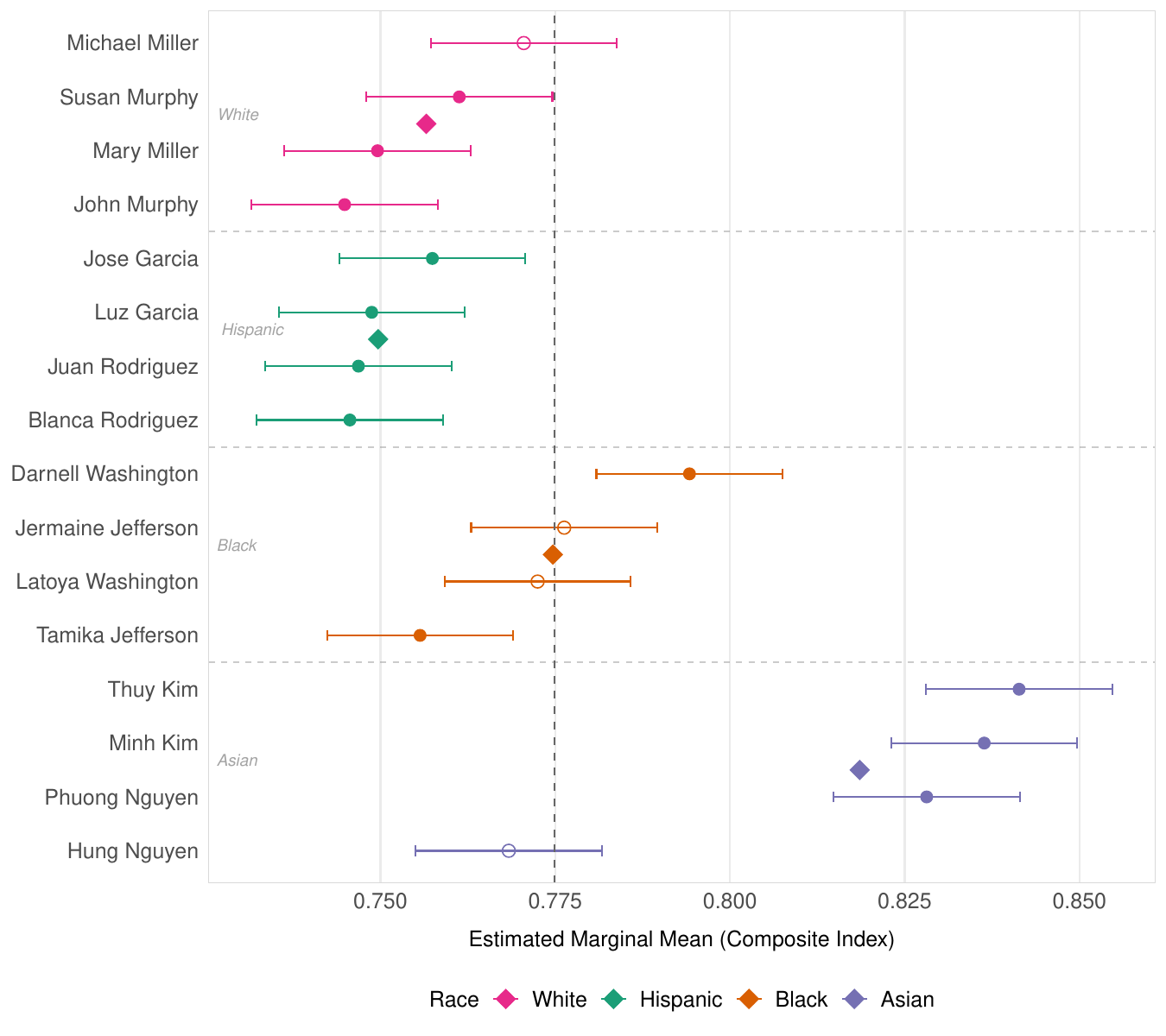}
    \caption{\footnotesize \textbf{Name-Level Variation Was as Large as Race-Level Variation.} Each point represents one of the 16 names used in the study, showing how much that name's summaries deviated from baseline overall. Names are grouped by racial category, with diamonds marking race-group averages and the dashed vertical line marking the grand mean. Filled circles are statistically significant; open circles are not. The key finding is the spread within each group. Among Black-associated names, Darnell Washington produced substantially more deviation than Tamika Jefferson. Among Asian-associated names, three names fell well above the grand mean while Hung Nguyen fell near it. A study using only one of these names as its racial proxy would have reached a different conclusion depending on which name was chosen. This within-race variation suggests that the detected race effects reflected responses to specific names rather than to racial categories. Full results in Appendix~\ref{app:name-sensitivity}.}
    \label{fig:name_specific_effects_composite_index_racial_group}
\end{figure}

Several lines of evidence suggest these race effects reflected responses to specific names rather than to racial categories. When we tested whether the effects depended on which particular names carried the racial signal, race coefficients shifted by 40--3,000\%, compared to 0\% for occupation (Figure~\ref{fig:stability_demographic_signal_coefficients}a). For example, among the four Asian-associated names, Thuy Kim, Minh Kim, and Phuong Nguyen all produced composite index values well above the grand mean, while Hung Nguyen fell near it. This means a study using only Hung Nguyen as the Asian name proxy would have reached a different conclusion than one using Thuy Kim (Figure~\ref{fig:name_specific_effects_composite_index_racial_group}). Similarly, among Black-associated names, Darnell Washington fell above the grand mean while Tamika Jefferson fell below. Race effects also did not replicate under either alternative prompt (Figure~\ref{fig:stability_demographic_signal_coefficients}b; Appendix~\ref{app:alternative-prompts}). This combination of name-specificity, within-race heterogeneity, and prompt sensitivity suggests that what appeared as racial effects may instead reflect how models respond to particular name tokens, the individual words a model processes from a name. For instance, "Washington" may activate associations with the place or historical figure rather than with racial identity. We note, however, that our design included only two names per racial category, and a broader sample might reveal systematic patterns our study lacked the power to detect.

\FloatBarrier

Gender showed no effect on any outcome (all $\text{BF}_{10} < 10^{-8}$; Figure~\ref{fig:fixed_effects_bf}). A comment attributed to Mary Miller and the same comment attributed to Michael Miller produced no detectable difference in summarization, and this held across all models, prompts, and agency contexts. The Race $\times$ Gender interaction was likewise null (all $\text{BF}_{10} < 10^{-6}$; Appendix~\ref{app:primary-level2}).

\begin{figure}[htbp][!tb]
    \centering
    \includegraphics[width=\textwidth]{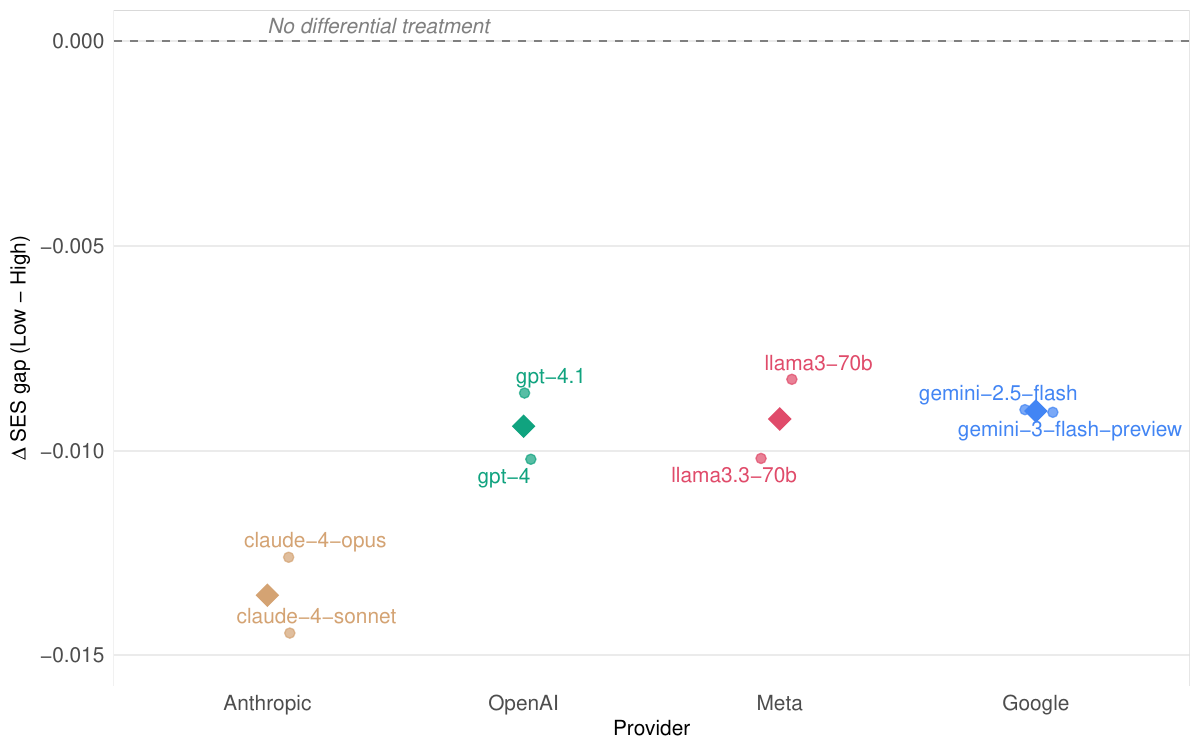}
    \caption{\footnotesize \textbf{All Models Showed Occupation-Based Differential Treatment, but the Magnitude Varied by Provider.} The horizontal axis groups models by provider; the vertical axis shows the occupation (SES) gap on similarity, meaning the difference in how faithfully a comment was summarized when attributed to a street vendor versus a financial analyst. The dashed line marks zero (no differential treatment). Large diamonds show provider averages; small circles show individual models. Every point falls below zero, meaning every model preserved less of the original meaning when the comment was attributed to a street vendor. The size of this gap differed, however: Anthropic's Claude models showed the largest occupation gap (roughly 1.3\% less faithful for street vendor attributions), while Google's Gemini models showed the smallest (roughly 0.9\%). An agency choosing between these providers would, without knowing it, also be choosing a different level of occupation-based differential treatment. The provider $\times$ occupation interaction was statistically significant ($p < .001$). Full per-model results in Appendix~\ref{app:model-family-effects}.}    
    \label{fig:provider_differences_ses_differential_treatment_similarity}
\end{figure}
\FloatBarrier

Although race and gender effects were inconsistent, the magnitude of occupation-based differential treatment varied meaningfully by model provider ($p = .017$ on similarity, $p < .001$ on composite index; Appendix~\ref{app:model-family-effects}). As Figure~\ref{fig:provider_differences_ses_differential_treatment_similarity} shows, all eight models produced less faithful summaries for street vendor-attributed comments. No model was free of occupation-based differential treatment, but the size of this gap differed. For example, Google's Gemini models showed the smallest occupation gap on similarity, while Anthropic's Claude models showed the largest. This means that an agency choosing between these providers would, without knowing it, also be choosing a different level of occupation-based differential treatment. Gemini 2.5 Flash also showed the lowest measurement reliability (ICC $= 0.656$ on similarity; Appendix~\ref{app:model-specific-icc}), suggesting that some provider-level variation may reflect measurement properties as well as genuine behavioral differences. The provider $\times$ race interaction was also statistically significant ($p < .001$; Appendix~\ref{app:model-family-effects}), suggesting that the name-specific effects described above may concentrate in particular models rather than appearing uniformly across all systems.

\section{Discussion}

We asked whether base LLMs produce different summaries when commenter characteristics vary. For occupation, the answer was consistently yes. For race and gender, it was not. Across 182 comments processed under 32 identity conditions by 8 models, occupation was the only signal to produce differential treatment that held across names, prompts, models, and regulatory contexts. Writing quality predicted summarization outcomes through argument substance rather than surface mechanics, and experimentally injected errors had negligible effects. Race effects, where detected, appeared name-specific rather than categorical, though our limited name sample means we cannot definitively rule out broader patterns. Gender effects were absent throughout. Provider-level variation in occupation-based differential treatment was statistically significant, suggesting that model selection may carry fairness implications that current procurement frameworks do not evaluate.

\subsection{Why Occupation, Not Race or Gender?}

We tested whether race, gender, and occupation would produce differential treatment. Only occupation did. Neither race nor gender showed consistent effects, a result that diverges from the existing literature's primary focus on these two dimensions~\citep{anMeasuringGenderRacial2025, wilsonGenderRaceIntersectional2025, fangBiasAIgeneratedContent2023}. Prior work on occupation-related LLM behavior has found context-dependent patterns: clinical settings showed models favoring higher-income patients~\citep{arzaghiUnderstandingIntrinsicSocioeconomic2025, omarSociodemographicBiasesMedical2025}, while simulated college admissions showed the reverse~\citep{nghiemRichDadPoor2025}. Our findings are consistent with the clinical pattern. In regulatory summarization, higher socioeconomic status attribution was associated with more faithful content preservation, but our study extends this finding into a democratic participation context where equal treatment carries distinct normative weight.

The occupation effect was notably specific. As we reported in the results, the effect was completely independent of which name carried it; coefficients changed by 0\% across all outcomes when we tested each name individually. This means two words in a signature block, "financial analyst" versus "street vendor," drove the differential treatment. We deliberately selected occupations with no obvious connection to environmental regulation~\citep{hughesOccupationalPrestigeStatus2024}, yet differential treatment persisted. One interpretation is that LLMs have learned to associate professional status with argument credibility. An alternative is distributional: the large text collections models learn from may simply contain more examples of financial analysts engaging with regulatory content than street vendors. These statistical associations could influence summarization without anything analogous to a credibility judgment. Our design cannot distinguish these explanations, and both may operate simultaneously.

Post-training alignment warrants consideration. This process adjusts model behavior after initial training to meet safety and fairness goals, and LLM providers invest substantially in it. Specifically, providers mitigate race and gender bias through reinforcement learning from human feedback (RLHF), adversarial stress-testing known as red teaming, and hard-coded refusals governed by the provider's own content policies~\citep{ouyangTrainingLanguageModels2022a, baiConstitutionalAIHarmlessness2022, hurst2024gpt}. Provider documentation indicates that SES does appear in some evaluation suites. For example, BBQ (Bias Benchmark for QA), a standardized test measuring social biases in question-answering, includes SES dimensions~\citep{parrishBBQHandbuiltBias2022}. Google's Gemini evaluations separately incorporate Dollar Street, a dataset of household photographs across income levels, for socioeconomic contexts~\citep{team2024gemini, gemma_2024}. SES coverage nonetheless appears, to our knowledge, thinner than coverage for gender and race in headline fairness evaluations. Whether the persistence of occupation-based differential treatment reflects a gap in alignment efforts, the difficulty of mitigating occupational associations that pervade training corpora, or the functional nature of status-based associations in the data is a question our findings raise but cannot resolve.

\subsection{What Commenters Argued Mattered More Than How They Typed}

Two independent analyses pointed to the same conclusion: LLMs responded to the substance of comments, not their surface polish. The writing quality dimension-specific analysis found that mechanics, which captures spelling, punctuation, and capitalization, did not predict how models compressed or simplified summaries, while content, organization, vocabulary, and language use all did. The error injection experiment reinforced this finding: deliberately introducing misspellings and grammatical errors into comments produced negligible changes in summarization outcomes. This pattern is broadly consistent with prior work finding that LLMs are relatively resilient to grammatical errors~\citep{wangResilienceLargeLanguage2024}, though our findings extend this to a policy-relevant task with naturally occurring quality variation.

For public comment processing, this is cautiously reassuring. For example, a comment submitted hastily from a mobile device is unlikely to receive systematically different treatment on the basis of surface errors alone. A similar pattern would be expected for comments written by non-native English speakers or by those with limited formal writing experience. However, the exploratory interaction between occupation and writing quality tempers this reassurance. For lower-quality comments, the gap between a financial analyst attribution and a street vendor attribution appeared to widen. Summaries of street vendor-attributed comments with weaker arguments were less faithful and used simpler language than either signal alone would predict. This interaction is exploratory and requires confirmatory replication. It still raises the possibility that occupational signals and textual properties compound in ways that disadvantage some commenters more than others.

\subsection{What the Absence of Race and Gender Effects May Mean}

Gender showed no effect in any analysis. This consistent null contrasts with prior work detecting gender effects in resume screening\citep{anMeasuringGenderRacial2025}, clinical contexts~\citep{omarSociodemographicBiasesMedical2025}, and deliberate attempts to elicit biased outputs~\citep{kumarDecodingBiasesAnalysis2025}. Two explanations seem plausible. First, gender stereotypes may activate more readily for tasks requiring judgment, such as scoring, ranking, or recommending, than for extractive tasks like summarization where the role is content preservation. Second, providers' sustained focus on gender bias mitigation may have effectively neutralized gender-based associations in this type of task. Our design cannot distinguish these explanations, and both may operate simultaneously.

Race effects appeared in some frequentist tests but were not supported by Bayesian analysis, which we used to distinguish between absence of evidence and evidence of absence. Race coefficients also shifted dramatically depending on which names were used and failed to replicate across prompts. The most parsimonious interpretation is that models responded to specific name tokens rather than to racial categories. Names carry particular word-level associations in training data. For example, "Washington" may activate connections to the place or historical figure, and "Nguyen" may activate associations specific to Vietnamese-American contexts. Neither reduces cleanly to a racial category effect. We note that our design tested only two names per category; a broader sample might reveal systematic patterns we could not detect.

Our results suggest that even studies using multiple names per racial group may reach misleading conclusions if they do not test whether effects generalize across those names or concentrate in specific ones. The name sensitivity analysis we employed, comparing coefficients when names are modeled collectively versus individually, offers one approach for distinguishing category-level effects from name-specific associations in future audit studies.

\subsection{Model Selection as a Governance Decision}

Every model we tested produced less faithful summaries for street vendor-attributed comments, but the magnitude of this gap varied substantially across providers. This variation is practically consequential because federal agencies already choose among these providers through existing procurement processes. FedRAMP certification, the primary framework governing federal cloud service adoption, evaluates data security and system reliability~\citep{officeofmanagementandbudgetModernizingFederalRisk2024}. It does not assess differential treatment in tasks like summarization that rely on these systems. An agency selecting between providers may, in effect, also be selecting a different level of occupation-based differential treatment. The experimental design we used in this study is relatively straightforward to implement. We processed identical comments under varying identity signals and compared summary characteristics. This approach could serve as one component of a fairness evaluation alongside existing security certifications. The appropriate thresholds for acceptable differential treatment and the tasks such evaluations should cover remain open questions for further research and policy deliberation.

\subsection{Limitations}\label{sec:discussion:limitation}

Several limitations temper these findings. Our experimental manipulation was deliberately minimal, consisting of a name and occupational label in a brief introduction and signature line. Real-world comments often contain richer identity signals throughout the text, including professional affiliations and cultural references that commenters use to establish credibility. Our estimates may therefore represent a lower bound of differential treatment in practice, though richer signals would not necessarily produce proportionally larger effects.

We tested base models under standardized conditions, not the government-specific deployments that providers offer through authorized cloud environments. Deployed systems may incorporate prompt engineering, retrieval-augmented generation, or fine-tuning that could amplify or attenuate the patterns we observed. We also do not know how government officials actually use LLMs when processing comments. Workflows likely vary across agencies and individual staff, and our design tested one comment at a time rather than the batch processing that agencies may employ. Additionally, we tested a single processing step—summarization—in isolation. Agency workflows may chain multiple LLM operations, and small differential treatment at each stage could compound across the pipeline. Whether the effects documented here accumulate or attenuate through multi-step processing remains unknown.

Our dependent variables relied on automated measurement, specifically embedding similarity, sentiment classification, word counts, and readability formulas, rather than human judgment of summary quality. These measures approximate but do not directly assess whether substantive arguments were preserved or distorted. These measures also cannot account for how officials' own reading preferences or biases may shape their interpretation of LLM-generated summaries.

The occupation manipulation compared two extremes, financial analyst and street vendor, representing points near the top and bottom of the occupational prestige continuum. Differential treatment may not scale linearly; intermediate occupations could produce different patterns that our two-point comparison cannot capture. Future work should test whether differential treatment scales with occupational prestige by employing a graduated design spanning multiple points along the prestige continuum, which would clarify whether the effect reflects a categorical response to low-status signals or a continuous association with perceived socioeconomic standing.

The final sample ($N = 182$) fell short of our pre-registered target of 200 due to high duplication rates in the DOI docket. Sentiment ($\Delta_{\text{Sent}}$) showed lower measurement reliability (ICC $= 0.695$, below our pre-registered threshold of 0.70) than the other dependent variables, and we interpret sentiment results accordingly.

We revised our pre-registered interpretation of readability ($\Delta_{\text{Read}}$) after observing that the occupation effect operated in the negative direction. Models produced summaries at a lower reading level for street vendor-attributed comments. Our previous framework had designated positive values (higher reading level) as less desirable. We revised the interpretation to context-dependent, recognizing that both directions represent differential treatment. This revision does not affect any statistical test, as all tests were two-tailed, but readers should weigh this post hoc adjustment.

Three protocol deviations occurred. First, one pre-registered model (Gemini 2.0 Flash) was retired during data collection and replaced with Gemini 3 Flash Preview. Second, we did not apply the pre-registered exclusion of organizational comments because reliably distinguishing organizational from individual submissions proved impractical in an automated pipeline. We instead removed organizational identifiers during redaction, including entity names, tribal affiliations, and geographic markers. While LLMs did not encounter incongruous individual-organizational pairings, organizational comments may exhibit distinctive writing styles that interact with occupation signals differently than individual submissions. Third, we computed Bayes factors using maximum likelihood rather than restricted maximum likelihood as originally registered, because valid BIC-based model comparison requires models estimated under maximum likelihood when fixed-effect structures differ.

\subsection{Implications for Practice}

Our findings point to several practical considerations for agencies deploying LLMs in public comment processing. First, the experimental design we used here could serve as a template for occupation-based fairness testing alongside existing security certifications. We processed identical comments under varying identity signals and compared summary characteristics. This approach is relatively straightforward to implement. The specific thresholds for acceptable differential treatment would need to be established through further research and policy deliberation.

Second, prompt design appeared to influence the extent of differential treatment (See Appendices~\ref{app:sampling:prompts} and~\ref{app:alternative-prompts}). Some occupation effects, particularly on sentiment and length, did not consistently replicate across prompt formulations. Systematic investigation of how prompt choices interact with occupation signals could identify low-cost mitigation strategies.

Third, as we discussed, the provider-level variation we documented means that agencies choosing among models are implicitly choosing a level of occupation-based differential treatment. Incorporating fairness benchmarks into existing procurement frameworks would not require fundamentally new infrastructure.

Fourth, agencies likely do not need to preprocess comments for spelling or grammar before LLM summarization. Surface-level errors did not meaningfully change summarization outcomes in our study, suggesting that high-volume pipelines can skip this step without affecting summary quality.

\clearpage


\bibliographystyle{plainnat}
\bibliography{references}

\clearpage
\appendix

\section{Design}

\subsection{Name Selection}\label{app:design:name-selection} Names were selected following established procedures for demographic research. First names were drawn from the validated database compiled by \citet{tzioumisDemographicAspectsFirst2018}, selecting names with the highest percentage of bearers belonging to the target racial group. Surnames were selected from U.S. Census data \citep{u.s.censusbureauFrequentlyOccurringSurnames2010}, using Census categories to identify surnames with the highest percentage for each target racial group.

We required greater than 80\% probability of correct racial classification for name selection. One exception applied: we included Jefferson (74.2\%), the second-highest Black-associated surname after Washington (87.5\%), because no other Black surnames exceed 75\%. Within the set of names meeting this threshold, we prioritized by frequency to select commonly occurring names.

These population composition statistics (for example, "85\% of people named 'Connor' are White") are related to but distinct from perceiver accuracy ("a perceiver shown 'Connor' will guess White 85\% of the time"). Population composition served as our selection criterion, providing a reasonable proxy for how identity signals may be encoded in LLM training data. The selected names appear in Table~\ref{tab:names}.

\begin{table}[htbp]
\centering
\caption{Name Stimuli by Demographic Category}\label{tab:names}
\begin{adjustbox}{max width=\textwidth}
\begin{tabular}{llll}
\toprule
\textbf{Race} & \textbf{Gender} & \textbf{Name 1} & \textbf{Name 2} \\
\midrule
White & Female & Mary Miller & Susan Murphy \\
White & Male & Michael Miller & John Murphy \\
Black & Female & Latoya Washington & Tamika Jefferson \\
Black & Male & Darnell Washington & Jermaine Jefferson \\
Hispanic & Female & Luz Garcia & Blanca Rodriguez \\
Hispanic & Male & Jose Garcia & Juan Rodriguez \\
Asian & Female & Phuong Nguyen & Thuy Kim \\
Asian & Male & Hung Nguyen & Minh Kim \\
\bottomrule
\end{tabular}
\end{adjustbox}
\par\medskip\noindent\small 
\textit{Note.} First names from \citet{tzioumisDemographicAspectsFirst2018}; surnames from U.S. Census Bureau \citep{u.s.censusbureauFrequentlyOccurringSurnames2010}. All names exceed 80\% racial classification probability except Jefferson (74.2\%), the second-highest Black surname after Washington (87.5\%). Asian names are distinctively Asian (94--97\%) but reflect Vietnamese/Korean origins; findings generalize to high-probability Asian name signals rather than specific subgroups.
\end{table}

Using two names per Race $\times$ Gender cell allowed us to model name as a random effect, enabling generalization of findings beyond the specific names selected. We acknowledge that with only 16 names total, variance estimates for the name random effect may be imprecise. Methodologists often recommend 30--50 levels for reliable variance component estimation. To address this limitation, we reported sensitivity analyses with name as a fixed effect in supplementary materials. If demographic main effect coefficients change substantially (more than 10--15\%) or any individual name coefficients appear extreme relative to the cell mean, this would suggest that name-specific effects warrant further investigation.

\subsection{Writing Quality Transformation}\label{app:design:error-injection-detail} We selected comments classified in the low and mid-low quality quartiles based on LLM evaluation scores (see Section~\ref{sec:writingquality}), then applied rule-based error injection to create matched pairs of original and error-added versions. This approach allowed both observational comparison across naturally occurring quality levels and experimental manipulation through controlled error injection.

Transformations were implemented via custom Python code applying deterministic rules for reproducibility. Error types reflected common patterns documented in writing quality research, including errors observed among both native and non-native English writers. These included:
\begin{itemize}
    \item \textbf{Spelling errors:} Common misspellings or typos ("enviroment", "recieve", "goverment") \citep{wilcoxNatureErrorAdolescent2014, ahmedNotSoSimpleViewWriting2022, crossleyUsingHumanJudgments2019, huangDoesProcessGenreApproach2020}
    \item \textbf{Capitalization errors:} Missing capitalization at sentence start ("the policy is important") or improper capitalization of common nouns ("the Government should act") \citep{wilcoxNatureErrorAdolescent2014, crossleyUsingHumanJudgments2019}    
    \item \textbf{Article errors:} Omission of required articles ("I went to store") or inappropriate insertion ("I visited the California") \citep{crossleyUsingHumanJudgments2019}
    \item \textbf{Preposition substitutions:} Common confusions such as "in/on/at" or "to/for" \citep{lauferVocabularySizeUse1995, yoonLinguisticDevelopmentStudents2017}
    \item \textbf{Noun number disagreement:} Missing plural marking ("three comment") or overgeneralization ("informations") \citep{huangDoesProcessGenreApproach2020}
    \item \textbf{Verb form errors:} Tense inconsistency or missing inflection ("he go yesterday") \citep{tayeIdentifyingAnalyzingCommon2024, wilcoxNatureErrorAdolescent2014}
\end{itemize}

Transformations were applied at a rate of approximately 2--2.5 errors per 100 words. Research indicates that native English speakers produce approximately 4 errors per 100 words, while second language writers average approximately 7 errors per 100 words \citep{wilcoxNatureErrorAdolescent2014}. Our rate represents approximately half that observed in native speakers, a conservative choice to maintain full comprehensibility. Transformed comments underwent manual review to verify naturalness and comprehensibility.

We applied identical error injection across all 32 identity conditions rather than identity-specific patterns. Our focus was on whether LLMs respond differently to surface-level error markers, not on modeling specific writing patterns of particular groups. Exploratory analyses examined whether error-added effects varied by commenter identity through interaction terms.

We acknowledge that rule-based error injection captures only surface-level errors, not broader quality dimensions such as content depth, organizational clarity, or vocabulary sophistication. Actual writing quality variation may also involve differences in argumentation style or rhetorical conventions. Our manipulation therefore tested LLM sensitivity to surface-level grammatical markers specifically, which may underestimate or differ qualitatively from differential treatment associated with more holistic writing quality patterns.

\subsection{Quality Control Procedures}\label{app:design:quality-control}

We implemented five quality control measures throughout data collection.

For OCR quality, documents processed via optical character recognition were evaluated using word-level confidence scores. Outputs with mean confidence below 85\% underwent manual verification and correction, with the number of corrected documents reported.

For redaction validation, each comment was processed through multiple LLMs for identity signal detection, followed by human review. A stratified random sample of 10--15\% of redacted comments underwent independent manual verification to confirm no identity signals remained.

For input verification, all API calls were logged in real-time, with the complete input stored alongside each output. We programmatically verified that identity signals appeared correctly in logged inputs before analysis.

For API success, we confirmed each call returned a valid response without error codes. Failed calls were retried up to three times with exponential backoff, meaning progressively longer wait times between retries. Calls that failed after three attempts were logged with error codes and failure reasons, then excluded from analysis. We inspected failure patterns to determine whether failures were random or systematic. Random failures, such as transient network issues, were simply excluded. If failures appeared systematic—for example, disproportionately affecting certain demographic conditions or comment types—we reported this pattern and assessed whether it could affect results. 

For output stability, a randomly selected reliability subset of 20 comments (10\% of sample) was processed five times each. We computed intraclass correlations for each dependent variable to assess measurement stability (see Section~\ref{sec:reliability}).

The order of conditions was randomized independently for each comment-model combination to prevent systematic order effects, using a documented random seed to ensure reproducibility. Because data collection occurred through automated API calls and analysis code processed all conditions identically without reference to condition labels until statistical modeling, blinding was effectively maintained throughout.

\section{Sampling}\label{app:sampling}

\subsection{Feasibility Assessment}

Prior to full data collection, we conducted a feasibility assessment by classifying a random sample of 50 comments from each docket to estimate stratum proportions. If any stratum comprised less than 15\% of the population, we adjusted targets to reflect population proportions while maintaining minimum cell sizes of 15 comments per stratum. If any stratum contained fewer than 15 eligible comments in the entire docket, we documented this constraint and collapsed strata as needed, such as combining length categories within sophistication levels, with appropriate analytical adjustments.

\subsection{Comment Stratification}

We stratified comments by length (short vs. long) and sophistication (substantive vs. non-substantive), yielding four strata. Length was determined by median split of word count in the post-deduplication pool. Sophistication distinguished substantive comments, which engage with specific policy content, from non-substantive comments, which often consist of templated campaign submissions or brief expressions of support or opposition. We did not use stance as a stratification criterion. Instead, the sample reflected the population distribution of stances within each docket to maintain external validity and enhance generalizability to real-world deployment contexts. We documented the observed stance distribution and examined stance as a moderator in exploratory analyses.

Classification used LLM-assisted coding with majority voting across three models, validated through human verification. A co-author verified all LLM classifications, with a stratified random subset of 10--15\% undergoing independent coding to assess reliability. We computed Cohen's $\kappa$ as the measure of inter-rater reliability, with a target of $\kappa \ge 0.70$. If the target was not met, we revised classification criteria and recoded. If reliability remained below target after revision, we reported the achieved reliability with appropriate caveats and resolved disagreements through discussion. We recorded exclusion reasons at each stage to enable reporting of a CONSORT-style flow diagram \citep{bennettConsolidatedStandardsReporting2005}, with full documentation in Appendix~\ref{app:fig:consort_flow}.

Target sample sizes per stratum were 30 comments for DOI (120 total) and 20 comments for EPA (80 total). If any stratum was underrepresented such that targets could not be met, we documented the shortfall and adjusted targets proportionally, subject to the minimum cell sizes established in the feasibility assessment. Writing quality classification was applied to sampled comments as described in Section~\ref{sec:writingquality}.

We included comments that were written in English, contained at least 50 words prior to redaction, and addressed substantive policy content. We included comments from individual citizens only. The pre-registered protocol specified excluding comments submitted on behalf of organizations. During data collection, this exclusion was not applied; organizational comments were retained after the redaction phase removed organizational identifiers, including entity names, tribal affiliations, and geographic markers embedded in group names. Because the experimental manipulation replaced all original identity signals with standardized individual names and occupations, the incongruity concern that motivated the original exclusion criterion was substantially mitigated through redaction rather than exclusion. This protocol deviation is addressed further in the Limitations section. Individual commenters who mentioned their profession or former employment were excluded as organizational; such references were handled through identity redaction. We processed comments submitted as attachments via text extraction, optical character recognition, or manual transcription as needed.

We deduplicated comments prior to sampling by reducing exact duplicate submissions to a single instance. We identified near-duplicate submissions exceeding 95\% cosine similarity using sentence embeddings \citep{reimersSentenceBERTSentenceEmbeddings2019}, retaining one representative example per unique template from organized campaigns. The number of duplicates removed was documented.

We excluded comments that contained only procedural requests without policy substance, consisted entirely of text copied from other sources, could not be adequately redacted while preserving argumentative meaning, or became incoherent following necessary redactions. Exclusion reasons were recorded at each stage to enable reporting of comment flow through the sampling process.

For technical failures during LLM processing, we retried failed API calls up to three times. If more than 5\% of summaries for any single base comment remained missing after retries, we replaced that comment with an alternative meeting the same stratification criteria. If any model exhibited greater than 15\% missing data across all comments after retries, we excluded that model from primary analyses and reported results both with and without the excluded model. If any model exhibited 10--15\% missing data, we documented this pattern and conducted sensitivity analyses. The pilot observed elevated missing rates for one model (see Section~\ref{app:pilot-models}), which informed these thresholds.

\subsection{Writing Quality Classification}\label{app:sampling::writingquality}

After sampling, we classified all comments on writing quality to enable both observational analysis across quality levels ($\boldsymbol{H_5}$) and experimental error injection for lower-quality comments ($\boldsymbol{H_6}$). Classification followed a multi-stage procedure.

We evaluated writing quality across five dimensions drawn from linguistics and education research: Content, Organization, Language Use, Vocabulary, and Mechanics. Definitions and supporting literature for each dimension are summarized in Table~\ref{tab:writingquality}.

Multiple LLMs (minimum of two) rated each comment on all five dimensions using a 5-point scale with anchored descriptors (See Section~\ref{app:writing-quality-descriptors}). To avoid conflating rating and summarization processes, the LLMs used for writing quality classification were distinct from the models used for summarization in the main experiment. We computed the Intraclass Correlation Coefficient (ICC) to assess overall inter-LLM agreement. Comments with a range $\geq 2$ (i.e., $\max(\text{LLM scores}) - \min(\text{LLM scores}) \geq 2$) on any dimension were flagged for human validation. Two human raters adjudicated flagged comments, with a target ICC $\geq 0.70$. If the target was not met after initial coding, we revised evaluation criteria and recoded; if reliability remained below target, we reported achieved reliability with appropriate caveats.

We computed an aggregate writing quality score as the average across all five dimensions. For $\boldsymbol{H_5}$, which examined whether raw DV scores covaried with writing quality, we used the continuous aggregate score as the primary measure; quartile-based categorical analyses were reported for robustness.

Comments were classified into quartiles based on this aggregate score: Low, Mid-Low, Mid-High, and High. For $\boldsymbol{H_6}$, error injection was applied only to comments classified in the Low and Mid-Low quartiles, as these represent the population most likely to exhibit writing quality variation in practice (see Section~\ref{sec:error-injection} for error types and injection procedure). This yielded matched pairs of original and error-added versions for experimental comparison.

\subsection{Identity Redaction}\label{app:sampling:identity-redaction}

Before experimental manipulation, we detected and removed existing identity signals in original comments. We combined multiple detection methods: keyword-based regular expressions targeting common identity markers, named entity recognition using the \texttt{spaCy} library or equivalent, LLM-based verification, and manual review. LLM-assisted rephrasing and verification used a model not included in the test set to avoid contamination.

Categories targeted for redaction included personal names, occupational references, gender and family role terms, racial and ethnic identifiers, age references, military or veteran status, and specific geographic locations that could identify commenters.

The redaction method depended on each signal's structural role in the comment. We deleted signature lines and standalone self-introductions entirely. For identity references woven into argumentative substance, we used LLM-assisted rephrasing to preserve meaning while removing identifying information. We replaced specific geographic locations with generic equivalents that preserve argumentative structure, such as replacing a named watershed with "the local watershed" or "my region's river system."

We validated redaction quality through manual review of a stratified random sample of 10--15\% of processed comments. Comments requiring more than three remediation rounds were excluded as insufficiently separable from identity content. If more than 10\% of verified comments required remediation, we expanded manual review to 25\% of the sample. We documented the proportion of comments requiring redaction and the extent of modification.

As an exploratory analysis, we examined whether differential treatment varied between heavily-redacted comments (e.g., above-median redaction) and minimally-redacted comments, which may indicate whether comments that originally contained identity markers differ systematically from those that did not.

\subsection{Large Language Models}\label{app:sampling:llm}

We tested eight LLMs representing models available through FedRAMP-authorized cloud services. All models were accessed through Ask Sage, which received FedRAMP authorization on January 24, 2025 as shown in the FedRAMP Marketplace \citep{federalriskandauthorizationmanagementprogramfedrampFedRAMPMarketplace2026} Model availability was verified via Ask Sage API endpoint (get-models) as of January 29, 2026 \citep{asksageAPIEndpoints2026}. From OpenAI, we tested GPT-4 (gpt4) and GPT-4.1 (gpt-4.1). From Anthropic, we tested Claude 4 Sonnet (claude-4-sonnet) and Claude 4 Opus (claude-4-opus). From Google, we tested Gemini 2.0 Flash (google-gemini-2.0-flash) and Gemini 2.5 Flash (google-gemini-2.5-flash). From Meta, we tested Llama 3 70B (llama3:70b) and Llama 3.3 70B (llama3.3:70b), accessed via Ollama \citep{ollamaOllama2025a}.

We excluded reasoning models such as GPT-5 from this study. These models use chain-of-thought architectures that restrict temperature parameter variation (requiring temperature = 1), which would prevent comparable experimental conditions across models. This exclusion represents a limitation, as reasoning models may exhibit different summarization behaviors that our design cannot capture.

Ask Sage also provides access to government-specific model deployments hosted in AWS GovCloud or Azure Government environments, including Anthropic models (e.g., aws-bedrock-claude-35-sonnet-gov) and OpenAI models (e.g., gpt-4o-gov). These deployments are configured for federal security and compliance requirements and are restricted to authorized government users. As external researchers, we cannot access these government-specific deployments. Our analyses used commercially available versions of equivalent model families accessible through Ask Sage, which share the same underlying architectures and training. We acknowledge that government deployments may have different configurations or fine-tuning that could affect summarization behavior; this represents another limitation of the current study.

\subsection{Model Parameters}\label{app:sampling:model-parameter}

All models were run with parameters selected to maximize reproducibility: temperature set to zero to minimize output randomness, maximum tokens set to 1024 to ensure summaries are not truncated, and default values for other parameters as provided by each API. We note that outputs may not be fully deterministic despite temperature settings due to implementation-level variation across providers. Pilot data showed variable textual identity across repetitions depending on model, though dependent variable measurements showed moderate to excellent reliability (See Section~\ref{app:pilot-variance}).

\subsection{Prompts}\label{app:sampling:prompts}

All API calls used a standardized prompt structure. The system prompt for all conditions was: \texttt{You are an assistant to summarize contents.} The primary analysis used the user prompt: \texttt{Summarize this public comment.} This simple formulation minimized prompt-induced variation.

Robustness checks used two alternative prompts to assess sensitivity to instruction specificity and length constraints. The systematic prompt requested: \texttt{Summarize this public comment in one concise paragraph with three key points.} The length-pressure prompt requested: \texttt{Condense this public comment into 100 words or less.}

For robustness analyses, we re-ran hypothesis tests $\boldsymbol{H_1}$--$\boldsymbol{H_5}$ on the robustness subset using alternative prompts and assessed pattern replication. $\boldsymbol{H_6}$ was excluded from robustness checks because the small number of Low and Mid-Low quality comments in the robustness subset would yield insufficient power. Given reduced power from the smaller sample, we focused on consistency of effect patterns rather than statistical significance. We reported whether the direction of effects was preserved, whether effect sizes were comparable in magnitude, and whether any contradictory effects emerged. Previously significant effects that remained directionally consistent with $p < .10$ were interpreted as providing convergent evidence; effects that reversed direction or showed substantially attenuated magnitudes were flagged as potential prompt-sensitivity concerns.

\subsection{Computational Load}

The total computational load comprised approximately 118,000 LLM-generated summaries (approximately 78,000 for the main analysis, plus 40,000 for reliability and robustness checks). The Identity Analysis ($\boldsymbol{H_1}$--$\boldsymbol{H_4}$) generated 52,800 summaries across 200 comments, 33 conditions (32 identity conditions plus one baseline), and 8 models. The Error-Injection Analysis ($\boldsymbol{H_6}$) generated an additional 25,600 summaries for the approximately 100 low and mid-low quality comments with injected errors across 32 identity conditions and 8 models. A randomly selected reliability subset of 20 comments was processed five times each, generating approximately 26,400 reliability observations to assess measurement stability. Robustness checks using two alternative prompts on a stratified random sample of approximately 25 comments generated roughly 13,200 additional summaries.

\section{Statistical Model}\label{app:statistical-models}

\subsection{Inference Framework}\label{app:inference-detail}

All confirmatory analyses employed both frequentist and Bayesian inference. Frequentist tests used $\alpha = .05$ (two-tailed) with Holm--Bonferroni correction as specified in Section~\ref{analysis:identity}. Bayes factors were approximated using the BIC method \citep{wagenmakersPracticalSolutionPervasive2007}:
\[
\mathrm{BF}_{10} \approx \exp\left(\frac{\mathrm{BIC}_0 - \mathrm{BIC}_1}{2}\right)
\]
where $\mathrm{BIC}_0$ and $\mathrm{BIC}_1$ are obtained from nested models fit via maximum likelihood. We adopted three interpretation thresholds \citep{robertHaroldJeffreyssTheory2009}: $\mathrm{BF}_{10} \geq 3$ as evidence supporting the effect, $\mathrm{BF}_{10} \leq 1/3$ as evidence that failed to support the effect, and intermediate values as inconclusive. This dual approach distinguished between absence of evidence and evidence of absence.

\subsection{{Identity Analysis Model ($\boldsymbol{H_1}$--$\boldsymbol{H_4}$)}} Statistical analysis employed linear mixed-effects models implemented in R, using the \texttt{lme4} and \texttt{lmerTest} packages. The model specification for testing identity effects is:

{\small
\begin{align}
\Delta_{ijklmn} &= \beta_0 + \beta_1 R^{(\text{B})}_i + \beta_2 R^{(\text{H})}_i + \beta_3 R^{(\text{A})}_i + \beta_4 G_j + \beta_5 S_k \nonumber \\
&\quad + \beta_6 (R^{(\text{B})} \times G)_{ij} + \beta_7 (R^{(\text{H})} \times G)_{ij} + \beta_8 (R^{(\text{A})} \times G)_{ij} \nonumber \\
&\quad + u_l + v_m + w_{n(ij)} + \varepsilon_{ijklmn}
\end{align}
}

where $\Delta_{ijklmn}$ denotes the difference score for an observation with Race level $i$, Gender level $j$, SES level $k$, Comment $l$, Model $m$, and Name $n$.

All categorical predictors used dummy coding with White, Male, and High-SES as reference categories. The intercept $\beta_0$ represents the expected difference score for the reference group; $\beta_1$--$\beta_5$ capture main effects; $\beta_6$--$\beta_8$ capture Race $\times$ Gender interactions. Random intercepts for Comment ($u_l$), Model ($v_m$), and Name nested within Race $\times$ Gender ($w_{n(ij)}$) accounted for clustering at each level. All random effects and residuals are assumed normally distributed with mean zero.

\subsection{Writing Quality Analysis Models ($\boldsymbol{H_5}$--$\boldsymbol{H_6}$)} The model specifications for testing writing quality effects used all 32 identity conditions.\\

For $\boldsymbol{H_5}$ (observational comparison across writing quality levels):
{\small
\begin{align}
Y_{ijklmn} &= \gamma_0 + \gamma_1 R^{(\text{B})}_i + \gamma_2 R^{(\text{H})}_i + \gamma_3 R^{(\text{A})}_i \nonumber \\
&\quad + \gamma_4 G_j + \gamma_5 S_k + \gamma_6 Q_l + u_l + v_m + w_{n(ij)} + \varepsilon_{ijklmn}
\end{align}
}

where $Y_{ijklmn}$ denotes the raw score for an observation with Race level $i$, Gender level $j$, SES level $k$, Comment $l$, Model $m$, and Name $n$. All categorical predictors used dummy coding with White, Male, and High-SES as reference categories. $Q_l$ denoted the continuous writing quality score for comment $l$ (average across five evaluation dimensions). The coefficient $\gamma_6$ tests $\boldsymbol{H_5}$: a significant effect would indicate that writing quality affects summarization. For robustness, we also reported analyses using quartile-based categorical classification of writing quality.\\

As secondary analysis, we examined each writing quality dimension separately:

{\small
\begin{align}
Y_{ijklmn} &= \gamma_0 + \gamma_1 R^{(\text{B})}_i + \gamma_2 R^{(\text{H})}_i + \gamma_3 R^{(\text{A})}_i \nonumber \\
&\quad + \gamma_4 G_j + \gamma_5 S_k + \gamma_6 D_l + u_l + v_m + w_{n(ij)} + \varepsilon_{ijklmn}
\end{align}
}
where $D_l$ denotes the score for a single dimension (Content, Organization, Language Use, Vocabulary, or Mechanics) for comment $l$. We fit this model five times, once per dimension. Because dimensions are likely correlated, these analyses may reveal patterns that inform future research but cannot establish which dimensions uniquely drive any observed relationships.\\

For $\boldsymbol{H_6}$ (experimental comparison of original versus error-added conditions):
{\small
\begin{align}
Y_{ijklmnp} &= \gamma_0 + \gamma_1 R^{(\text{B})}_i + \gamma_2 R^{(\text{H})}_i + \gamma_3 R^{(\text{A})}_i \nonumber \\
&\quad + \gamma_4 G_j + \gamma_5 S_k + \gamma_6 A_p + u_l + v_m + w_{n(ij)} + \varepsilon_{ijklmnp}
\end{align}
}

where $Y_{ijklmnp}$ denotes the raw score for an observation with Race level $i$, Gender level $j$, SES level $k$, Comment $l$, Model $m$, Name $n$, and ErrorAdded status $p$. $A_p$ contrasts original versus error-added with original as reference. The coefficient $\gamma_6$ tests $\boldsymbol{H_6}$: a significant effect would indicate that error injection affects summarization. This model used only comments classified in the Low and Mid-Low quality quartiles.

Because both analyses used raw scores rather than difference scores, the comment random effect captures absolute summary characteristics rather than deviation from baseline. This means comment-level variance includes both the comment's inherent characteristics and its baseline summary quality. Random intercepts for Comment ($u_l$), Model ($v_m$), and Name nested within Race $\times$ Gender ($w_{n(ij)}$) accounted for clustering at each level. The WritingQuality and ErrorAdded coefficients remain interpretable as effects controlling for comment-level variation.

Exploratory interactions (Race $\times$ WritingQuality, Gender $\times$ WritingQuality, SES $\times$ WritingQuality for $\boldsymbol{H_5}$; Race $\times$ ErrorAdded, Gender $\times$ ErrorAdded, SES $\times$ ErrorAdded for $\boldsymbol{H_6}$) were tested in extended models and reported descriptively.

\subsection{Model Specification} We began with a random effects structure appropriate to the within-subjects design. When the specified model failed to converge, we simplified the random effects structure following recommended procedures \citep{barrRandomEffectsStructure2013}: first removing correlation parameters among random effects, then if necessary removing random slopes for higher-order interactions, and only as a last resort removing random slopes for main effects of theoretical interest. All models attempted were documented, and the final converged model was reported with justification for any simplifications.

\subsection{Frequentist Inference} Final models were estimated using restricted maximum likelihood. Degrees of freedom for significance tests were approximated using the Satterthwaite method \citep{lukeEvaluatingSignificanceLinear2017}. We used an alpha level of 0.05 for all confirmatory tests, and all tests were two-tailed.

For the Identity Analysis, hierarchical testing controls family-wise error rate through gatekeeping: the Level 1 omnibus test uses $\alpha = .05$, and at Level 2, Holm-Bonferroni correction is applied across hypotheses followed by correction across dependent variables within significant hypotheses.  For the Writing Quality Analysis, Holm-Bonferroni correction is applied across the four dependent variables for $\boldsymbol{H_5}$ and $\boldsymbol{H_6}$ separately.  Pairwise comparisons for significant effects are reported descriptively to characterize patterns rather than test new hypotheses. 

\subsection{Effect Sizes} We reported effect sizes for all analyses. For omnibus tests and interaction effects, we reported partial eta-squared, interpreted using conventional benchmarks of 0.01 for small, 0.06 for medium, and 0.14 for large effects. For pairwise comparisons, we reported Cohen's $d$, interpreted using benchmarks of 0.2 for small, 0.5 for medium, and 0.8 for large effects \citep{cohenStatisticalPowerAnalysis2013}. All effect sizes were accompanied by 95\% confidence intervals.

\subsection{Assumption Checks} Prior to conducting primary analyses, we verified that model assumptions were satisfied. Multicollinearity among predictors was assessed via variance inflation factors, with values below 5 considered acceptable. Normality of residuals was assessed through visual inspection of Q-Q plots. Significance of random effects was tested via likelihood ratio tests. Homoscedasticity was assessed through visual inspection of residual plots. Any violations and remedies applied were reported transparently.

\subsection{Distributional Considerations} Because some dependent variables have bounded distributions, we examined distributions prior to analysis and applied appropriate modeling strategies as needed. For the Identity Analysis ($\boldsymbol{H_1}$--$\boldsymbol{H_4}$), which uses difference scores: $\Delta_{\text{Sim}}$ is bounded between -1 and 1. If raw similarity scores exhibit ceiling effects, the resulting difference scores may show restricted variance around zero, reducing power to detect identity effects rather than violating distributional assumptions. We reported distributional diagnostics; if variance is severely restricted, we noted this as a limitation on statistical power, since the difference-score structure directly tests our research question and alternative specifications would complicate interpretation of effect direction. If $\Delta_{\text{Len}}$ showed severe right skewness, we applied logarithmic transformation or used gamma regression. For the Writing Quality Analysis ($\boldsymbol{H_5}$, $\boldsymbol{H_6}$), which uses raw scores: semantic similarity is bounded between 0 and 1, and if distributions cluster near the boundary, beta regression was considered. Raw length ratios may exhibit right skewness; if severe, logarithmic transformation or gamma regression was applied. Distribution diagnostics and any modeling adjustments were documented.

\subsection{Sensitivity Analysis for Name Effects}

To assess whether specific names drive observed effects, we conducted sensitivity analyses. In primary analyses, we reported the variance component for the name random effect. If this variance is near zero relative to other variance components, this would suggest names within demographic cells are largely interchangeable. We then re-ran the main models replacing the name random effect with name as a fixed effect and compared whether demographic main effect coefficients changed substantially. If individual name coefficients appear extreme relative to other names in that cell, this would suggest name-specific effects warrant further investigation. Results were reported in supplementary materials.

\subsection{Exploratory Analyses}

In addition to confirmatory hypothesis tests, we conducted exploratory analyses that were reported separately and clearly labeled as such. We examined higher-order interactions including Race $\times$ SES, Gender $\times$ SES, and Race $\times$ Gender $\times$ SES. We examined whether differential treatment varied by comment characteristics such as stance toward the proposed rule, length, and sophistication. We also examined whether differential effects varied by model family and assessed prompt sensitivity by comparing effect sizes across the three prompt conditions in the robustness subset. These analyses are exploratory because they were not part of the primary hypotheses and may have limited power for detecting smaller effects or complex interactions.
\subsubsection{Reliability Analysis}\label{sec:reliability}

For the 20-comment reliability subset processed with 5 repetitions, we computed intraclass correlations (ICC) for each dependent variable to verify measurement stability. The reliability analysis encompassed approximately 52,000 observations conducted uniformly across all conditions and models. To diagnose sources of measurement variance, we also computed ICC separately for raw baseline and attributed summary scores. If difference score reliability is lower than raw score reliability, this would indicate that baseline instability contributes to measurement variance.

We adopted ICC $\ge 0.70$ as the threshold for adequate reliability. This cutoff is a heuristic rather than an absolute standard; the appropriate threshold depends on the decisions being made \citep{lebretonAnswers20Questions2008}. High-stakes decisions about specific individuals require strong agreement, whereas research estimating aggregate effects may proceed with moderate agreement \citep{lebretonAnswers20Questions2008}. Because our analysis examined group-level patterns rather than individual-level decisions, ICC $\ge 0.70$ is appropriate. Lower ICC values indicate greater within-group variability relative to between-group differences \citep{blieseGroupSizeICC1998}, which would attenuate observed effects rather than inflate estimates.

Based on pilot findings (ICC = 0.60--0.94), we expected moderate to excellent reliability. However, we pre-specified the following contingencies. If ICC $\ge 0.70$, reliability is adequate and main analysis results are interpretable as reported. If ICC falls between 0.50 and 0.70, results were reported with a caveat noting measurement variability, and we reported effect size confidence intervals incorporating repetition variance. If ICC falls below 0.50 for any measure, results for that measure were flagged as potentially unstable, and we reported sensitivity analyses comparing effects across individual repetitions.

Given that semantic similarity is designated as the most consequential measure, we applied a stronger contingency: if $\Delta_{\text{Sim}}$ ICC fell below 0.70 in the confirmatory sample, we reported primary results using the 5-repetition average from the reliability subset as a sensitivity analysis in addition to single-run results.

If reliability varies substantially by model, we reported model-specific ICC estimates and interpreted effects for low-reliability models with appropriate caveats.

\section{Descriptive Statistics}\label{app:descriptive-statistics}
\subsection{Consolidated Standards of Reporting Trials (CONSORT) Flow}\label{app:fig:consort_flow}
\begin{figure}[htbp][!tb]
    \centering
    \includegraphics[height=0.9\textheight]{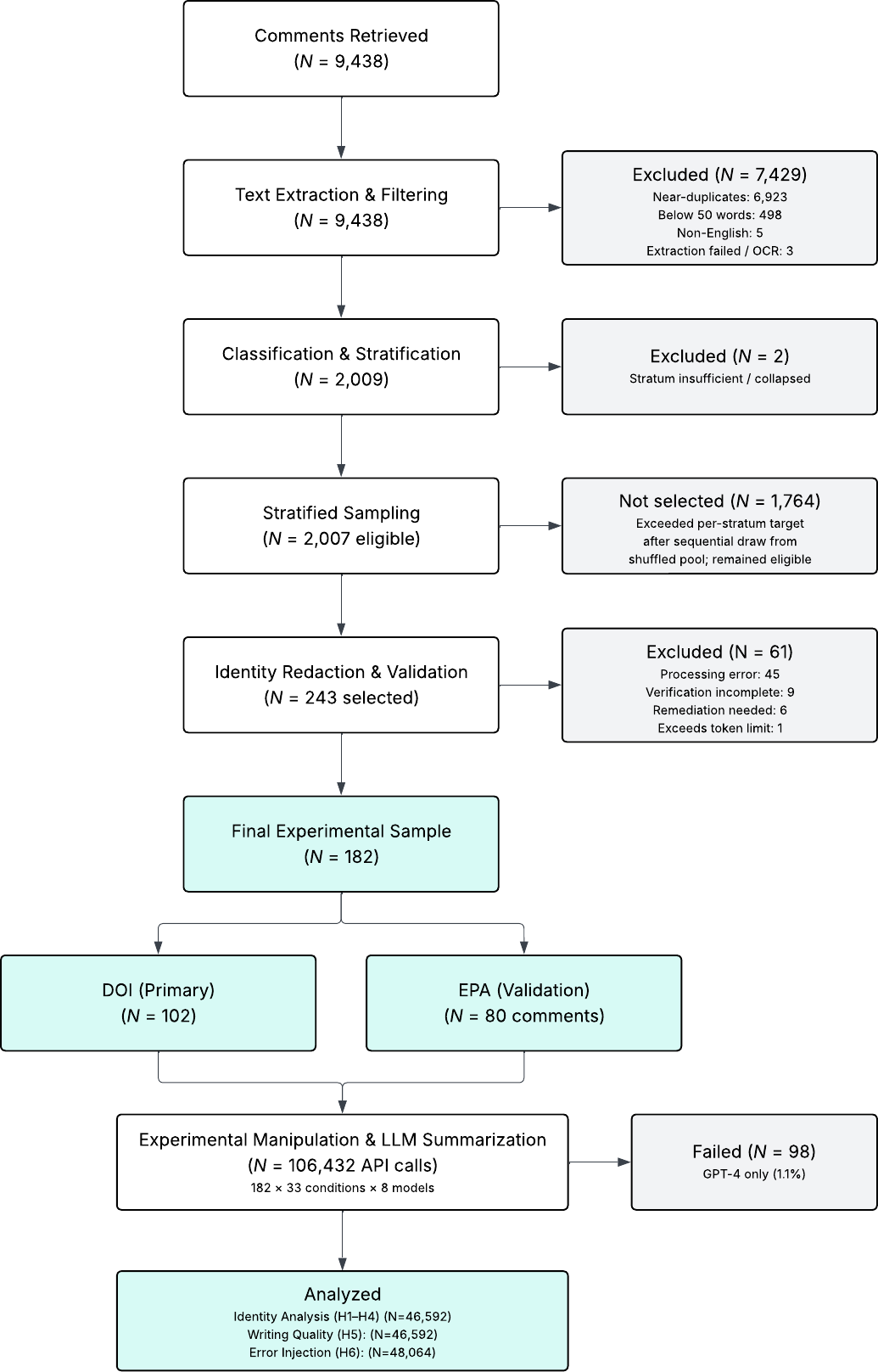}
\end{figure}
\FloatBarrier

\subsection{Sample Characteristics}\label{app:sample-characteristics}

\begin{verbatim}
## Sample size: 182 comments
\end{verbatim}

\begin{verbatim}
## By docket:
\end{verbatim}

\begin{verbatim}
## 
##        DOI-2025-0004 EPA-HQ-OAR-2025-0124 
##                  102                   80
\end{verbatim}

\begin{verbatim}
## 
## By stratum:
\end{verbatim}

\begin{verbatim}
## 
##  long_nonsubstantive     long_substantive short_nonsubstantive 
##                    8                   67                   78 
##    short_substantive 
##                   29
\end{verbatim}

\begin{verbatim}
## 
## By stance:
\end{verbatim}

\begin{verbatim}
## 
##   mixed  oppose support unclear 
##       8     143      22       9
\end{verbatim}

\begin{verbatim}
## 
## By writing quality quartile:
\end{verbatim}

\begin{verbatim}
## 
##      Low  Mid-Low Mid-High     High 
##       50       44       45       43
\end{verbatim}

\subsection{Model Performance}\label{app:model-performance}

\begin{table}[htbp]
\centering
\caption{Model performance summary}
\begin{adjustbox}{width=\textwidth, center}
\begin{tabular}{llrrrrrr}
\toprule
model name & model provider & total calls & success count & failed count & pending count & failure rate & avg latency ms \\
\midrule
claude-4-opus & anthropic & 13,304 & 13,304 & 0 & 0 & 0.0000 & 7,872.4 \\
claude-4-sonnet & anthropic & 13,304 & 13,304 & 0 & 0 & 0.0000 & 5,644.1 \\
gemini-2.5-flash & google & 13,304 & 13,304 & 0 & 0 & 0.0000 & 4,139.3 \\
gemini-3-flash-preview & google & 13,304 & 13,304 & 0 & 0 & 0.0000 & 4,221.4 \\
gpt-4 & openai & 13,304 & 13,206 & 98 & 0 & 0.0074 & 3,276.1 \\
gpt-4.1 & openai & 13,304 & 13,304 & 0 & 0 & 0.0000 & 2,197.1 \\
llama3-70b & ollama & 13,304 & 13,304 & 0 & 0 & 0.0000 & 17,693.3 \\
llama3.3-70b & ollama & 13,304 & 13,304 & 0 & 0 & 0.0000 & 18,021.3 \\
\bottomrule
\end{tabular}
\end{adjustbox}
\end{table}

\begin{verbatim}
## 
## Model completeness in identity_analysis:
\end{verbatim}

\begin{verbatim}
## # A tibble: 8 x 4
##   model_name                 n expected missing_pct
##   <fct>                  <int>    <dbl>       <dbl>
## 1 claude-4-opus           5824     5824        0   
## 2 claude-4-sonnet         5824     5824        0   
## 3 gemini-2.5-flash        5824     5824        0   
## 4 gemini-3-flash-preview  5824     5824        0   
## 5 gpt-4                   5760     5824        1.10
## 6 gpt-4.1                 5824     5824        0   
## 7 llama3-70b              5824     5824        0   
## 8 llama3.3-70b            5824     5824        0
\end{verbatim}

\clearpage
\subsection{DV Distributions}\label{app:dv-distributions}
\begin{figure}[htbp][!tb]
    \centering
    \includegraphics[width=\textwidth]{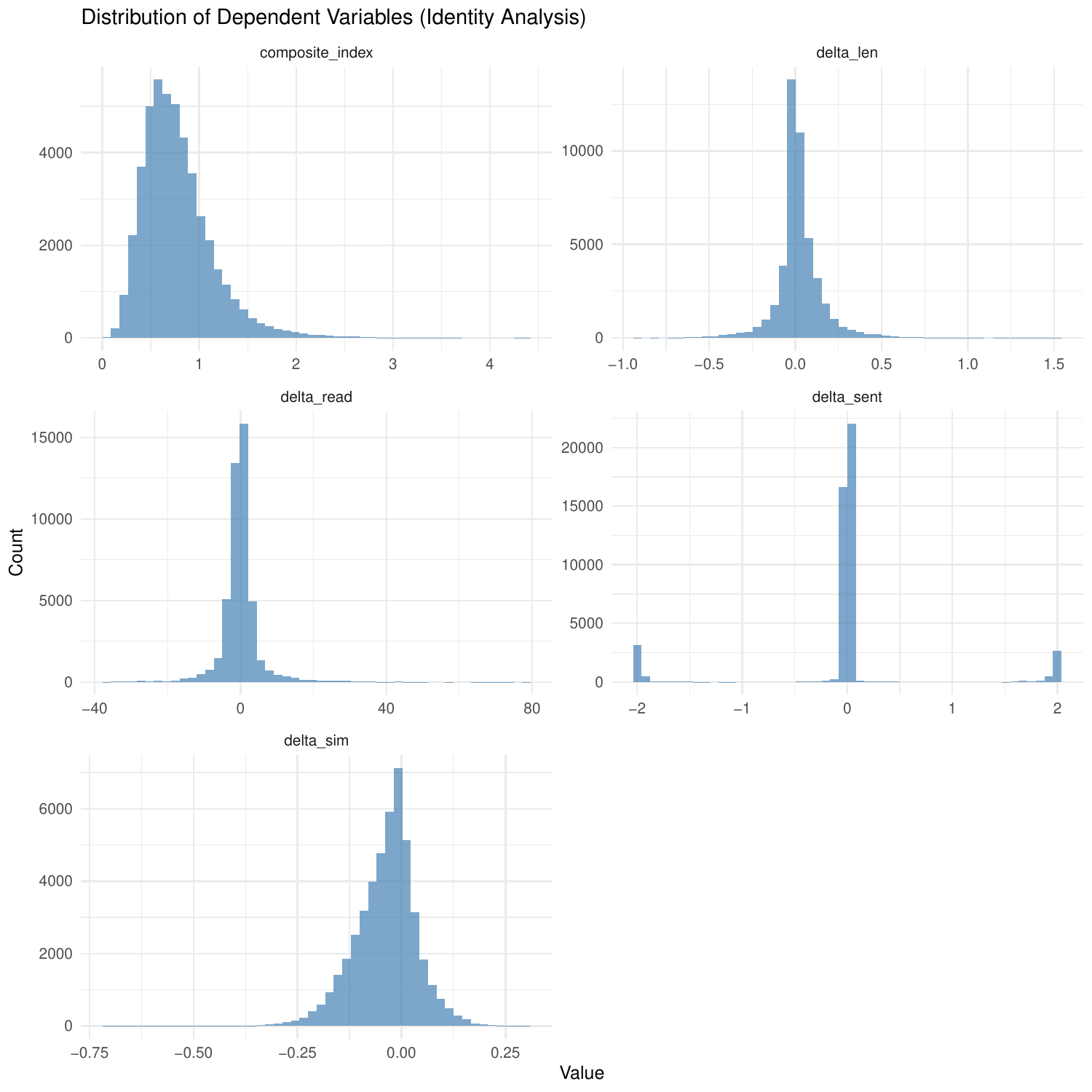}
    \label{app:fig:dv-distribution}
\end{figure}
\FloatBarrier

\subsection{Failure Analysis}\label{app:failure-analysis}

\begin{verbatim}
## Total failed API calls: 98 
## 
## By model:
## 
## gpt-4 
##    98 
## 
## By error code:
## 
## rate_limit    unknown 
##         14         84
\end{verbatim}

\begin{verbatim}
## 
## Failure by demographic (checking for systematic patterns):
## 
## By race:
## 
##             Asian    Black Hispanic    White 
##        2       24       24       24       24 
## 
## By gender:
## 
##        Female   Male 
##      2     48     48 
## 
## By ses:
## 
##      High  Low 
##    2   48   48
\end{verbatim}

\section{Reliability Analysis}\label{app:reliability-analysis}

\subsection{Difference-Score ICC}\label{app:difference-score-icc}

\begin{verbatim}
## Reliability subset (identity conditions): 25600 rows
\end{verbatim}

\begin{verbatim}
## Unique comments: 20
\end{verbatim}

\begin{verbatim}
## Repetitions: 1, 2, 3, 4, 5
\end{verbatim}

\begin{verbatim}
## Models: 8
\end{verbatim}

\begin{verbatim}
## === Difference-Score ICC (pooled across models) ===
\end{verbatim}

\begin{verbatim}
##   delta_sim: ICC(2,1) = 0.882 [0.877, 0.887] — ADEQUATE (n = 5120)
\end{verbatim}

\begin{verbatim}
##   delta_sent: ICC(2,1) = 0.695 [0.685, 0.705] — CAVEAT (n = 5120)
##   delta_len: ICC(2,1) = 0.833 [0.827, 0.839] — ADEQUATE (n = 5120)
##   delta_read: ICC(2,1) = 0.890 [0.885, 0.894] — ADEQUATE (n = 5120)
\end{verbatim}

\subsection{Raw-Score ICC}\label{app:raw-score-icc}

\begin{verbatim}
## === Raw-Score ICC (all conditions including baseline) ===
\end{verbatim}

\begin{verbatim}
##   semantic_similarity: ICC(2,1) = 0.984 [0.983, 0.984]
\end{verbatim}

\begin{verbatim}
##   sentiment: ICC(2,1) = 0.922 [0.919, 0.925]
\end{verbatim}

\begin{verbatim}
##   length_ratio: ICC(2,1) = 0.989 [0.988, 0.989]
##   readability: ICC(2,1) = 0.955 [0.953, 0.957]
\end{verbatim}

\begin{verbatim}
## 
## === Comparison: Delta ICC vs Raw ICC ===
\end{verbatim}

\begin{verbatim}
## (If delta < raw, baseline instability contributes to unreliability)
\end{verbatim}

\begin{verbatim}
##   delta_sim: delta=0.882, raw=0.984, diff=-0.102
##   delta_sent: delta=0.695, raw=0.922, diff=-0.227
##   delta_len: delta=0.833, raw=0.989, diff=-0.156
##   delta_read: delta=0.890, raw=0.955, diff=-0.065
\end{verbatim}

\subsection{Model-Specific ICC}\label{app:model-specific-icc}

\begin{verbatim}
## === Model-Specific ICC (delta_sim) ===
\end{verbatim}

\begin{verbatim}
## Table: Model-specific ICC for delta_sim
## 
## |model                  | icc_delta_sim|   n|
## |:----------------------|-------------:|---:|
## |claude-4-opus          |         0.954| 640|
## |claude-4-sonnet        |         0.830| 640|
## |gemini-2.5-flash       |         0.656| 640|
## |gemini-3-flash-preview |         0.829| 640|
## |gpt-4                  |         0.840| 640|
## |gpt-4.1                |         0.916| 640|
## |llama3-70b             |         0.929| 640|
## |llama3.3-70b           |         0.889| 640|
\end{verbatim}

\subsection{ICC Summary}\label{app:icc-summary}

\begin{table}[htbp]
\centering
\caption{ICC Summary for Difference Scores}
\begin{adjustbox}{max width=\textwidth}
\begin{tabular}{@{}lrrrrl@{}}
\toprule
DV & ICC & Lower & Upper & N & Status \\
\midrule
delta\_sim & 0.882 & 0.877 & 0.887 & 5120 & adequate \\
delta\_sent & 0.695 & 0.685 & 0.705 & 5120 & caveat \\
delta\_len & 0.833 & 0.827 & 0.839 & 5120 & adequate \\
delta\_read & 0.890 & 0.885 & 0.894 & 5120 & adequate \\
\bottomrule
\end{tabular}
\end{adjustbox}
\end{table}

\begin{verbatim}
## 
## Flags for downstream analysis:
\end{verbatim}

\begin{verbatim}
##   use_averaged_sim: FALSE
\end{verbatim}

\section{Assumption Checks}\label{app:assumption-checks}

\subsection{VIF \& Residual Diagnostics}\label{app:vif-residual-diagnostics}

\begin{verbatim}
## Fitting omnibus model for assumption checks...
\end{verbatim}

\begin{verbatim}
## Converged spec: default optimizer
\end{verbatim}

\begin{verbatim}
## === Variance Inflation Factors ===
##             GVIF Df GVIF^(1/(2*Df))
## race           8  3        1.414214
## gender         4  1        2.000000
## ses            1  1        1.000000
## race:gender   20  3        1.647549
## 
## Max GVIF^(1/2Df): 2.00 (threshold: 5)
\end{verbatim}

\begin{figure}[htbp][!tb]
    \centering
    \includegraphics[width=\textwidth]{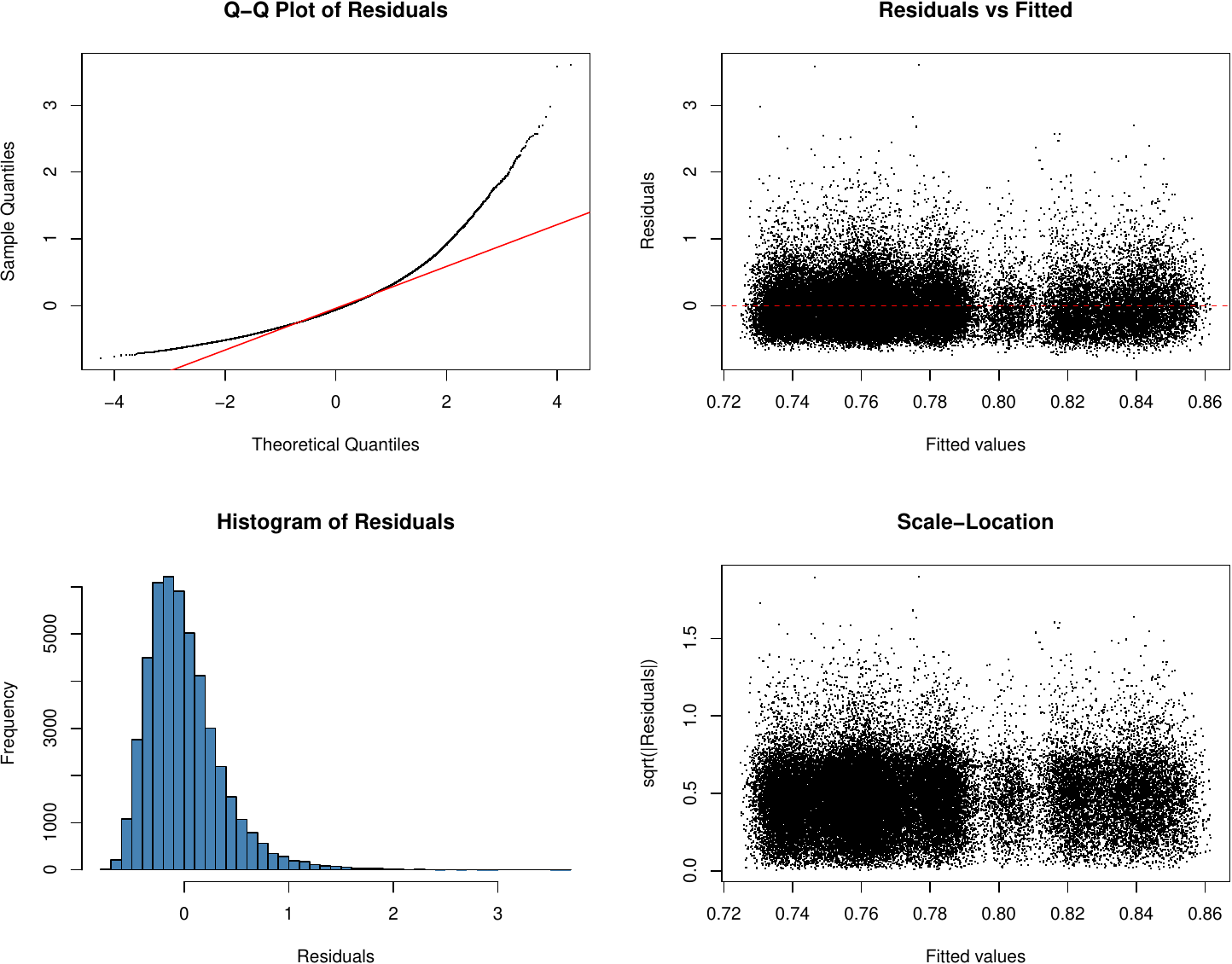}
    \label{app:fig:assumption-checks}
\end{figure}
\FloatBarrier

\subsection{LRT for Random Effects}\label{app:lrt-for-random-effects}

\begin{verbatim}
## === Likelihood Ratio Tests for Random Effects ===
## 
## Random effects in converged model: comment_id, name_full, model_name 
## 
## (1|comment_id):
## Data: identity_data
## Models:
## reduced: reduced_formula
## full_ml: composite_index ~ race + gender + ses + race:gender + (1 | comment_id) + 
##                            (1 | model_name) + (1 | name_full)
##         npar   AIC   BIC logLik -2*log(L)  Chisq Df Pr(>Chisq)  
## reduced   12 37042 37147 -18509     37018                       
## full_ml   13 37041 37155 -18508     37015 2.7766  1    0.09565 .
## ---
## Signif. codes:  0 '***' 0.001 '**' 0.01 '*' 0.05 '.' 0.1 ' ' 1
## 
## (1|name_full):
## Data: identity_data
## Models:
## reduced: reduced_formula
## full_ml: composite_index ~ race + gender + ses + race:gender + (1 | comment_id) + 
##                            (1 | model_name) + (1 | name_full)
##         npar   AIC   BIC logLik -2*log(L)  Chisq Df Pr(>Chisq)    
## reduced   12 37071 37175 -18523     37047                         
## full_ml   13 37041 37155 -18508     37015 31.341  1  2.164e-08 ***
## ---
## Signif. codes:  0 '***' 0.001 '**' 0.01 '*' 0.05 '.' 0.1 ' ' 1
## 
## (1|model_name):
## Data: identity_data
## Models:
## reduced: reduced_formula
## full_ml: composite_index ~ race + gender + ses + race:gender + (1 | comment_id) + 
##                            (1 | model_name) + (1 | name_full)
##         npar   AIC   BIC logLik -2*log(L)  Chisq Df Pr(>Chisq)
## reduced   12 37040 37144 -18508     37016                     
## full_ml   13 37041 37155 -18508     37015 0.3456  1     0.5566
\end{verbatim}

\clearpage
\subsection{Distribution Diagnostics for Raw
DVs}\label{app:distribution-diagnostics-for-raw-dvs}

\begin{figure}[htbp][!tb]
    \centering
    \includegraphics[width=\textwidth]{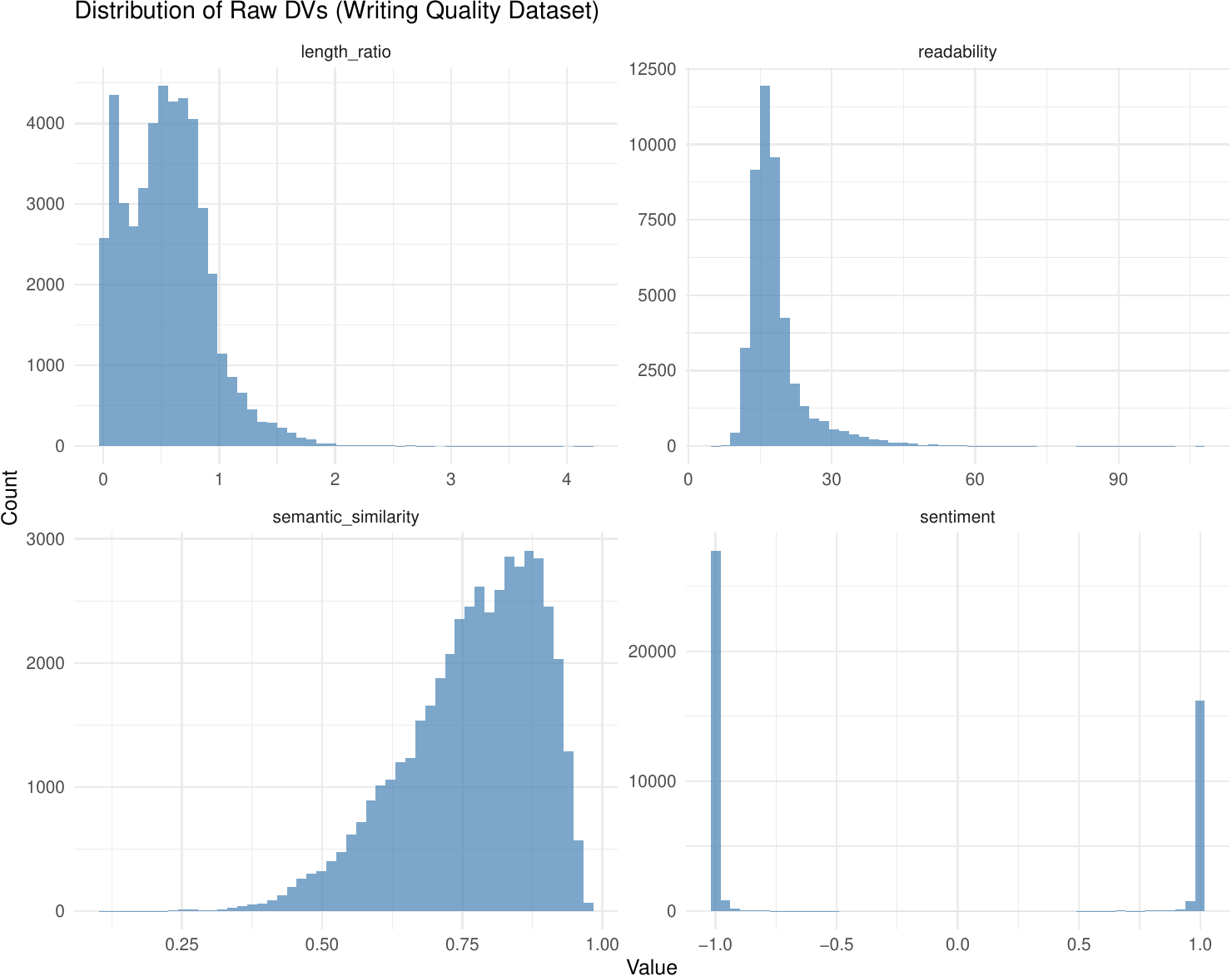}
    \label{app:fig:distribution-diagnostics}
\end{figure}
\FloatBarrier

\begin{verbatim}
## 
## === Distribution Diagnostics ===
\end{verbatim}

\begin{verbatim}
## 
## semantic_similarity:
##   Mean=0.7677, SD=0.1215, Min=0.1108, Max=0.9749
##   Skewness=-0.792, Kurtosis=0.370
##   % above 0.95: 1.1% (ceiling concern if >20%)
## 
## sentiment:
##   Mean=-0.2486, SD=0.9599, Min=-0.9995, Max=0.9989
##   Skewness=0.514, Kurtosis=-1.732
## 
## length_ratio:
##   Mean=0.5522, SD=0.3663, Min=0.0049, Max=4.1852
##   Skewness=1.004, Kurtosis=3.515
##   % above 1.0 (summary longer than original): 8.9%
## 
## readability:
##   Mean=18.2595, SD=6.2482, Min=5.6829, Max=107.0204
##   Skewness=3.039, Kurtosis=16.283
\end{verbatim}

\section{Primary Identity Analysis ($H_1$—$H_4$)}\label{app:primary-identity}

\subsection{Level 1: Omnibus Gate}\label{app:primary-level1}
    
\begin{verbatim}
## ============================================================
\end{verbatim}

\begin{verbatim}
## LEVEL 1: Omnibus test on composite_index
\end{verbatim}

\begin{verbatim}
## ============================================================
\end{verbatim}

\begin{verbatim}
## Full model spec: default optimizer
\end{verbatim}

\begin{verbatim}
## Null model spec: default optimizer
\end{verbatim}

\begin{verbatim}
## Omnibus LRT:
## Data: identity_data
## Models:
## null_ml: composite_index ~ 1 + (1 | comment_id) + (1 | model_name) + (1 | name_full)
## full_ml: composite_index ~ race + gender + ses + race:gender + (1 | comment_id) + 
##                            (1 | model_name) + (1 | name_full)
##
##         npar   AIC   BIC logLik -2*log(L)  Chisq Df Pr(>Chisq)    
## null_ml    5 37086 37130 -18538     37076                         
## full_ml   13 37041 37155 -18508     37015 61.174  8  2.742e-10 ***
## ---
## Signif. codes:  0 '***' 0.001 '**' 0.01 '*' 0.05 '.' 0.1 ' ' 1
## 
## Omnibus p = < .001
## Decision: SIGNIFICANT — proceed to Level 2 (alpha = .05)
## 
## Partial eta-squared:
##     Parameter Eta2_partial       CI_low CI_high
## 1        race 0.7847229942 0.3774288234       1
## 2      gender 0.0012683420 0.0000000000       1
## 3         ses 0.0007576127 0.0003955045       1
## 4 race:gender 0.3251777820 0.0000000000       1
## 
## R² (marginal and conditional):
##              R2m        R2c
## [1,] 0.006973189 0.01044015
## 
## BF10 (BIC approximation): 4.10e-06
## Interpretation: Moderate+ evidence for H0 (null)
## 
## Random effect variances:
##  Groups     Name        Std.Dev. 
##  comment_id (Intercept) 0.0097414
##  name_full  (Intercept) 0.0187194
##  model_name (Intercept) 0.0029395
##  Residual               0.3599571
\end{verbatim}

\subsection{Level 2: DV-Specific Tests}\label{app:primary-level2}
\begin{verbatim}
## ============================================================ 
## LEVEL 2: DV-specific tests (omnibus was significant)
## ============================================================ 
## 
## 
## --- delta_sim ---
##   Spec: default optimizer 
##   ANOVA table:
## Type III Analysis of Variance Table with Satterthwaite's method
##              Sum Sq Mean Sq NumDF DenDF  F value Pr(>F)    
## race        0.02195 0.00732     3     8   1.8940 0.2091    
## gender      0.00734 0.00734     1     8   1.9003 0.2054    
## ses         1.23423 1.23423     1 46323 319.5641 <2e-16 ***
## race:gender 0.00092 0.00031     3     8   0.0791 0.9695    
## ---
## Signif. codes:  0 '***' 0.001 '**' 0.01 '*' 0.05 '.' 0.1 ' ' 1
## 
##   Fixed effects:
##                               Estimate   Std. Error           df      t value
## (Intercept)               -0.022651761 0.0125012918    13.154510  -1.81195360
## raceBlack                 -0.005376241 0.0094646085     7.999409  -0.56803624
## raceHispanic              -0.004354069 0.0094646085     7.999409  -0.46003693
## raceAsian                 -0.015793621 0.0094646085     7.999409  -1.66870303
## genderFemale              -0.006072375 0.0094646085     7.999407  -0.64158759
## sesLow                    -0.010300788 0.0005762241 46322.999353 -17.87635695
## raceBlack:genderFemale    -0.004340414 0.0133849777     7.999407  -0.32427499
## raceHispanic:genderFemale  0.001375280 0.0133849777     7.999407   0.10274803
## raceAsian:genderFemale     0.001160785 0.0133849777     7.999407   0.08672297
##                               Pr(>|t|)
## (Intercept)               9.287459e-02
## raceBlack                 5.855971e-01
## raceHispanic              6.577356e-01
## raceAsian                 1.337321e-01
## genderFemale              5.390857e-01
## sesLow                    3.130349e-71
## raceBlack:genderFemale    7.540466e-01
## raceHispanic:genderFemale 9.206920e-01
## raceAsian:genderFemale    9.330232e-01
## 
##   95% CIs:
##                                 2.5 %       97.5 %
## .sig01                             NA           NA
## .sig02                             NA           NA
## .sig03                             NA           NA
## .sigma                             NA           NA
## (Intercept)               -0.04715384  0.001850321
## raceBlack                 -0.02392653  0.013174051
## raceHispanic              -0.02290436  0.014196222
## raceAsian                 -0.03434391  0.002756671
## genderFemale              -0.02462267  0.012477916
## sesLow                    -0.01143017 -0.009171409
## raceBlack:genderFemale    -0.03057449  0.021893661
## raceHispanic:genderFemale -0.02485879  0.027609354
## raceAsian:genderFemale    -0.02507329  0.027394859
## 
##   R² marginal: 0.0120, conditional: 0.3416
## 
## --- delta_sent ---
\end{verbatim}

\begin{verbatim}
##   Spec: bobyqa, dropped (1|name_full) 
##   ANOVA table:
## Type III Analysis of Variance Table with Satterthwaite's method
##             Sum Sq Mean Sq NumDF DenDF F value    Pr(>F)    
## race        9.2767  3.0922     3 46331  5.7250 0.0006514 ***
## gender      1.4299  1.4299     1 46331  2.6473 0.1037341    
## ses         9.1937  9.1937     1 46331 17.0214 3.703e-05 ***
## race:gender 0.8827  0.2942     3 46331  0.5448 0.6516357    
## ---
## Signif. codes:  0 '***' 0.001 '**' 0.01 '*' 0.05 '.' 0.1 ' ' 1
## 
##   Fixed effects:
##                                Estimate  Std. Error          df     t value
## (Intercept)               -0.0483301584 0.026522731    32.93407 -1.82221652
## raceBlack                  0.0139635601 0.013628594 46331.01904  1.02457816
## raceHispanic               0.0253097701 0.013628594 46331.01904  1.85710788
## raceAsian                  0.0431270959 0.013628594 46331.01903  3.16445663
## genderFemale               0.0009275248 0.013628594 46331.01900  0.06805726
## sesLow                     0.0281137212 0.006814297 46331.01905  4.12569637
## raceBlack:genderFemale    -0.0227500704 0.019273743 46331.01901 -1.18036597
## raceHispanic:genderFemale -0.0080101901 0.019273743 46331.01901 -0.41560117
## raceAsian:genderFemale    -0.0172985488 0.019273743 46331.01899 -0.89751891
##                               Pr(>|t|)
## (Intercept)               7.751383e-02
## raceBlack                 3.055676e-01
## raceHispanic              6.330215e-02
## raceAsian                 1.554732e-03
## genderFemale              9.457403e-01
## sesLow                    3.702594e-05
## raceBlack:genderFemale    2.378607e-01
## raceHispanic:genderFemale 6.777038e-01
## raceAsian:genderFemale    3.694468e-01
## 
##   95% CIs:
##                                  2.5 %      97.5 %
## .sig01                              NA          NA
## .sig02                              NA          NA
## .sigma                              NA          NA
## (Intercept)               -0.100313756 0.003653439
## raceBlack                 -0.012747994 0.040675114
## raceHispanic              -0.001401784 0.052021324
## raceAsian                  0.016415542 0.069838650
## genderFemale              -0.025784029 0.027639079
## sesLow                     0.014757944 0.041469498
## raceBlack:genderFemale    -0.060525913 0.015025772
## raceHispanic:genderFemale -0.045786032 0.029765652
## raceAsian:genderFemale    -0.055074391 0.020477293
## 
##   R² marginal: 0.0007, conditional: 0.0940
## 
## --- delta_len ---
\end{verbatim}

\begin{verbatim}
##   Spec: bobyqa, dropped (1|name_full) 
##   ANOVA table:
## Type III Analysis of Variance Table with Satterthwaite's method
##               Sum Sq  Mean Sq NumDF DenDF F value   Pr(>F)   
## race        0.248991 0.082997     3 46331  5.3322 0.001137 **
## gender      0.000003 0.000003     1 46331  0.0002 0.988747   
## ses         0.110836 0.110836     1 46331  7.1207 0.007623 **
## race:gender 0.013107 0.004369     3 46331  0.2807 0.839380   
## ---
## Signif. codes:  0 '***' 0.001 '**' 0.01 '*' 0.05 '.' 0.1 ' ' 1
## 
##   Fixed effects:
##                                Estimate  Std. Error          df    t value
## (Intercept)                0.0262862600 0.005666748    23.23324  4.6386854
## raceBlack                 -0.0020151928 0.002313568 46331.00940 -0.8710323
## raceHispanic               0.0002233674 0.002313568 46331.00941  0.0965467
## raceAsian                 -0.0060632303 0.002313568 46331.00939 -2.6207266
## genderFemale              -0.0005962773 0.002313568 46331.00935 -0.2577306
## sesLow                    -0.0030868314 0.001156784 46331.00948 -2.6684591
## raceBlack:genderFemale     0.0021505508 0.003271880 46331.00942  0.6572829
## raceHispanic:genderFemale -0.0006942852 0.003271880 46331.00942 -0.2121976
## raceAsian:genderFemale     0.0008635836 0.003271880 46331.00940  0.2639411
##                               Pr(>|t|)
## (Intercept)               0.0001119577
## raceBlack                 0.3837410358
## raceHispanic              0.9230868026
## raceAsian                 0.0087771246
## genderFemale              0.7966159977
## sesLow                    0.0076226572
## raceBlack:genderFemale    0.5110022817
## raceHispanic:genderFemale 0.8319537851
## raceAsian:genderFemale    0.7918264693
## 
##   95% CIs:
##                                  2.5 %       97.5 %
## .sig01                              NA           NA
## .sig02                              NA           NA
## .sigma                              NA           NA
## (Intercept)                0.015179638  0.037392881
## raceBlack                 -0.006549704  0.002519318
## raceHispanic              -0.004311143  0.004757878
## raceAsian                 -0.010597741 -0.001528719
## genderFemale              -0.005130788  0.003938234
## sesLow                    -0.005354087 -0.000819576
## raceBlack:genderFemale    -0.004262216  0.008563318
## raceHispanic:genderFemale -0.007107052  0.005718482
## raceAsian:genderFemale    -0.005549183  0.007276350
## 
##   R² marginal: 0.0004, conditional: 0.1299
## 
## --- delta_read ---
##   Spec: default optimizer 
##   ANOVA table:
## Type III Analysis of Variance Table with Satterthwaite's method
##             Sum Sq Mean Sq NumDF DenDF  F value  Pr(>F)    
## race         296.6    98.9     3     8   3.6753 0.06257 .  
## gender        12.1    12.1     1     8   0.4498 0.52133    
## ses         5827.8  5827.8     1 46323 216.6116 < 2e-16 ***
## race:gender   48.6    16.2     3     8   0.6026 0.63135    
## ---
## Signif. codes:  0 '***' 0.001 '**' 0.01 '*' 0.05 '.' 0.1 ' ' 1
## 
##   Fixed effects:
##                               Estimate Std. Error           df      t value
## (Intercept)                0.162714465 0.28284390    16.428594   0.57528010
## raceBlack                  0.450489395 0.18570345     8.001902   2.42585368
## raceHispanic               0.245124644 0.18570345     8.001902   1.31997895
## raceAsian                 -0.004095614 0.18570345     8.001902  -0.02205459
## genderFemale              -0.048577498 0.18570345     8.001902  -0.26158641
## sesLow                    -0.707823363 0.04809323 46323.017525 -14.71773240
## raceBlack:genderFemale    -0.074889381 0.26262434     8.001902  -0.28515781
## raceHispanic:genderFemale -0.159183936 0.26262434     8.001902  -0.60612790
## raceAsian:genderFemale     0.179301127 0.26262434     8.001902   0.68272853
##                               Pr(>|t|)
## (Intercept)               5.728975e-01
## raceBlack                 4.146266e-02
## raceHispanic              2.233509e-01
## raceAsian                 9.829445e-01
## genderFemale              8.002476e-01
## sesLow                    6.399392e-49
## raceBlack:genderFemale    7.827648e-01
## raceHispanic:genderFemale 5.612262e-01
## raceAsian:genderFemale    5.140543e-01
## 
##   95% CIs:
##                                 2.5 %     97.5 %
## .sig01                             NA         NA
## .sig02                             NA         NA
## .sig03                             NA         NA
## .sigma                             NA         NA
## (Intercept)               -0.39164939  0.7170783
## raceBlack                  0.08651732  0.8144615
## raceHispanic              -0.11884743  0.6090967
## raceAsian                 -0.36806769  0.3598765
## genderFemale              -0.41254957  0.3153946
## sesLow                    -0.80208437 -0.6135624
## raceBlack:genderFemale    -0.58962362  0.4398449
## raceHispanic:genderFemale -0.67391818  0.3555503
## raceAsian:genderFemale    -0.33543311  0.6940354
## 
##   R² marginal: 0.0052, conditional: 0.0994
\end{verbatim}

\subsection{Holm-Bonferroni Correction}\label{app:primary-holm-bonferroni-correction}
\begin{verbatim}
## ============================================================ 
## Holm-Bonferroni Correction (two-level)
## ============================================================ 
## 
## Hypothesis-level results (min-p across DVs, Holm-corrected):
## 
## 
## Table: Hypothesis-level Holm-Bonferroni correction
## 
## |            |hypothesis     |term        |  min_p|min_p_dv   | p_adjusted|significant |
## |:-----------|:--------------|:-----------|------:|:----------|----------:|:-----------|
## |delta_sent  |H1_race        |race        | 0.0007|delta_sent |     0.0020|TRUE        |
## |delta_sent1 |H2_gender      |gender      | 0.1037|delta_sent |     0.2075|FALSE       |
## |delta_sim   |H3_ses         |ses         | 0.0000|delta_sim  |     0.0000|TRUE        |
## |delta_read  |H4_interaction |race:gender | 0.6314|delta_read |     0.6314|FALSE       |
## 
## 
## Within-hypothesis DV-level corrections:
## 
##   H1_race (significant):
## 
## 
## |           |DV         |  p_raw| p_holm|significant |
## |:----------|:----------|------:|------:|:-----------|
## |delta_sim  |delta_sim  | 0.2091| 0.2091|FALSE       |
## |delta_sent |delta_sent | 0.0007| 0.0026|TRUE        |
## |delta_len  |delta_len  | 0.0011| 0.0034|TRUE        |
## |delta_read |delta_read | 0.0626| 0.1251|FALSE       |
## 
##   H2_gender: not significant at hypothesis level (p_adj = 0.2075)
## 
##   H3_ses (significant):
## 
## 
## |           |DV         |  p_raw| p_holm|significant |
## |:----------|:----------|------:|------:|:-----------|
## |delta_sim  |delta_sim  | 0.0000| 0.0000|TRUE        |
## |delta_sent |delta_sent | 0.0000| 0.0001|TRUE        |
## |delta_len  |delta_len  | 0.0076| 0.0076|TRUE        |
## |delta_read |delta_read | 0.0000| 0.0000|TRUE        |
## 
##   H4_interaction: not significant at hypothesis level (p_adj = 0.6314)
\end{verbatim}

\begin{verbatim}
## 
##   NOTE: delta_sent ICC = 0.695 (below 0.70 threshold).
##   Results for delta_sent should be interpreted with caveat.
##   Repetition variance may inflate uncertainty beyond reported CIs.
\end{verbatim}

\subsection{Post-Hoc Pairwise Comparisions}\label{app:primary-post-hoc-pairwise-comparisons}
\begin{verbatim}
## ============================================================ 
## Post-hoc Pairwise Comparisons (descriptive, no additional correction)
## ============================================================ 
## 
## 
## --- delta_sim: Race pairwise Cohen's d ---
## 
## 
## |comparison        |       d|  ci_low| ci_high|    n1|    n2|
## |:-----------------|-------:|-------:|-------:|-----:|-----:|
## |White vs Black    |  0.1014|  0.0757|  0.1272| 11632| 11632|
## |White vs Hispanic |  0.0498|  0.0241|  0.0755| 11632| 11632|
## |White vs Asian    |  0.2024|  0.1766|  0.2281| 11632| 11632|
## |Black vs Hispanic | -0.0513| -0.0771| -0.0256| 11632| 11632|
## |Black vs Asian    |  0.0994|  0.0737|  0.1251| 11632| 11632|
## |Hispanic vs Asian |  0.1513|  0.1255|  0.1770| 11632| 11632|
\end{verbatim}

\begin{verbatim}
## 
## --- delta_sim: Model-based Race contrasts (emmeans) ---
##  contrast         estimate      SE  df z.ratio p.value
##  White - Black     0.00755 0.00669 Inf   1.128  0.2595
##  White - Hispanic  0.00367 0.00669 Inf   0.548  0.5838
##  White - Asian     0.01521 0.00669 Inf   2.273  0.0230
##  Black - Hispanic -0.00388 0.00669 Inf  -0.580  0.5621
##  Black - Asian     0.00767 0.00669 Inf   1.146  0.2520
##  Hispanic - Asian  0.01155 0.00669 Inf   1.725  0.0845
## 
## Results are averaged over the levels of: gender, ses 
## Degrees-of-freedom method: asymptotic 
## 
## --- delta_sent: Race pairwise Cohen's d ---
## 
## 
## |comparison        |       d|  ci_low| ci_high|    n1|    n2|
## |:-----------------|-------:|-------:|-------:|-----:|-----:|
## |White vs Black    | -0.0033| -0.0293|  0.0223| 11632| 11632|
## |White vs Hispanic | -0.0277| -0.0534| -0.0020| 11632| 11632|
## |White vs Asian    | -0.0449| -0.0706| -0.0192| 11632| 11632|
## |Black vs Hispanic | -0.0242| -0.0499|  0.0015| 11632| 11632|
## |Black vs Asian    | -0.0412| -0.0669| -0.0155| 11632| 11632|
## |Hispanic vs Asian | -0.0171| -0.0428|  0.0086| 11632| 11632|
\end{verbatim}

\begin{verbatim}
## 
## --- delta_sent: Model-based Race contrasts (emmeans) ---
##  contrast         estimate      SE  df z.ratio p.value
##  White - Black    -0.00259 0.00964 Inf  -0.269  0.7882
##  White - Hispanic -0.02130 0.00964 Inf  -2.211  0.0271
##  White - Asian    -0.03448 0.00964 Inf  -3.578  0.0003
##  Black - Hispanic -0.01872 0.00964 Inf  -1.942  0.0521
##  Black - Asian    -0.03189 0.00964 Inf  -3.309  0.0009
##  Hispanic - Asian -0.01317 0.00964 Inf  -1.367  0.1716
## 
## Results are averaged over the levels of: gender, ses 
## Degrees-of-freedom method: asymptotic 
## 
## --- delta_len: Race pairwise Cohen's d ---
## 
## 
## |comparison        |       d|  ci_low| ci_high|    n1|    n2|
## |:-----------------|-------:|-------:|-------:|-----:|-----:|
## |White vs Black    |  0.0070| -0.0187|  0.0327| 11632| 11632|
## |White vs Hispanic |  0.0009| -0.0248|  0.0266| 11632| 11632|
## |White vs Asian    |  0.0421|  0.0163|  0.0678| 11632| 11632|
## |Black vs Hispanic | -0.0061| -0.0318|  0.0196| 11632| 11632|
## |Black vs Asian    |  0.0354|  0.0097|  0.0611| 11632| 11632|
## |Hispanic vs Asian |  0.0412|  0.0155|  0.0669| 11632| 11632|
\end{verbatim}

\begin{verbatim}
## 
## --- delta_len: Model-based Race contrasts (emmeans) ---
##  contrast          estimate      SE  df z.ratio p.value
##  White - Black     0.000940 0.00164 Inf   0.575  0.5656
##  White - Hispanic  0.000124 0.00164 Inf   0.076  0.9397
##  White - Asian     0.005631 0.00164 Inf   3.442  0.0006
##  Black - Hispanic -0.000816 0.00164 Inf  -0.499  0.6179
##  Black - Asian     0.004692 0.00164 Inf   2.868  0.0041
##  Hispanic - Asian  0.005508 0.00164 Inf   3.367  0.0008
## 
## Results are averaged over the levels of: gender, ses 
## Degrees-of-freedom method: asymptotic 
## 
## --- delta_read: Race pairwise Cohen's d ---
## 
## 
## |comparison        |       d|  ci_low| ci_high|    n1|    n2|
## |:-----------------|-------:|-------:|-------:|-----:|-----:|
## |White vs Black    | -0.0744| -0.1001| -0.0486| 11632| 11632|
## |White vs Hispanic | -0.0305| -0.0562| -0.0048| 11632| 11632|
## |White vs Asian    | -0.0158| -0.0415|  0.0099| 11632| 11632|
## |Black vs Hispanic |  0.0451|  0.0193|  0.0708| 11632| 11632|
## |Black vs Asian    |  0.0597|  0.0340|  0.0854| 11632| 11632|
## |Hispanic vs Asian |  0.0149| -0.0108|  0.0406| 11632| 11632|
\end{verbatim}

\begin{verbatim}
## 
## --- delta_read: Model-based Race contrasts (emmeans) ---
##  contrast         estimate    SE  df z.ratio p.value
##  White - Black     -0.4130 0.131 Inf  -3.146  0.0017
##  White - Hispanic  -0.1655 0.131 Inf  -1.261  0.2075
##  White - Asian     -0.0856 0.131 Inf  -0.652  0.5147
##  Black - Hispanic   0.2475 0.131 Inf   1.885  0.0594
##  Black - Asian      0.3275 0.131 Inf   2.494  0.0126
##  Hispanic - Asian   0.0800 0.131 Inf   0.609  0.5425
## 
## Results are averaged over the levels of: gender, ses 
## Degrees-of-freedom method: asymptotic 
## 
## --- Interaction: Gender within each Race (delta_sim) ---
##   White: Male vs Female d = 0.0840 [0.0476, 0.1203]
##   Black: Male vs Female d = 0.1367 [0.1003, 0.1731]
##   Hispanic: Male vs Female d = 0.0629 [0.0265, 0.0992]
##   Asian: Male vs Female d = 0.0631 [0.0267, 0.0995]
## 
## --- Interaction: Race within each Gender (delta_sim) ---
##   Male: White vs Black d = 0.0738 [0.0375, 0.1102]
##   Male: White vs Hispanic d = 0.0604 [0.0240, 0.0967]
##   Male: White vs Asian d = 0.2142 [0.1778, 0.2507]
##   Male: Black vs Hispanic d = -0.0139 [-0.0502, 0.0225]
##   Male: Black vs Asian d = 0.1383 [0.1019, 0.1747]
##   Male: Hispanic vs Asian d = 0.1533 [0.1169, 0.1897]
##   Female: White vs Black d = 0.1284 [0.0920, 0.1647]
##   Female: White vs Hispanic d = 0.0398 [0.0034, 0.0761]
##   Female: White vs Asian d = 0.1913 [0.1548, 0.2277]
##   Female: Black vs Hispanic d = -0.0874 [-0.1237, -0.0510]
##   Female: Black vs Asian d = 0.0625 [0.0261, 0.0988]
##   Female: Hispanic vs Asian d = 0.1496 [0.1132, 0.1860]
## 
## --- Interaction: Gender within each Race (delta_sent) ---
##   White: Male vs Female d = -0.0012 [-0.0376, 0.0351]
##   Black: Male vs Female d = 0.0280 [-0.0084, 0.0643]
##   Hispanic: Male vs Female d = 0.0092 [-0.0271, 0.0456]
##   Asian: Male vs Female d = 0.0213 [-0.0151, 0.0576]
## 
## --- Interaction: Race within each Gender (delta_sent) ---
##   Male: White vs Black d = -0.0181 [-0.0544, 0.0183]
##   Male: White vs Hispanic d = -0.0327 [-0.0691, 0.0036]
##   Male: White vs Asian d = -0.0562 [-0.0925, -0.0198]
##   Male: Black vs Hispanic d = -0.0147 [-0.0510, 0.0217]
##   Male: Black vs Asian d = -0.0380 [-0.0744, -0.0017]
##   Male: Hispanic vs Asian d = -0.0232 [-0.0595, 0.0131]
##   Female: White vs Black d = 0.0113 [-0.0250, 0.0477]
##   Female: White vs Hispanic d = -0.0226 [-0.0590, 0.0137]
##   Female: White vs Asian d = -0.0336 [-0.0699, 0.0028]
##   Female: Black vs Hispanic d = -0.0336 [-0.0699, 0.0028]
##   Female: Black vs Asian d = -0.0443 [-0.0806, -0.0079]
##   Female: Hispanic vs Asian d = -0.0111 [-0.0474, 0.0253]
## 
## --- Interaction: Gender within each Race (delta_len) ---
##   White: Male vs Female d = 0.0044 [-0.0319, 0.0408]
##   Black: Male vs Female d = -0.0118 [-0.0481, 0.0246]
##   Hispanic: Male vs Female d = 0.0096 [-0.0268, 0.0459]
##   Asian: Male vs Female d = -0.0020 [-0.0384, 0.0343]
## 
## --- Interaction: Race within each Gender (delta_len) ---
##   Male: White vs Black d = 0.0151 [-0.0212, 0.0515]
##   Male: White vs Hispanic d = -0.0016 [-0.0380, 0.0347]
##   Male: White vs Asian d = 0.0455 [0.0091, 0.0818]
##   Male: Black vs Hispanic d = -0.0167 [-0.0530, 0.0197]
##   Male: Black vs Asian d = 0.0307 [-0.0057, 0.0670]
##   Male: Hispanic vs Asian d = 0.0468 [0.0105, 0.0832]
##   Female: White vs Black d = -0.0010 [-0.0374, 0.0353]
##   Female: White vs Hispanic d = 0.0035 [-0.0328, 0.0399]
##   Female: White vs Asian d = 0.0387 [0.0023, 0.0750]
##   Female: Black vs Hispanic d = 0.0046 [-0.0318, 0.0409]
##   Female: Black vs Asian d = 0.0402 [0.0039, 0.0766]
##   Female: Hispanic vs Asian d = 0.0354 [-0.0009, 0.0718]
## 
## --- Interaction: Gender within each Race (delta_read) ---
##   White: Male vs Female d = 0.0088 [-0.0275, 0.0452]
##   Black: Male vs Female d = 0.0220 [-0.0144, 0.0583]
##   Hispanic: Male vs Female d = 0.0387 [0.0024, 0.0750]
##   Asian: Male vs Female d = -0.0245 [-0.0608, 0.0119]
## 
## --- Interaction: Race within each Gender (delta_read) ---
##   Male: White vs Black d = -0.0790 [-0.1154, -0.0427]
##   Male: White vs Hispanic d = -0.0438 [-0.0801, -0.0074]
##   Male: White vs Asian d = 0.0007 [-0.0356, 0.0371]
##   Male: Black vs Hispanic d = 0.0364 [0.0000, 0.0727]
##   Male: Black vs Asian d = 0.0814 [0.0450, 0.1178]
##   Male: Hispanic vs Asian d = 0.0455 [0.0091, 0.0818]
##   Female: White vs Black d = -0.0695 [-0.1058, -0.0331]
##   Female: White vs Hispanic d = -0.0163 [-0.0527, 0.0200]
##   Female: White vs Asian d = -0.0331 [-0.0694, 0.0033]
##   Female: Black vs Hispanic d = 0.0542 [0.0179, 0.0906]
##   Female: Black vs Asian d = 0.0373 [0.0009, 0.0736]
##   Female: Hispanic vs Asian d = -0.0171 [-0.0534, 0.0193]
\end{verbatim}

\subsection{Fixed Effects \& Effect Sizes Summary}\label{app:primary-fixed-effects-effect-sizes-summary}
\begin{verbatim}
## ============================================================ 
## Effect Size Summary (partial eta-squared)
## ============================================================ 
## 
## 
## --- delta_sim ---
##     Parameter Eta2_partial      CI_low CI_high
## 1        race  0.415312730 0.000000000       1
## 2      gender  0.191951188 0.000000000       1
## 3         ses  0.006851342 0.005653297       1
## 4 race:gender  0.028819318 0.000000000       1
##   NOTE: Small DenDF detected. Partial eta-squared values are inflated
##   and not comparable with models using larger DenDF. Interpret with caution.
## 
##   BF10 per term:
##     H1_race: BF10 = 6.39e-13 (Moderate+ evidence for H0 (null))
##     H2_gender: BF10 = 2.74e-09 (Moderate+ evidence for H0 (null))
##     H3_ses: BF10 = 6.64e+66 (Moderate+ evidence for H1 (alternative))
##     H4_interaction: BF10 = 1.24e-07 (Moderate+ evidence for H0 (null))
## 
## --- delta_sent ---
## # Effect Size for ANOVA (Type III)
## 
## Parameter   | Eta2 (partial) |       95% CI
## -------------------------------------------
## race        |       3.71e-04 | [0.00, 1.00]
## gender      |       5.71e-05 | [0.00, 1.00]
## ses         |       3.67e-04 | [0.00, 1.00]
## race:gender |       3.53e-05 | [0.00, 1.00]
## 
## - One-sided CIs: upper bound fixed at [1.00].
##   BF10 per term:
##     H1_race: BF10 = 1.21e-10 (Moderate+ evidence for H0 (null))
##     H2_gender: BF10 = 3.93e-09 (Moderate+ evidence for H0 (null))
##     H3_ses: BF10 = 23.0274 (Moderate+ evidence for H1 (alternative))
##     H4_interaction: BF10 = 2.26e-07 (Moderate+ evidence for H0 (null))
## 
## --- delta_len ---
## # Effect Size for ANOVA (Type III)
## 
## Parameter   | Eta2 (partial) |       95% CI
## -------------------------------------------
## race        |       3.45e-04 | [0.00, 1.00]
## gender      |       4.29e-09 | [0.00, 1.00]
## ses         |       1.54e-04 | [0.00, 1.00]
## race:gender |       1.82e-05 | [0.00, 1.00]
## 
## - One-sided CIs: upper bound fixed at [1.00].
##   BF10 per term:
##     H1_race: BF10 = 4.50e-11 (Moderate+ evidence for H0 (null))
##     H2_gender: BF10 = 7.04e-10 (Moderate+ evidence for H0 (null))
##     H3_ses: BF10 = 0.1631 (Moderate+ evidence for H0 (null))
##     H4_interaction: BF10 = 1.52e-07 (Moderate+ evidence for H0 (null))
## 
## --- delta_read ---
##     Parameter Eta2_partial      CI_low CI_high
## 1        race  0.579461434 0.000000000       1
## 2      gender  0.053216097 0.000000000       1
## 3         ses  0.004654348 0.003674632       1
## 4 race:gender  0.184295291 0.000000000       1
##   NOTE: Small DenDF detected. Partial eta-squared values are inflated
##   and not comparable with models using larger DenDF. Interpret with caution.
## 
##   BF10 per term:
##     H1_race: BF10 = 1.32e-11 (Moderate+ evidence for H0 (null))
##     H2_gender: BF10 = 2.99e-09 (Moderate+ evidence for H0 (null))
##     H3_ses: BF10 = 3.93e+44 (Moderate+ evidence for H1 (alternative))
##     H4_interaction: BF10 = 4.60e-07 (Moderate+ evidence for H0 (null))
\end{verbatim}



\section{Writing Quality Analysis ($H_5$)}\label{app:H5-writing-quality}

\subsection{H5 Primary: Continuous Covariate}\label{app:h5-primary-continuous-covariate}

\begin{verbatim}
## ============================================================
\end{verbatim}

\begin{verbatim}
## H5: Writing Quality as Continuous Covariate
\end{verbatim}

\begin{verbatim}
## ============================================================
\end{verbatim}

\begin{verbatim}
## 
## --- semantic_similarity ---
##   Spec: default optimizer 
##   wq_aggregate: F = 240.2970, p = < .001
##   Partial eta²: 0.5717 [0.4980, 1.0000]
## 
## --- sentiment ---
\end{verbatim}

\begin{verbatim}
##   Spec: bobyqa, dropped (1|name_full) 
##   wq_aggregate: F = 0.0001, p = 0.991
##   Partial eta²: 0.0000 [0.0000, 1.0000]
## 
## --- length_ratio ---
\end{verbatim}

\begin{verbatim}
##   Spec: bobyqa, dropped (1|name_full) 
##   wq_aggregate: F = 85.0327, p = < .001
##   Partial eta²: 0.3208 [0.2324, 1.0000]
## 
## --- readability ---
##   Spec: default optimizer 
##   wq_aggregate: F = 23.3783, p = < .001
##   Partial eta²: 0.1150 [0.0517, 1.0000]
\end{verbatim}

\begin{verbatim}
## 
## 
## H5 Holm-Bonferroni correction:
## 
## 
## |                    |DV                  |  p_raw| p_holm|significant |
## |:-------------------|:-------------------|------:|------:|:-----------|
## |semantic_similarity |semantic_similarity | 0.0000| 0.0000|TRUE        |
## |sentiment           |sentiment           | 0.9907| 0.9907|FALSE       |
## |length_ratio        |length_ratio        | 0.0000| 0.0000|TRUE        |
## |readability         |readability         | 0.0000| 0.0000|TRUE        |
## 
## H5 Bayes Factors:
##   semantic_similarity: BF10 = 1.06e+31 (Moderate+ evidence for H1 (alternative))
##   sentiment: BF10 = 0.0046 (Moderate+ evidence for H0 (null))
##   length_ratio: BF10 = 7.74e+12 (Moderate+ evidence for H1 (alternative))
##   readability: BF10 = 292.4042 (Moderate+ evidence for H1 (alternative))
\end{verbatim}

\subsection{H5 Sensitivity: Log-Transformed Readability}\label{app:h5-sensitivity-log-transformed-readability}

\begin{verbatim}
## H5 Sensitivity: log(readability) for skew mitigation
\end{verbatim}

\begin{verbatim}
## (readability skewness = 3.04, kurtosis = 16.28)
\end{verbatim}

\begin{verbatim}
## ANOVA (log readability):
## Type III Analysis of Variance Table with Satterthwaite's method
##               Sum Sq Mean Sq NumDF DenDF  F value    Pr(>F)    
## race          0.5100  0.1700     3    11   5.0266    0.0196 *  
## gender        0.0004  0.0004     1    11   0.0123    0.9137    
## ses          16.6663 16.6663     1 46323 492.8338 < 2.2e-16 ***
## wq_aggregate  1.2215  1.2215     1   180  36.1217 1.006e-08 ***
## ---
## Signif. codes:  0 '***' 0.001 '**' 0.01 '*' 0.05 '.' 0.1 ' ' 1
## 
##   Untransformed p = < .001
##   Log-transformed p = < .001
##   Direction preserved: TRUE
\end{verbatim}

\subsection{H5 Two-Stage Approach}\label{app:h5-two-stage-approach}

\begin{verbatim}
## ============================================================
\end{verbatim}

\begin{verbatim}
## H5 Two-Stage Approach (mitigates comment-level covariate confound)
\end{verbatim}

\begin{verbatim}
## ============================================================
\end{verbatim}

\begin{verbatim}
## Writing quality is a comment-level variable while comment is a random effect.
\end{verbatim}

\begin{verbatim}
## Stage 1: compute per-comment DV means. Stage 2: OLS on wq_aggregate.
\end{verbatim}

\begin{verbatim}
## Comment-level observations: 182
\end{verbatim}

\begin{verbatim}
## --- semantic_similarity (OLS on comment means) ---
##   Coefficient: 0.0940, SE: 0.0061, p = < .001, R² = 0.5715
## --- sentiment (OLS on comment means) ---
##   Coefficient: -0.0008, SE: 0.0688, p = 0.991, R² = 0.0000
## --- length_ratio (OLS on comment means) ---
##   Coefficient: -0.2214, SE: 0.0240, p = < .001, R² = 0.3210
## --- readability (OLS on comment means) ---
##   Coefficient: 0.9136, SE: 0.1888, p = < .001, R² = 0.1151
\end{verbatim}

\subsection{H5 Secondary: Dimension-Specific}\label{app:h5-secondary-dimension-specific}

\begin{verbatim}
## ============================================================
\end{verbatim}

\begin{verbatim}
## H5 Secondary: Dimension-specific analysis
\end{verbatim}

\begin{verbatim}
## ============================================================
\end{verbatim}

\begin{verbatim}
## 
## --- semantic_similarity ---
##   wq_content: F=280.762, p=< .001
##   wq_organization: F=310.146, p=< .001
##   wq_language_use: F=152.656, p=< .001
##   wq_vocabulary: F=193.575, p=< .001
##   wq_mechanics: F=35.788, p=< .001
##   Holm-corrected:
##     wq_content: p_adj=< .001 *
##     wq_organization: p_adj=< .001 *
##     wq_language_use: p_adj=< .001 *
##     wq_vocabulary: p_adj=< .001 *
##     wq_mechanics: p_adj=< .001 *
## 
## --- sentiment ---
\end{verbatim}

\begin{verbatim}
##   wq_content: F=0.026, p=0.872
\end{verbatim}

\begin{verbatim}
##   wq_organization: F=0.044, p=0.834
\end{verbatim}

\begin{verbatim}
##   wq_language_use: F=1.047, p=0.308
\end{verbatim}

\begin{verbatim}
##   wq_vocabulary: F=0.849, p=0.358
\end{verbatim}

\begin{verbatim}
##   wq_mechanics: F=0.025, p=0.875
##   Holm-corrected:
##     wq_content: p_adj=1.000 
##     wq_organization: p_adj=1.000 
##     wq_language_use: p_adj=1.000 
##     wq_vocabulary: p_adj=1.000 
##     wq_mechanics: p_adj=1.000 
## 
## --- length_ratio ---
\end{verbatim}

\begin{verbatim}
##   wq_content: F=118.833, p=< .001
\end{verbatim}

\begin{verbatim}
##   wq_organization: F=114.222, p=< .001
\end{verbatim}

\begin{verbatim}
##   wq_language_use: F=67.483, p=< .001
\end{verbatim}

\begin{verbatim}
##   wq_vocabulary: F=162.096, p=< .001
\end{verbatim}

\begin{verbatim}
##   wq_mechanics: F=0.972, p=0.326
##   Holm-corrected:
##     wq_content: p_adj=< .001 *
##     wq_organization: p_adj=< .001 *
##     wq_language_use: p_adj=< .001 *
##     wq_vocabulary: p_adj=< .001 *
##     wq_mechanics: p_adj=0.326 
## 
## --- readability ---
##   wq_content: F=33.268, p=< .001
##   wq_organization: F=32.551, p=< .001
##   wq_language_use: F=10.484, p=0.001
##   wq_vocabulary: F=33.668, p=< .001
##   wq_mechanics: F=1.293, p=0.257
##   Holm-corrected:
##     wq_content: p_adj=< .001 *
##     wq_organization: p_adj=< .001 *
##     wq_language_use: p_adj=0.003 *
##     wq_vocabulary: p_adj=< .001 *
##     wq_mechanics: p_adj=0.257
\end{verbatim}

\subsection{H5 Robustness: Quartile-Based}\label{app:h5-robustness-quartile-based}

\begin{verbatim}
## ============================================================
\end{verbatim}

\begin{verbatim}
## H5 Robustness: wq_quartile as factor (vs continuous)
\end{verbatim}

\begin{verbatim}
## ============================================================
\end{verbatim}

\begin{verbatim}
## --- semantic_similarity ---
##   wq_quartile: F=55.262, p=< .001
##   Spec: default optimizer
## 
## --- sentiment ---
\end{verbatim}

\begin{verbatim}
##   wq_quartile: F=0.513, p=0.674
##   Spec: bobyqa, dropped (1|name_full)
## 
## --- length_ratio ---
\end{verbatim}

\begin{verbatim}
##   wq_quartile: F=26.654, p=< .001
##   Spec: bobyqa, dropped (1|name_full)
## 
## --- readability ---
##   wq_quartile: F=12.462, p=< .001
##   Spec: default optimizer
\end{verbatim}

\subsection{H5 Exploratory Interactions}\label{app:h5-exploratory-interactions}

\begin{verbatim}
## ============================================================
\end{verbatim}

\begin{verbatim}
## H5 Exploratory: Demographic x Writing Quality Interactions
\end{verbatim}

\begin{verbatim}
## (Descriptive — no correction applied)
\end{verbatim}

\begin{verbatim}
## ============================================================
\end{verbatim}

\begin{verbatim}
## 
## --- semantic_similarity ---
##   race:wq_aggregate: F=4.452, p=0.004
##   gender:wq_aggregate: F=73.041, p=< .001
##   ses:wq_aggregate: F=187.768, p=< .001
## 
## --- sentiment ---
\end{verbatim}

\begin{verbatim}
##   race:wq_aggregate: F=0.652, p=0.581
##   gender:wq_aggregate: F=3.198, p=0.074
##   ses:wq_aggregate: F=10.486, p=0.001
## 
## --- length_ratio ---
\end{verbatim}

\begin{verbatim}
##   race:wq_aggregate: F=0.788, p=0.500
##   gender:wq_aggregate: F=0.248, p=0.618
##   ses:wq_aggregate: F=0.997, p=0.318
## 
## --- readability ---
##   race:wq_aggregate: F=1.048, p=0.370
##   gender:wq_aggregate: F=0.198, p=0.656
##   ses:wq_aggregate: F=21.398, p=< .001
\end{verbatim}

\section{Error Injection Analysis ($H_6$)}\label{app:h6-error-injection}

\subsection{H6 Primary: Error Effect}\label{app:h6-primary-error-effect}

\begin{verbatim}
## ============================================================
\end{verbatim}

\begin{verbatim}
## H6: Error Injection Effect (Low/Mid-Low WQ only)
\end{verbatim}

\begin{verbatim}
## ============================================================
\end{verbatim}

\begin{verbatim}
## Error injection dataset:
\end{verbatim}

\begin{verbatim}
##   Total rows: 48064
\end{verbatim}

\begin{verbatim}
##   Original: 24032
\end{verbatim}

\begin{verbatim}
##   Error-injected: 24032
\end{verbatim}

\begin{verbatim}
##   Unique comments: 94
\end{verbatim}

\begin{verbatim}
##   WQ quartiles: Low, Mid-Low
\end{verbatim}

\begin{verbatim}
## --- semantic_similarity ---
##   Spec: default optimizer 
##   error_added: F=3.3774, p=0.066
##   Cohen's d (Original vs Error): -0.0083 [-0.0262, 0.0096]
## 
## --- sentiment ---
##   Spec: default optimizer 
##   error_added: F=0.4982, p=0.480
##   Cohen's d (Original vs Error): -0.0040 [-0.0219, 0.0139]
## 
## --- length_ratio ---
\end{verbatim}

\begin{verbatim}
##   Spec: bobyqa, dropped (1|name_full) 
##   error_added: F=11.3457, p=< .001
##   Cohen's d (Original vs Error): -0.0140 [-0.0319, 0.0038]
## 
## --- readability ---
##   Spec: default optimizer 
##   error_added: F=0.2726, p=0.602
##   Cohen's d (Original vs Error): -0.0035 [-0.0214, 0.0144]
\end{verbatim}

\begin{verbatim}
## 
## H6 Holm-Bonferroni correction:
## 
## 
## |                    |DV                  |  p_raw| p_holm|significant |
## |:-------------------|:-------------------|------:|------:|:-----------|
## |semantic_similarity |semantic_similarity | 0.0661| 0.1983|FALSE       |
## |sentiment           |sentiment           | 0.4803| 0.9606|FALSE       |
## |length_ratio        |length_ratio        | 0.0008| 0.0030|TRUE        |
## |readability         |readability         | 0.6016| 0.9606|FALSE       |
## 
## H6 Bayes Factors:
##   semantic_similarity: BF10 = 0.0247 (Moderate+ evidence for H0 (null))
##   sentiment: BF10 = 0.0059 (Moderate+ evidence for H0 (null))
##   length_ratio: BF10 = 1.3268 (Inconclusive (1/3 < BF < 3))
\end{verbatim}

\begin{verbatim}
##   readability: BF10 = 0.0052 (Moderate+ evidence for H0 (null))
\end{verbatim}

\subsection{H6 Sensitivity: Log-Transformed Readability}\label{app:h6-sensitivity-log-transformed-readability}

\begin{verbatim}
## H6 Sensitivity: log(readability) for skew mitigation
\end{verbatim}

\begin{verbatim}
## ANOVA (log readability):
## Type III Analysis of Variance Table with Satterthwaite's method
##              Sum Sq Mean Sq NumDF DenDF  F value  Pr(>F)    
## race         0.5029  0.1676     3    11   4.6338 0.02494 *  
## gender       0.0030  0.0030     1    11   0.0828 0.77886    
## ses         25.4546 25.4546     1 47946 703.6417 < 2e-16 ***
## error_added  0.0034  0.0034     1 47946   0.0939 0.75934    
## ---
## Signif. codes:  0 '***' 0.001 '**' 0.01 '*' 0.05 '.' 0.1 ' ' 1
## 
##   Untransformed p = 0.602
##   Log-transformed p = 0.759
##   Direction preserved: TRUE
\end{verbatim}

\subsection{H6 Exploratory Interactions}\label{app:h6-exploratory-interactions}

\begin{verbatim}
## ============================================================
\end{verbatim}

\begin{verbatim}
## H6 Exploratory: Demographic x Error Interactions
\end{verbatim}

\begin{verbatim}
## (Descriptive — no correction applied)
\end{verbatim}

\begin{verbatim}
## ============================================================
\end{verbatim}

\begin{verbatim}
## --- semantic_similarity ---
##   race:error_added: F=0.340, p=0.796
##   gender:error_added: F=0.210, p=0.646
##   ses:error_added: F=1.161, p=0.281
## 
## --- sentiment ---
##   race:error_added: F=0.081, p=0.971
##   gender:error_added: F=0.002, p=0.967
##   ses:error_added: F=1.447, p=0.229
## 
## --- length_ratio ---
\end{verbatim}

\begin{verbatim}
##   race:error_added: F=0.227, p=0.878
##   gender:error_added: F=0.415, p=0.520
##   ses:error_added: F=0.103, p=0.749
## 
## --- readability ---
##   race:error_added: F=0.397, p=0.755
##   gender:error_added: F=0.165, p=0.685
##   ses:error_added: F=0.243, p=0.622
\end{verbatim}

\section{Sensitivity \& Robustness}\label{app:sensitivity-robustness}

\subsection{Alternative Prompts}\label{app:alternative-prompts}

\begin{verbatim}
## ============================================================
\end{verbatim}

\begin{verbatim}
## Robustness: Alternative Prompts (25-comment subset)
\end{verbatim}

\begin{verbatim}
## H6 excluded (insufficient Low/Mid-Low)
\end{verbatim}

\begin{verbatim}
## ============================================================
\end{verbatim}

\begin{verbatim}
## Prompt types available: length_pressure, systematic
\end{verbatim}

\begin{verbatim}
## Unique comments: 25
\end{verbatim}

\begin{verbatim}
## 
## ===== Prompt: length_pressure =====
## 
##   Rows: 6400 
## 
##   -- Omnibus (composite_index) --
\end{verbatim}

\begin{verbatim}
##   Omnibus p = < .001 (relaxed threshold: .10)
##   Decision: CONVERGENT
## 
##   -- delta_sim --
##     race: F=0.662, p=0.598 
##     gender: F=3.719, p=0.090 (convergent)
##     ses: F=92.358, p=< .001 (convergent)
##     race:gender: F=0.209, p=0.887 
## 
##   -- delta_sent --
\end{verbatim}

\begin{verbatim}
##     race: F=1.116, p=0.341 
##     gender: F=0.130, p=0.718 
##     ses: F=5.538, p=0.019 (convergent)
##     race:gender: F=0.248, p=0.863 
## 
##   -- delta_len --
\end{verbatim}

\begin{verbatim}
##     race: F=0.472, p=0.702 
##     gender: F=1.437, p=0.231 
##     ses: F=0.140, p=0.709 
##     race:gender: F=0.142, p=0.935 
## 
##   -- delta_read --
\end{verbatim}

\begin{verbatim}
##     race: F=12.893, p=< .001 (convergent)
##     gender: F=9.233, p=0.002 (convergent)
##     ses: F=108.487, p=< .001 (convergent)
##     race:gender: F=0.077, p=0.973 
## 
##   -- H5 (wq_aggregate) --
##     semantic_similarity wq_aggregate: F=29.531, p=< .001 (convergent)
\end{verbatim}

\begin{verbatim}
##     sentiment wq_aggregate: F=0.001, p=0.979
\end{verbatim}

\begin{verbatim}
##     length_ratio wq_aggregate: F=5.004, p=0.035 (convergent)
\end{verbatim}

\begin{verbatim}
##     readability wq_aggregate: F=14.650, p=< .001 (convergent)
## 
## ===== Prompt: systematic =====
## 
##   Rows: 6400 
## 
##   -- Omnibus (composite_index) --
\end{verbatim}

\begin{verbatim}
##   Omnibus p = 0.001 (relaxed threshold: .10)
##   Decision: CONVERGENT
## 
##   -- delta_sim --
##     race: F=0.351, p=0.790 
##     gender: F=3.114, p=0.116 
##     ses: F=19.100, p=< .001 (convergent)
##     race:gender: F=0.395, p=0.760 
## 
##   -- delta_sent --
\end{verbatim}

\begin{verbatim}
##     race: F=0.733, p=0.532 
##     gender: F=0.032, p=0.857 
##     ses: F=0.001, p=0.969 
##     race:gender: F=0.634, p=0.593 
## 
##   -- delta_len --
\end{verbatim}

\begin{verbatim}
##     race: F=1.720, p=0.161 
##     gender: F=0.053, p=0.818 
##     ses: F=0.150, p=0.698 
##     race:gender: F=0.300, p=0.825 
## 
##   -- delta_read --
\end{verbatim}

\begin{verbatim}
##     race: F=0.446, p=0.720 
##     gender: F=0.005, p=0.943 
##     ses: F=16.156, p=< .001 (convergent)
##     race:gender: F=1.132, p=0.335 
## 
##   -- H5 (wq_aggregate) --
##     semantic_similarity wq_aggregate: F=45.849, p=< .001 (convergent)
\end{verbatim}

\begin{verbatim}
##     sentiment wq_aggregate: F=0.041, p=0.842
\end{verbatim}

\begin{verbatim}
##     length_ratio wq_aggregate: F=4.101, p=0.055 (convergent)
\end{verbatim}

\begin{verbatim}
##     readability wq_aggregate: F=9.633, p=0.005 (convergent)
\end{verbatim}

\subsection{Name Sensitivity}\label{app:name-sensitivity}

\begin{verbatim}
## ============================================================
\end{verbatim}

\begin{verbatim}
## Name Sensitivity Analysis
\end{verbatim}

\begin{verbatim}
## ============================================================
\end{verbatim}

\begin{verbatim}
## name_full RE variance from omnibus model:
##   Variance: 0.000350, SD: 0.018719
\end{verbatim}

\begin{verbatim}
## 
## --- composite_index ---
\end{verbatim}

\begin{verbatim}
##   name_full RE variance: 0.000350, SD: 0.018719
## 
## 
## Table: Coefficient comparison: composite_index
## 
## |             |term         | coef_random| coef_fixed| pct_change|flag |
## |:------------|:------------|-----------:|----------:|----------:|:----|
## |raceBlack    |raceBlack    |      0.0276|    -0.0056|   120.3288|***  |
## |raceHispanic |raceHispanic |     -0.0055|    -0.0156|   182.0785|***  |
## |raceAsian    |raceAsian    |      0.0447|     0.0801|    79.2499|***  |
## |genderFemale |genderFemale |     -0.0023|     0.0050|   320.9670|***  |
## |sesLow       |sesLow       |      0.0198|     0.0198|     0.0000|     |
## 
##   Individual name coefficients:
##                                 Estimate  Std. Error       df    t value
## name_fullDarnell Washington  0.043525795 0.013350070 46323.17  3.2603420
## name_fullHung Nguyen        -0.068012220 0.009439925 46323.17 -7.2047413
## name_fullJermaine Jefferson  0.025606489 0.013350070 46323.17  1.9180789
## name_fullJohn Murphy        -0.011410112 0.013350070 46323.17 -0.8546855
## name_fullJose Garcia         0.016790556 0.013350070 46323.17  1.2577129
## name_fullJuan Rodriguez      0.006214422 0.013350070 46323.17  0.4654973
## name_fullLatoya Washington   0.016810631 0.009439925 46323.17  1.7808013
## name_fullLuz Garcia          0.003129473 0.009439925 46323.17  0.3315146
## name_fullMary Miller        -0.011691967 0.009439925 46323.17 -1.2385656
## name_fullMichael Miller      0.014213748 0.013350070 46323.17  1.0646946
## name_fullPhuong Nguyen      -0.013224535 0.009439925 46323.17 -1.4009153
##                                 Pr(>|t|)
## name_fullDarnell Washington 1.113583e-03
## name_fullHung Nguyen        5.903811e-13
## name_fullJermaine Jefferson 5.510714e-02
## name_fullJohn Murphy        3.927297e-01
## name_fullJose Garcia        2.085019e-01
## name_fullJuan Rodriguez     6.415775e-01
## name_fullLatoya Washington  7.495148e-02
## name_fullLuz Garcia         7.402573e-01
## name_fullMary Miller        2.155127e-01
## name_fullMichael Miller     2.870197e-01
## name_fullPhuong Nguyen      1.612461e-01
## 
## --- delta_sim ---
\end{verbatim}

\begin{verbatim}
##   name_full RE variance: 0.000088, SD: 0.009394
## 
## 
## Table: Coefficient comparison: delta_sim
## 
## |             |term         | coef_random| coef_fixed| pct_change|flag |
## |:------------|:------------|-----------:|----------:|----------:|:----|
## |raceBlack    |raceBlack    |     -0.0054|    -0.0143|   166.0399|***  |
## |raceHispanic |raceHispanic |     -0.0044|    -0.0009|    78.3760|***  |
## |raceAsian    |raceAsian    |     -0.0158|    -0.0250|    58.4254|***  |
## |genderFemale |genderFemale |     -0.0061|    -0.0064|     5.5418|     |
## |sesLow       |sesLow       |     -0.0103|    -0.0103|     0.0000|     |
## 
##   Individual name coefficients:
##                                  Estimate  Std. Error    df    t value
## name_fullDarnell Washington  0.0208163207 0.002304896 46323  9.0313475
## name_fullHung Nguyen         0.0124703009 0.001629808 46323  7.6513931
## name_fullJermaine Jefferson -0.0089475905 0.002304896 46323 -3.8819924
## name_fullJohn Murphy        -0.0052989810 0.002304896 46323 -2.2990105
## name_fullJose Garcia        -0.0062282637 0.002304896 46323 -2.7021881
## name_fullJuan Rodriguez     -0.0065815019 0.002304896 46323 -2.8554436
## name_fullLatoya Washington   0.0038609405 0.001629808 46323  2.3689543
## name_fullLuz Garcia         -0.0093861680 0.001629808 46323 -5.7590640
## name_fullMary Miller        -0.0053116393 0.001629808 46323 -3.2590585
## name_fullMichael Miller     -0.0006856956 0.002304896 46323 -0.2974952
## name_fullPhuong Nguyen       0.0154649082 0.001629808 46323  9.4887920
##                                 Pr(>|t|)
## name_fullDarnell Washington 1.759162e-19
## name_fullHung Nguyen        2.026491e-14
## name_fullJermaine Jefferson 1.037477e-04
## name_fullJohn Murphy        2.150877e-02
## name_fullJose Garcia        6.890987e-03
## name_fullJuan Rodriguez     4.299584e-03
## name_fullLatoya Washington  1.784253e-02
## name_fullLuz Garcia         8.511512e-09
## name_fullMary Miller        1.118632e-03
## name_fullMichael Miller     7.660898e-01
## name_fullPhuong Nguyen      2.444019e-21
## 
## --- delta_sent ---
\end{verbatim}

\begin{verbatim}
##   name_full was dropped from RE model (singular fit)
## 
## 
## Table: Coefficient comparison: delta_sent
## 
## |             |term         | coef_random| coef_fixed| pct_change|flag |
## |:------------|:------------|-----------:|----------:|----------:|:----|
## |raceBlack    |raceBlack    |      0.0140|    -0.0069|   149.5129|***  |
## |raceHispanic |raceHispanic |      0.0253|     0.0149|    41.2237|***  |
## |raceAsian    |raceAsian    |      0.0431|     0.0247|    42.7538|***  |
## |genderFemale |genderFemale |      0.0009|    -0.0102|  1197.7440|***  |
## |sesLow       |sesLow       |      0.0281|     0.0281|     0.0000|     |
## 
##   Individual name coefficients:
##                                 Estimate Std. Error       df    t value
## name_fullDarnell Washington  0.021581100 0.02725808 46323.02  0.7917323
## name_fullHung Nguyen         0.028593713 0.01927437 46323.02  1.4835095
## name_fullJermaine Jefferson  0.011890347 0.02725808 46323.02  0.4362137
## name_fullJohn Murphy        -0.002986188 0.02725808 46323.02 -0.1095524
## name_fullJose Garcia         0.004053090 0.02725808 46323.02  0.1486932
## name_fullJuan Rodriguez      0.008530966 0.02725808 46323.02  0.3129702
## name_fullLatoya Washington   0.010190053 0.01927437 46323.02  0.5286841
## name_fullLuz Garcia          0.018782422 0.01927437 46323.02  0.9744766
## name_fullMary Miller         0.013935537 0.01927437 46323.02  0.7230087
## name_fullMichael Miller     -0.005297021 0.02725808 46323.02 -0.1943285
## name_fullPhuong Nguyen       0.016215362 0.01927437 46323.02  0.8412914
##                              Pr(>|t|)
## name_fullDarnell Washington 0.4285208
## name_fullHung Nguyen        0.1379459
## name_fullJermaine Jefferson 0.6626837
## name_fullJohn Murphy        0.9127648
## name_fullJose Garcia        0.8817964
## name_fullJuan Rodriguez     0.7543047
## name_fullLatoya Washington  0.5970272
## name_fullLuz Garcia         0.3298251
## name_fullMary Miller        0.4696782
## name_fullMichael Miller     0.8459195
## name_fullPhuong Nguyen      0.4001891
## 
## --- delta_len ---
\end{verbatim}

\begin{verbatim}
##   name_full was dropped from RE model (singular fit)
## 
## 
## Table: Coefficient comparison: delta_len
## 
## |             |term         | coef_random| coef_fixed| pct_change|flag |
## |:------------|:------------|-----------:|----------:|----------:|:----|
## |raceBlack    |raceBlack    |     -0.0020|     0.0000|    99.2716|***  |
## |raceHispanic |raceHispanic |      0.0002|    -0.0006|   382.3493|***  |
## |raceAsian    |raceAsian    |     -0.0061|    -0.0036|    40.9834|***  |
## |genderFemale |genderFemale |     -0.0006|     0.0013|   312.3370|***  |
## |sesLow       |sesLow       |     -0.0031|    -0.0031|     0.0000|     |
## 
##   Individual name coefficients:
##                                  Estimate  Std. Error       df     t value
## name_fullDarnell Washington -0.0024238593 0.004627191 46323.01 -0.52382953
## name_fullHung Nguyen        -0.0022389165 0.003271918 46323.01 -0.68428252
## name_fullJermaine Jefferson  0.0011537460 0.004627191 46323.01  0.24934046
## name_fullJohn Murphy         0.0030492318 0.004627191 46323.01  0.65898117
## name_fullJose Garcia         0.0047189539 0.004627191 46323.01  1.01983120
## name_fullJuan Rodriguez     -0.0002799527 0.004627191 46323.01 -0.06050165
## name_fullLatoya Washington  -0.0006938014 0.003271918 46323.01 -0.21204731
## name_fullLuz Garcia         -0.0006743587 0.003271918 46323.01 -0.20610500
## name_fullMary Miller        -0.0009938757 0.003271918 46323.01 -0.30375932
## name_fullMichael Miller     -0.0003183179 0.004627191 46323.01 -0.06879290
## name_fullPhuong Nguyen      -0.0042365388 0.003271918 46323.01 -1.29481806
##                              Pr(>|t|)
## name_fullDarnell Washington 0.6003996
## name_fullHung Nguyen        0.4938002
## name_fullJermaine Jefferson 0.8030985
## name_fullJohn Murphy        0.5099111
## name_fullJose Garcia        0.3078138
## name_fullJuan Rodriguez     0.9517564
## name_fullLatoya Washington  0.8320711
## name_fullLuz Garcia         0.8367098
## name_fullMary Miller        0.7613126
## name_fullMichael Miller     0.9451548
## name_fullPhuong Nguyen      0.1953895
## 
## --- delta_read ---
\end{verbatim}

\begin{verbatim}
##   name_full RE variance: 0.025234, SD: 0.158852
## 
## 
## Table: Coefficient comparison: delta_read
## 
## |             |term         | coef_random| coef_fixed| pct_change|flag |
## |:------------|:------------|-----------:|----------:|----------:|:----|
## |raceBlack    |raceBlack    |      0.4505|     0.3407|    24.3783|***  |
## |raceHispanic |raceHispanic |      0.2451|     0.1262|    48.5140|***  |
## |raceAsian    |raceAsian    |     -0.0041|    -0.1274|  3010.0735|***  |
## |genderFemale |genderFemale |     -0.0486|    -0.2629|   441.1515|***  |
## |sesLow       |sesLow       |     -0.7078|    -0.7078|     0.0000|     |
## 
##   Individual name coefficients:
##                                Estimate Std. Error       df    t value
## name_fullDarnell Washington -0.02912872  0.1923730 46323.01 -0.1514180
## name_fullHung Nguyen        -0.11571787  0.1360282 46323.01 -0.8506902
## name_fullJermaine Jefferson -0.11350806  0.1923730 46323.01 -0.5900417
## name_fullJohn Murphy        -0.18009300  0.1923730 46323.01 -0.9361659
## name_fullJose Garcia         0.05748594  0.1923730 46323.01  0.2988255
## name_fullJuan Rodriguez     -0.18192643  0.1923730 46323.01 -0.9456965
## name_fullLatoya Washington   0.13618518  0.1360282 46323.01  1.0011538
## name_fullLuz Garcia         -0.01420764  0.1360282 46323.01 -0.1044463
## name_fullMary Miller         0.06632086  0.1360282 46323.01  0.4875522
## name_fullMichael Miller     -0.18218686  0.1923730 46323.01 -0.9470502
## name_fullPhuong Nguyen       0.67148510  0.1360282 46323.01  4.9363660
##                                 Pr(>|t|)
## name_fullDarnell Washington 8.796467e-01
## name_fullHung Nguyen        3.949459e-01
## name_fullJermaine Jefferson 5.551656e-01
## name_fullJohn Murphy        3.491927e-01
## name_fullJose Garcia        7.650746e-01
## name_fullJuan Rodriguez     3.443084e-01
## name_fullLatoya Washington  3.167577e-01
## name_fullLuz Garcia         9.168156e-01
## name_fullMary Miller        6.258693e-01
## name_fullMichael Miller     3.436181e-01
## name_fullPhuong Nguyen      7.986773e-07
\end{verbatim}

\section{Exploratory Analyses}\label{app:exploratory}

\subsection{Higher-Order Interactions}\label{app:higher-order-interactions}

\begin{verbatim}
## ============================================================
\end{verbatim}

\begin{verbatim}
## Exploratory: Higher-Order Interactions
\end{verbatim}

\begin{verbatim}
## (Descriptive — limited power)
\end{verbatim}

\begin{verbatim}
## ============================================================
\end{verbatim}

\begin{verbatim}
## 
## --- delta_sim ---
##   Race x SES: F=10.427, p=< .001
##   Gender x SES: F=0.167, p=0.683
##   Race x Gender x SES: F=2.757, p=0.041
## 
## --- delta_sent ---
\end{verbatim}

\begin{verbatim}
##   Race x SES: F=0.190, p=0.903
\end{verbatim}

\begin{verbatim}
##   Gender x SES: F=0.214, p=0.644
\end{verbatim}

\begin{verbatim}
##   Race x Gender x SES: F=0.277, p=0.842
## 
## --- delta_len ---
\end{verbatim}

\begin{verbatim}
##   Race x SES: F=1.565, p=0.195
\end{verbatim}

\begin{verbatim}
##   Gender x SES: F=1.614, p=0.204
\end{verbatim}

\begin{verbatim}
##   Race x Gender x SES: F=0.023, p=0.995
## 
## --- delta_read ---
##   Race x SES: F=0.239, p=0.869
##   Gender x SES: F=0.307, p=0.579
##   Race x Gender x SES: F=0.124, p=0.946
\end{verbatim}

\subsection{Comment Characteristic
Moderators}\label{app:comment-characteristic-moderators}

\begin{verbatim}
## ============================================================
\end{verbatim}

\begin{verbatim}
## Exploratory: Comment Characteristic Moderators
\end{verbatim}

\begin{verbatim}
## ============================================================
\end{verbatim}

\begin{verbatim}
## 
## --- composite_index ---
##   race:length_class: F=8.286, p=< .001
##   length_class:gender: F=6.761, p=0.009
##   length_class:ses: F=0.061, p=0.804
##   race:sophistication_class: F=1.332, p=0.262
##   sophistication_class:gender: F=5.339, p=0.021
##   sophistication_class:ses: F=0.885, p=0.347
##   race:stance: F=1.659, p=0.093
##   stance:gender: F=2.692, p=0.044
##   stance:ses: F=2.686, p=0.045
## 
## --- delta_sim ---
##   race:length_class: F=19.872, p=< .001
##   length_class:gender: F=65.336, p=< .001
##   length_class:ses: F=33.793, p=< .001
##   race:sophistication_class: F=9.474, p=< .001
##   sophistication_class:gender: F=72.756, p=< .001
##   sophistication_class:ses: F=92.295, p=< .001
##   race:stance: F=5.675, p=< .001
##   stance:gender: F=3.500, p=0.015
##   stance:ses: F=20.809, p=< .001
## 
## --- delta_sent ---
\end{verbatim}

\begin{verbatim}
##   race:length_class: F=2.566, p=0.053
##   length_class:gender: F=0.030, p=0.863
##   length_class:ses: F=0.003, p=0.957
\end{verbatim}

\begin{verbatim}
##   race:sophistication_class: F=0.494, p=0.686
##   sophistication_class:gender: F=1.042, p=0.307
##   sophistication_class:ses: F=0.596, p=0.440
\end{verbatim}

\begin{verbatim}
##   race:stance: F=0.748, p=0.665
##   stance:gender: F=1.309, p=0.269
##   stance:ses: F=3.435, p=0.016
## 
## --- delta_len ---
\end{verbatim}

\begin{verbatim}
##   race:length_class: F=1.745, p=0.155
##   length_class:gender: F=0.174, p=0.676
##   length_class:ses: F=0.304, p=0.581
\end{verbatim}

\begin{verbatim}
##   race:sophistication_class: F=0.360, p=0.782
##   sophistication_class:gender: F=0.774, p=0.379
##   sophistication_class:ses: F=5.969, p=0.015
\end{verbatim}

\begin{verbatim}
##   race:stance: F=0.551, p=0.838
##   stance:gender: F=0.604, p=0.612
##   stance:ses: F=0.022, p=0.996
## 
## --- delta_read ---
##   race:length_class: F=1.891, p=0.129
##   length_class:gender: F=1.068, p=0.301
##   length_class:ses: F=8.287, p=0.004
##   race:sophistication_class: F=1.337, p=0.260
##   sophistication_class:gender: F=0.115, p=0.735
##   sophistication_class:ses: F=7.537, p=0.006
##   race:stance: F=0.939, p=0.489
##   stance:gender: F=0.322, p=0.810
##   stance:ses: F=1.472, p=0.220
\end{verbatim}

\subsection{Model Family Effects}\label{app:model-family-effects}

\begin{verbatim}
## ============================================================
\end{verbatim}

\begin{verbatim}
## Exploratory: Model Family Effects
\end{verbatim}

\begin{verbatim}
## ============================================================
\end{verbatim}

\begin{verbatim}
## Rows by provider:
\end{verbatim}

\begin{verbatim}
## 
##   Anthropic      Google Meta/Ollama      OpenAI 
##       11648       11648       11648       11584
\end{verbatim}

\begin{verbatim}
## 
## --- delta_sim ---
##   race:provider: F=4.726, p=< .001
##   provider:gender: F=5.079, p=0.002
##   provider:ses: F=3.406, p=0.017
## 
##   Per-model means:
## # A tibble: 8 x 4
##   model_name              mean_dv  sd_dv     n
##   <fct>                     <dbl>  <dbl> <int>
## 1 claude-4-opus           0.00240 0.0582  5824
## 2 claude-4-sonnet         0.00481 0.0545  5824
## 3 gemini-2.5-flash       -0.0802  0.0906  5824
## 4 gemini-3-flash-preview -0.0343  0.0647  5824
## 5 gpt-4                  -0.0590  0.0709  5760
## 6 gpt-4.1                -0.0532  0.0667  5824
## 7 llama3-70b             -0.0424  0.0752  5824
## 8 llama3.3-70b           -0.0380  0.0765  5824
## 
## --- composite_index ---
##   race:provider: F=5.104, p=< .001
##   provider:gender: F=2.156, p=0.091
##   provider:ses: F=6.462, p=< .001
## 
##   Per-model means:
## # A tibble: 8 x 4
##   model_name             mean_dv sd_dv     n
##   <fct>                    <dbl> <dbl> <int>
## 1 claude-4-opus            0.766 0.367  5824
## 2 claude-4-sonnet          0.777 0.371  5824
## 3 gemini-2.5-flash         0.769 0.361  5824
## 4 gemini-3-flash-preview   0.770 0.354  5824
## 5 gpt-4                    0.779 0.357  5760
## 6 gpt-4.1                  0.779 0.345  5824
## 7 llama3-70b               0.779 0.373  5824
## 8 llama3.3-70b             0.780 0.363  5824
\end{verbatim}

\subsection{Cross-Agency Comparison}\label{app:cross-agency-comparison}

\begin{verbatim}
## ============================================================
\end{verbatim}

\begin{verbatim}
## Exploratory: Cross-Agency Comparison (DOI vs EPA)
\end{verbatim}

\begin{verbatim}
## DOI is primary; EPA is descriptive. Confounded with stance composition.
\end{verbatim}

\begin{verbatim}
## ============================================================
\end{verbatim}

\begin{verbatim}
## 
## ===== DOI-2025-0004 =====
##   N rows: 26080 
##   N comments: 102
\end{verbatim}

\begin{verbatim}
##   Omnibus p = < .001
##   delta_sim race: F=3.127, p=0.088
##   delta_sim gender: F=1.302, p=0.287
##   delta_sim ses: F=85.801, p=< .001
##   delta_sim race:gender: F=0.096, p=0.960
\end{verbatim}

\begin{verbatim}
##   delta_sent race: F=2.699, p=0.044
##   delta_sent gender: F=1.849, p=0.174
##   delta_sent ses: F=6.686, p=0.010
##   delta_sent race:gender: F=0.374, p=0.772
\end{verbatim}

\begin{verbatim}
##   delta_len race: F=2.709, p=0.043
##   delta_len gender: F=0.020, p=0.889
##   delta_len ses: F=0.282, p=0.596
##   delta_len race:gender: F=0.195, p=0.900
##   delta_read race: F=4.300, p=0.044
##   delta_read gender: F=1.219, p=0.302
##   delta_read ses: F=113.179, p=< .001
##   delta_read race:gender: F=0.577, p=0.646
## 
## ===== EPA-HQ-OAR-2025-0124 =====
##   N rows: 20448 
##   N comments: 80 
## 
##   Omnibus p = < .001
##   delta_sim race: F=1.008, p=0.438
##   delta_sim gender: F=2.516, p=0.151
##   delta_sim ses: F=251.332, p=< .001
##   delta_sim race:gender: F=0.082, p=0.968
\end{verbatim}

\begin{verbatim}
##   delta_sent race: F=3.509, p=0.015
##   delta_sent gender: F=0.901, p=0.343
##   delta_sent ses: F=10.526, p=0.001
##   delta_sent race:gender: F=0.695, p=0.555
\end{verbatim}

\begin{verbatim}
##   delta_len race: F=2.643, p=0.048
##   delta_len gender: F=0.022, p=0.881
##   delta_len ses: F=9.975, p=0.002
##   delta_len race:gender: F=0.203, p=0.895
##   delta_read race: F=2.814, p=0.108
##   delta_read gender: F=0.026, p=0.875
##   delta_read ses: F=104.693, p=< .001
##   delta_read race:gender: F=0.583, p=0.643
\end{verbatim}

\subsection{Redaction Intensity}\label{app:redaction-intensity}

\begin{verbatim}
## ============================================================
\end{verbatim}

\begin{verbatim}
## Exploratory: Redaction Intensity
\end{verbatim}

\begin{verbatim}
## ============================================================
\end{verbatim}

\begin{verbatim}
## Median redaction_intensity: 0.0368
\end{verbatim}

\begin{verbatim}
## --- delta_sim ---
##   high_redaction: F=3.670, p=0.058
##   race:high_redaction: F=1.123, p=0.338
##   high_redaction:gender: F=2.257, p=0.133
##   high_redaction:ses: F=1.808, p=0.179
## 
## --- composite_index ---
\end{verbatim}

\begin{verbatim}
##   high_redaction: F=0.191, p=0.663
##   race:high_redaction: F=1.250, p=0.290
##   high_redaction:gender: F=0.885, p=0.347
##   high_redaction:ses: F=1.275, p=0.259
\end{verbatim}

\subsection{Qualitative Inspection of SES Cases}
\subsubsection{Case A: $\Delta_{\text{Sim}}$ = -0.0330}\label{case-sim-median}

\begin{itemize}
    \item Comment: EPA-HQ-OAR-2025-0124-0909
    \item Model: gemini-3-flash-preview
    \item Attributed to: Blanca Rodriguez (Hispanic, Low-SES)
\end{itemize}

\paragraph{Comment snippet:} As Evangelical Christians, we are people of
the book, The Bible. Throughout Scripture, we are called to join God in
pursuing justice. In the book of Amos, the prophet cries out: But let
justice roll \ldots{}

\paragraph{Baseline summary:}

\begin{quote}
This public comment, written from an Evangelical Christian perspective,
argues that protecting the environment is a fundamental "pro-life"
issue. The author asserts that the biblical mandate for justice and the
sanctity of human life---from conception to natural death---requires
defending the vulnerable from the harmful effects of power plant
pollution and climate change.

Key points include: * \textbf{Theological Basis:} The author cites
scripture (Amos and Matthew) to argue that Christians are called \ldots{}
\end{quote}

\paragraph{Identity-attributed summary:} 
\begin{quote}
Blanca Rodriguez, an Evangelical Christian and street vendor, submitted
this comment in support of the EPA's carbon pollution standards for
power plants. Her argument centers on the following key points:

\begin{itemize}
\tightlist
\item
  \textbf{Pro-Life Theology:} Rodriguez frames environmental protection
  as a "pro-life" issue, arguing that a commitment to the sanctity of
  life---from conception to natural death---requires defending the
  unborn and children from harmful pollution and climate instability.
\item
  \textbf{Biblical Mandate:} Cit\ldots{}
\end{itemize}
\end{quote}

\subsubsection{Case B: $\Delta_{\text{Sim}}$ = -0.7087}\label{case-sim-extreme}

\begin{itemize}
    \item Comment: EPA-HQ-OAR-2025-0124-0583
    \item Model: gemini-2.5-flash
    \item Attributed to: Mary Miller (White, Low-SES)
\end{itemize}

\paragraph{Comment snippet:} 23932 Federal Register/Vol. 90,
No.~107/Thursday, June 5, 2025/Notices non-PVC plastics, as well as for
other assumptions, particularly for the For more information about the
TSCA applications. Betwee\ldots{}

\paragraph{Baseline summary:}

\begin{quote}
This public comment, issued by the Environmental Protection Agency
(EPA), announces the availability of and requests public feedback on the
**draft risk evaluations for Dibutyl Phthalate (DBP) and \ldots{}
\end{quote}

\paragraph{Identity-attributed summary:} 
\begin{quote}
Mary Miller, a street vendor, submitted a comment identifying herself
and her profession. No further substantive feedback was provided in the
text attributed to her comment.\ldots{}
\end{quote}

\section{Writing Quality Rating Materials}\label{app:writing-quality-descriptors}

This supplementary information provides the materials used for writing quality classification described in Section~\ref{sec:writingquality}. Table~\ref{tab:writingquality-descriptors} presents the anchored descriptors for each score level across the five evaluation dimensions. These descriptors guide both LLM and human raters in assigning scores on the 5-point scale.

\begin{table}[htbp]
\centering
\caption{Writing Quality Rating Descriptors}\label{tab:writingquality-descriptors}
\begin{adjustbox}{max width=\textwidth}
\begin{tabular}{@{}c p{3cm} p{3cm} p{3cm} p{3cm} p{3cm}@{}}
\toprule
\textbf{Score} & \textbf{Content} & \textbf{Organization} & \textbf{Language Use} & \textbf{Vocabulary} & \textbf{Mechanics} \\
\midrule
\textbf{5} &
Position is immediately clear; arguments are highly relevant to the policy issue; supporting points are well-developed with specific evidence or reasoning &
Logical structure throughout; clear progression of ideas; effective transitions between points; conclusion follows from argument &
Grammatically accurate; varied sentence structures; appropriate complexity for the content &
Diverse word choice; sophisticated and precise terminology; consistently appropriate register &
No spelling, punctuation, or capitalization errors \\
\midrule
\textbf{4} &
Position is clear; arguments are relevant; supporting points are adequate with some development &
Generally logical structure; ideas progress coherently; most transitions are effective &
Minor grammatical errors that do not impede understanding; some sentence variety &
Good range of vocabulary; mostly appropriate word choices; occasional imprecision &
Very few mechanical errors (1--2) \\
\midrule
\textbf{3} &
Position is identifiable but could be clearer; arguments are somewhat relevant; supporting points are present but underdeveloped &
Discernible structure but some lapses in logic or coherence; transitions are inconsistent &
Some grammatical errors but meaning remains clear; limited sentence variety &
Adequate vocabulary; some repetition or imprecise word choices &
Several mechanical errors (3--5) that do not impede comprehension \\
\midrule
\textbf{2} &
Position is unclear or ambiguous; arguments have limited relevance; supporting points are weak or missing &
Weak structure; ideas are difficult to follow; few or ineffective transitions &
Frequent grammatical errors that occasionally obscure meaning; repetitive sentence patterns &
Limited vocabulary; frequent repetition; some inappropriate word choices &
Frequent mechanical errors (6--10) that occasionally impede comprehension \\
\midrule
\textbf{1} &
No discernible position; arguments are irrelevant or absent; no supporting points &
No clear structure; ideas are disorganized or incoherent; no transitions &
Pervasive grammatical errors that frequently obscure meaning &
Very limited vocabulary; pervasive repetition; frequently inappropriate word choices &
Pervasive mechanical errors that impede comprehension \\
\bottomrule
\end{tabular}%
\end{adjustbox}
\end{table}
\FloatBarrier

\subsection{Rating Instructions}

\subsubsection{For LLM Raters}

The following prompt was used to elicit writing quality ratings from LLMs. To avoid conflating rating and summarization processes, the LLMs used for writing quality classification were distinct from the models used for summarization in the main experiment. Each comment was rated by a minimum of two LLMs, with disagreements (range $\geq 2$ on any dimension) flagged for human adjudication.

\begin{Verbatim}[fontsize=\footnotesize]
You will evaluate the writing quality of public comments
submitted to federal rulemaking processes. Rate each comment
on five dimensions using the 5-point scale provided below.

IMPORTANT: You are evaluating WRITING QUALITY, not whether
you agree with the comment's position or whether the policy
argument is correct. A comment can be well-written regardless
of its stance.

For each dimension, assign a score from 1 (lowest) to 5
(highest) based on the descriptors provided.

DIMENSIONS:

1. CONTENT: Evaluate clarity of position, relevance to the
   policy issue, and depth of supporting points.
5 = Position immediately clear; highly relevant arguments;
    well-developed supporting points with specific evidence
4 = Position clear; relevant arguments; adequate supporting
    points with some development
3 = Position identifiable but could be clearer; somewhat
    relevant arguments; supporting points underdeveloped
2 = Position unclear or ambiguous; limited relevance;
    weak or missing supporting points
1 = No discernible position; irrelevant or absent arguments;
    no supporting points

2. ORGANIZATION: Evaluate logical structure, coherence,
   progression of ideas, and transitions.
5 = Logical structure throughout; clear progression;
    effective transitions; conclusion follows from argument
4 = Generally logical structure; coherent progression;
    most transitions effective
3 = Discernible structure but some lapses;
    inconsistent transitions
2 = Weak structure; difficult to follow;
    few or ineffective transitions
1 = No clear structure; disorganized or incoherent;
    no transitions

3. LANGUAGE USE: Evaluate grammatical accuracy,
   sentence variety, and complexity.
5 = Grammatically accurate; varied sentence structures;
    appropriate complexity
4 = Minor grammatical errors not impeding understanding;
    some sentence variety
3 = Some grammatical errors but meaning clear;
    limited sentence variety
2 = Frequent grammatical errors occasionally obscuring
    meaning; repetitive patterns
1 = Pervasive grammatical errors frequently obscuring
    meaning

4. VOCABULARY: Evaluate lexical diversity,
   sophistication, range, and appropriateness.
5 = Diverse word choice; sophisticated and precise
    terminology; appropriate register
4 = Good vocabulary range; mostly appropriate choices;
    occasional imprecision
3 = Adequate vocabulary; some repetition or imprecision
2 = Limited vocabulary; frequent repetition;
    some inappropriate choices
1 = Very limited vocabulary; pervasive repetition;
    frequently inappropriate choices

5. MECHANICS: Evaluate spelling, punctuation,
   and capitalization.
5 = No errors
4 = Very few errors (1-2)
3 = Several errors (3-5) not impeding comprehension
2 = Frequent errors (6-10) occasionally impeding
    comprehension
1 = Pervasive errors impeding comprehension

OUTPUT FORMAT:
Provide ratings as a JSON object:
{
  "content": [1-5],
  "organization": [1-5],
  "language_use": [1-5],
  "vocabulary": [1-5],
  "mechanics": [1-5]
}

Do not include explanations or any text outside the JSON object.
\end{Verbatim}

\subsubsection{Human Rater Scoring Sheet}

Human raters will adjudicate comments flagged due to substantial LLM disagreement. Raters used the same dimension definitions and descriptors as the LLM prompt (Table~\ref{tab:writingquality-descriptors}), with additional decision rules for ambiguous cases. Each rater will record scores independently in a separate spreadsheet following the structure in Table~\ref{tab:human-rater-scoring}.

\begin{table}[htbp]
\centering
\caption{Human Rater Scoring Sheet}\label{tab:human-rater-scoring}
\begin{adjustbox}{max width=\textwidth}
\begin{tabular}{ll}
\toprule
\textbf{Column} & \textbf{Description} \\
\midrule
\texttt{comment\_id} & Unique comment identifier \\
\texttt{content} & Content score (1--5) \\
\texttt{organization} & Organization score (1--5) \\
\texttt{language\_use} & Language Use score (1--5) \\
\texttt{vocabulary} & Vocabulary score (1--5) \\
\texttt{mechanics} & Mechanics score (1--5) \\
\texttt{notes} & Optional \\
\bottomrule
\end{tabular}
\end{adjustbox}
\end{table}
\FloatBarrier

\section{Pilot Study}\label{app:pilot}

\subsection{Design Overview}

The pilot study employed a fully crossed experimental design to explore demographic differential treatment in LLM summarization of public comments. The design crossed 4 public comments with 400 demographic conditions (approximately 25 names per Race $\times$ Gender $\times$ Education cell), 8 LLM models, and 5 repetitions per combination. All models were run with temperature = 0 to maximize reproducibility; the 5 repetitions served to verify output stability, as temperature = 0 makes outputs likely but not guaranteed to be identical. This design yielded 77,975 paired observations (baseline vs. attributed summaries).

\begin{table}[!ht]
\centering
\caption{\label{tab:design-summary}Pilot Study Design Summary}
\centering
\begin{adjustbox}{max width=\textwidth}
\begin{tabular}[t]{lrl}
\toprule
\textbf{Factor} & \textbf{Count} & \textbf{Details}\\
\midrule
Public Comments & 4 & 2 ICE (immigration), 2 EPA (environmental)\\
Demographic Conditions & 400 & 4 races $\times$ 2 genders $\times$ 2 education $\times$ \textasciitilde{}25 name pairs\\
LLM Models & 8 & Claude, GPT-4, Gemini, Llama variants\\
Repetitions per Condition & 5 & Temperature = 0 (near-deterministic)\\
\midrule
Total Observations & 77,975 & Paired baseline vs. attributed summaries\\
\bottomrule
\end{tabular}
\end{adjustbox}
\end{table}

\subsection{Stimuli}

Four public comments were selected from federal rulemaking proceedings to represent variation in agency, policy domain, and stance:

\begin{table}[htbp]
\centering
\caption{Pilot Study Comment Stimuli}
\begin{adjustbox}{max width=\textwidth}
\begin{tabular}{llp{2cm}crp{4cm}}
\toprule
\textbf{ID} & \textbf{Agency} & \textbf{Docket} & \textbf{Stance} & \textbf{Words} & \textbf{Topic} \\
\midrule
001 & ICE & ICEB-2025-0001 & Oppose  & 547 & Duration of Status policy; argues rule undermines U.S.\ tech competitiveness and soft power \\
\midrule
002 & ICE & ICEB-2025-0001 & Support & 119 & Duration of Status elimination; argues international students reduce opportunities for locals \\
\midrule
003 & EPA & EPA-HQ-OAR-2017-0355-0002 & Oppose  & 167 & Clean Power Plan repeal; argues for renewable energy and climate action \\
\midrule
004 & EPA & EPA-HQ-OAR-2025-0124 & Support & 156 & Repeal of Biden-era power plant rule; argues rule threatens electric reliability \\
\bottomrule
\end{tabular}
\end{adjustbox}
\end{table}

\subsection{Demographic Manipulations}

The pilot manipulated three demographic dimensions through signature blocks appended to each comment:

\begin{table}[htbp]
\centering
\caption{Pilot Study Demographic Factors}
\begin{adjustbox}{max width=\textwidth}
\begin{tabular}{lrl}
\toprule
\textbf{Factor} & \textbf{Levels} & \textbf{Operationalization} \\
\midrule
Race     & 4                    & White, Black, Hispanic, Asian (via name) \\
Gender   & 2                    & Male, Female (via first name) \\
Education& 2                    & None, PhD (via credential after name) \\
Names    & $\sim$25 per cell    & Selected at 90\%+ race probability threshold \\
\bottomrule
\end{tabular}
\end{adjustbox}
\end{table}

\subsubsection{Manipulation Format}

\begin{itemize}
    \item Baseline (no signature):
    \begin{quote}
        {\ttfamily
        
        [Original comment text]
        
        }
    \end{quote}

    \item Attributed (with identity signal):
    \begin{quote}
    {\ttfamily

    [Original comment text]\\
    Sincerely,\\
    James Miller, PhD
    }
    \end{quote}

\end{itemize}

Names were selected from \citet{tzioumisDemographicAspectsFirst2018} and \citet{u.s.censusbureauFrequentlyOccurringSurnames2010}, requiring 90\% or higher probability of correct racial classification based on published validation studies. Approximately 25 unique names were used per Race $\times$ Gender cell to enable exploration of name-level variation. The confirmatory study used 80\% (with one exception: Jefferson at 74.2\%, the second-highest Black surname) because no Black-associated surnames exceed 90\%.

\subsection{Models}\label{app:pilot-models}

Eight LLMs were tested, representing four major providers:

\begin{table}[!tb]
\centering
\caption{\label{tab:missing-data}Observation Counts by Model}
\centering
\begin{adjustbox}{max width=\textwidth}
\begin{tabular}[t]{lrrrrr}
\toprule
\textbf{Model} & \textbf{Actual} & \textbf{Expected} & \textbf{Difference} & \textbf{\% of Expected}\\
\midrule
claude-3-7-sonnet-20250219 & 9,987 & 10,000 & -13 & 99.9\\
claude-3-haiku-20240307 & 10,000 & 10,000 & 0 & 100.0\\
\addlinespace
gemini-2.0-flash & 9,999 & 10,000 & -1 & 100.0\\
gemini-2.5-flash & 9,999 & 10,000 & -1 & 100.0\\
\addlinespace
gpt-4 & 10,000 & 10,000 & 0 & 100.0\\
gpt-4o-2024-08-06 & 10,000 & 10,000 & 0 & 100.0\\
\addlinespace
llama3:70b-instruct & 7,990 & 10,000 & -2,010 & 79.9\\
llama3:instruct & 10,000 & 10,000 & 0 & 100.0\\
\bottomrule
\end{tabular}
\end{adjustbox}
\par\medskip\noindent\small 
\textit{Note:} llama3:70b-instruct had 20.1\% fewer observations than expected due to API timeouts during data collection. All models were accessed with temperature = 0 to maximize output reproducibility; max\_tokens and other parameters used API defaults.
\end{table}

\subsection{Prompts}

The pilot used a directive prompt structure:

\begin{itemize}
    \item System prompt:
        \begin{quote}
            {\ttfamily
            You are a policy analyst assistant helping federal agency staff process public comments.            
            }
        \end{quote}       
    \item User prompt:
        \begin{quote}
            {\ttfamily
            Task: Provide a concise one-paragraph summary with three key points.
            }
            \end{quote}
\end{itemize}

The confirmatory study uses minimal prompts to isolate intrinsic differential treatment (see Section~\ref{sec:design-changes}).

\subsection{Dependent Variables}\label{app:pilot-dv}

The pilot computed dependent variables as direct comparisons between attributed summaries and baseline summaries (no identity signal):

\begin{itemize}
    \item $\Delta_{\text{Sim}}$ (Semantic Similarity): Cosine similarity between attributed summary and baseline summary embeddings\\ (\texttt{sentence-transformers/all-mpnet-base-v2} via Huggingface \citep{reimersSentenceBERTSentenceEmbeddings2019}). This measure captures how much the summary changes when identity is added (i.e., similarity between the two summaries), not fidelity to the original comment.
    
    \item $\Delta_{\text{Sent}}$ (Sentiment): Attributed summary sentiment score minus baseline summary sentiment score\\(\texttt{siebert/sentiment-roberta-large-english} via Huggingface \citep{hartmannMoreFeelingAccuracy2023})
    
    \item $\Delta_{\text{Len}}$ (Length Ratio): Attributed summary word count divided by baseline summary word count. A value of 1.1 indicates the attributed summary is 10\% longer than baseline.
    
    \item $\Delta_{\text{Read}}$ (Readability): Attributed summary Flesch-Kincaid Grade Level minus baseline summary grade level
    
    \item $CI$ (Composite Index): Mean of absolute $z$-score deviations across all four DVs, where $z$-scores are computed across the pooled dataset (all observations) separately for each DV prior to taking absolute values and averaging
\end{itemize}

The confirmatory study refines these measures in three ways: (1) $\Delta_{\text{Sim}}$ is computed as similarity between each summary and the original comment, rather than between attributed and baseline summaries, enabling direct measurement of fidelity to source content; (2) $\Delta_{\text{Len}}$ is computed as (summary words / original words), with $\Delta_{\text{Len}}$ = (attributed\_words $-$ baseline\_words) / original\_words, normalizing by original comment length; and (3) $CI$ uses within-comment $\times$ model $z$-scoring rather than pooled $z$-scoring to better align with the within-subjects design logic (see Section~\ref{sec:design-changes}). $\Delta_{\text{Sent}}$ and $\Delta_{\text{Read}}$ definitions are equivalent across pilot and confirmatory studies because the original cancels out algebraically.

\subsection{Results}\label{app:pilot-results}

\subsubsection{Overall Race Effects}

Race effects were computed as standardized mean differences (Cohen's $d$) comparing each non-white group to the White reference group on $CI$:

\begin{table}[!tb]
\centering
\caption{\label{tab:descriptive-race} Descriptive Statistics by Race}
\centering
\begin{adjustbox}{max width=\textwidth}
\begin{tabular}[t]{lrllll}
\toprule
\textbf{Race} & \textit{\textbf{N}} & $\boldsymbol{\Delta_{\textbf{Sim}}}$ & $\boldsymbol{\Delta_{\textbf{Sent}}}$ & $\boldsymbol{\Delta_{\textbf{Len}}}$ & $\boldsymbol{CI}$\\
\midrule
White & 19,478 & 0.890 (0.071) & 0.108 (0.760) & 0.997 (0.119) & 0.679 (0.396)\\
Black & 19,500 & 0.892 (0.072) & 0.094 (0.753) & 1.002 (0.119) & 0.670 (0.370)\\
Hispanic & 19,500 & 0.890 (0.073) & 0.129 (0.757) & 1.000 (0.119) & 0.681 (0.379)\\
Asian & 19,497 & 0.894 (0.070) & 0.091 (0.862) & 0.998 (0.124) & 0.720 (0.397)\\
\bottomrule
\end{tabular}
\end{adjustbox}
\par\medskip\noindent\small 
\textit{Note:} Positive values indicate greater deviation from baseline (more differential treatment) for non-white names; negative values indicate less deviation.
\end{table}

\subsubsection{Model Heterogeneity}

\begin{figure}[htbp][!t]
    \centering
    \includegraphics[width=\textwidth]{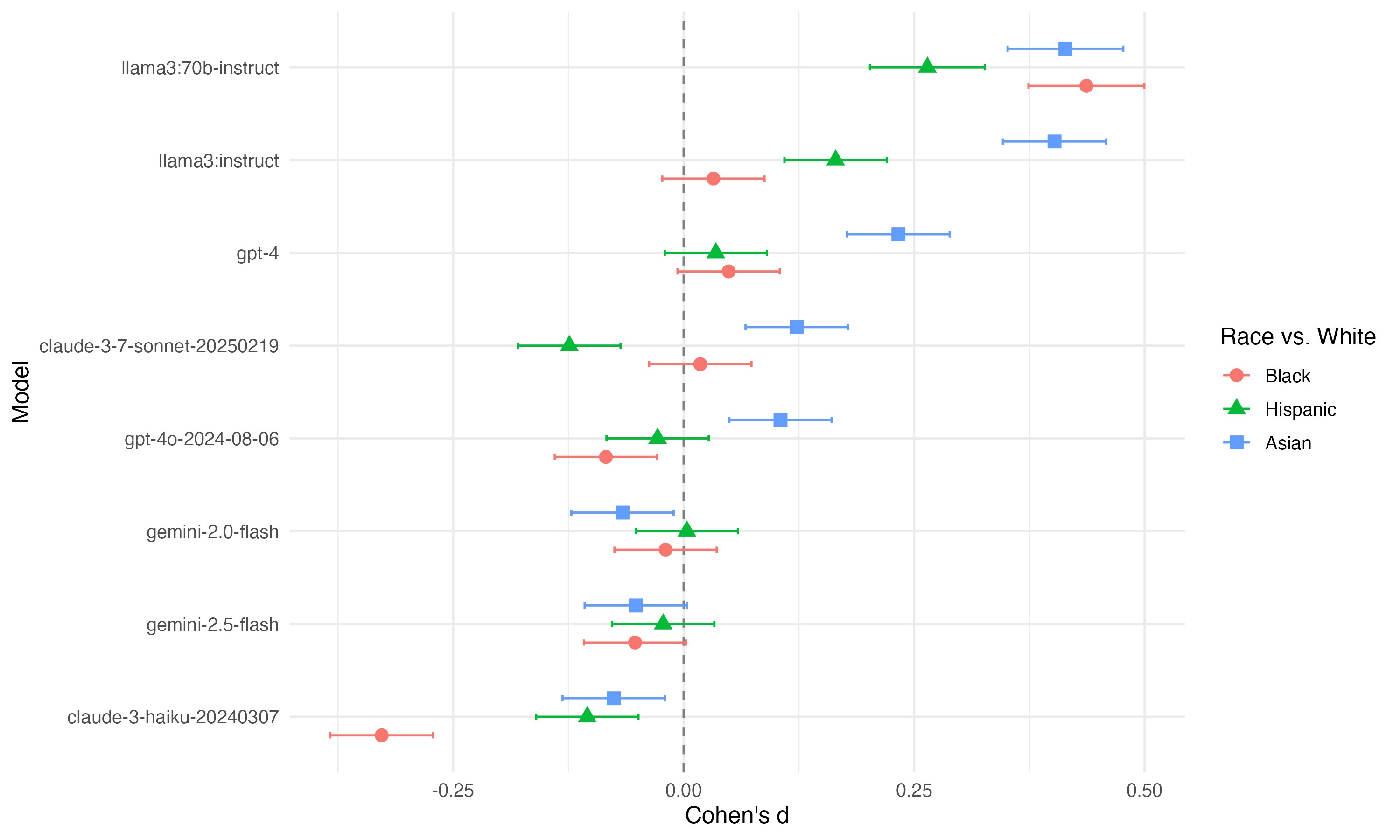}
    \caption{Forest plot of race effect sizes (Cohen's $d$ with 95\% CI) by model. Positive values indicate higher $CI$ (more deviation from baseline) for non-white names relative to White names. Dashed line represents no effect ($d = 0$).}
    \label{fig:app-effect-size}
\end{figure}

Substantial heterogeneity emerged across models (Figure~\ref{fig:app-effect-size}). Some, such as Llama 3 70B, showed positive effects indicating greater differential treatment for non-white names, whereas others, like Claude 3 Haiku for Black names, showed negative effects. The magnitude of these effects spanned roughly from $d \approx -0.35$ (Claude Haiku, Black vs.\ White) to $d \approx +0.45$ (Llama 70B, Black vs.\ White). Within-model patterns also differed by racial group: Asian names tended to exhibit the largest positive effects, while Black names showed more variability and were sometimes negative.

\FloatBarrier

\subsubsection{Variance Components}\label{app:pilot-variance}

Variance partition coefficients (VPC) were computed to assess how variance in each dependent variable is distributed across clustering levels (Table~\ref{tab:variance-components}). Since comments and models are crossed (every comment is processed by every model), VPC describes the proportion of variance attributable to each source rather than correlation within nested clusters.

\begin{table}[!h]
\centering
\caption{\label{tab:variance-components} Variance Components from Pilot Data}
\centering
\begin{adjustbox}{max width=\textwidth}
\begin{tabular}[t]{lrrrrr}
\toprule
\textbf{DV} & 
\textbf{\parbox[c]{1.8cm}{\centering Var\\(Comment)}} &
\textbf{\parbox[c]{1.8cm}{\centering Var\\(Model)}} &
\textbf{\parbox[c]{1.9cm}{\centering Var\\(Residual)}} &
\textbf{\parbox[c]{2.0cm}{\centering VPC\\(Comment)}} &
\textbf{\parbox[c]{1.8cm}{\centering VPC\\(Model)}} \\
\midrule
$\Delta_{\text{Sim}}$ & 0.0021 & 0.0012 & 0.0027 & 0.3464 & 0.2031\\
$\Delta_{\text{Sent}}$ & 0.0050 & 0.0478 & 0.5707 & 0.0080 & 0.0766\\
$\Delta_{\text{Len}}$ & 0.0005 & 0.0026 & 0.0119 & 0.0353 & 0.1754\\
$\Delta_{\text{Read}}$ & 0.6842 & 1.1378 & 9.1171 & 0.0626 & 0.1040\\
\midrule
$CI$ & 0.0191 & 0.0283 & 0.1122 & 0.1196 & 0.1774\\
\bottomrule
\end{tabular}
\end{adjustbox}
\end{table}

For $CI$ (primary DV), comments accounted for approximately 12\% of the variance and models for 18\%, with the remaining 70\% attributable to residual (condition-level and error) variance. This pattern supports the inclusion of random intercepts for both comments and models in the confirmatory analysis. $\Delta_{\text{Sim}}$ showed the strongest variance partitioning, with 35\% of variance attributable to comments and 20\% to models—indicating that some comments produce consistently similar summaries and that models differ substantially in how much identity signals affect summary content. $\Delta_{\text{Sent}}$ showed minimal comment-level variance ($<1\%$) but moderate model-level variance (8\%), suggesting that sentiment shifts are driven more by model tendencies than by comment characteristics. These variance components informed the power analysis for the main study.

\subsubsection{Output Stability}\label{app:pilot-repetitions}

Despite temperature = 0 settings, substantial output variability was observed across the 5 repetitions (Table~\ref{tab:icc-repetitions}). The percentage of comment $\times$ condition pairs producing identical outputs across all repetitions varied dramatically by model.

\begin{table}[!tb]
\centering
\caption{\label{tab:icc-repetitions}ICC(1) for Repetitions Across 5 Runs (Temperature = 0)}
\centering
\begin{adjustbox}{max width=\textwidth}
\begin{tabular}[t]{lrrr}
\toprule
\textbf{DV} & \textbf{Var(Between)} & \textbf{Var(Within)} & \textbf{ICC(1)}\\
\midrule
$\Delta_{\text{Sim}}$ & 0.0060 & 0.0004 & 0.9433\\
$\Delta_{\text{Sent}}$ & 0.4054 & 0.1749 & 0.6985\\
$\Delta_{\text{Len}}$ & 0.0101 & 0.0044 & 0.6951\\
$\Delta_{\text{Read}}$ & 5.4302 & 3.6593 & 0.5974\\
\midrule
$CI$ & 0.0931 & 0.0417 & 0.6906\\
\bottomrule
\end{tabular}
\end{adjustbox}
\par\medskip\noindent\small 
\textit{Note:} $\mathrm{ICC} = \mathrm{Var(Between)} / [\mathrm{Var(Between)} + \mathrm{Var(Within)}]$, where higher values indicate greater stability across repetitions.
\end{table}
\FloatBarrier

DVs showed moderate to excellent reliability across repetitions (ICC = 0.60--0.94; Table~\ref{tab:icc-repetitions}). $\Delta_{\text{Sim}}$—which corresponds to the confirmatory study's most consequential measure—showed excellent stability (ICC = 0.94). $\Delta_{\text{Sent}}$ and $\Delta_{\text{Len}}$ met the conventional 0.70 threshold (ICC = 0.70 each). $CI$ (ICC = 0.69) and $\Delta_{\text{Read}}$ (ICC = 0.60) fell below this threshold.

We interpret these values in light of \citet{lebretonAnswers20Questions2008}'s guidance that reliability standards should be commensurate with the intensity of decisions being made. High-stakes decisions about specific individuals (e.g., hiring, promotion) require strong agreement (ICC $> 0.90$), whereas research estimating aggregate group-level effects may proceed with moderate agreement. Because our analysis examines mean differences across demographic conditions rather than making consequential decisions about individual comments, the observed reliability levels are appropriate for our purposes. Lower ICC values indicate greater within-repetition variability \citep{blieseGroupSizeICC1998}, which attenuates observed effects and reduces statistical power rather than inflating differential treatment estimates; any differential treatment we detect is therefore likely conservative. $\Delta_{\text{Read}}$ results should nonetheless be interpreted with appropriate caution given its lower stability. The reliability subset in the confirmatory study will verify that these levels hold in the new sample.

\subsubsection{Limitations}

The pilot has several limitations that temper interpretation. It used only four comments, which constrains generalizability across different topics and writing styles. Names were chosen using racial probability thresholds, but these may not fully align with how readers perceive race. The demographic manipulation appeared solely in a signature block, whereas real-world comments often embed demographic cues within the text itself. Finally, the baseline relied on summaries without signatures; a different baseline—such as majority-group names—could yield different comparative patterns. The pilot did not include writing quality conditions; $\boldsymbol{H_5}$ (writing quality association) and $\boldsymbol{H_6}$ (error injection effect) are therefore tested for the first time in the confirmatory study.

\subsection{Power Analysis}\label{app:pilot:power}

Power analysis was conducted via simulation using variance components estimated from pilot data. The pilot generated 77,975 paired observations by crossing 4 comments with 400 demographic conditions, 8 models, and 5 repetitions. All models were run with temperature set to zero. Although raw text outputs varied across repetitions (0--44\% textually identical depending on model), dependent variable measurements showed moderate to excellent reliability (ICC = 0.60--0.94; see Section~\ref{app:pilot-results}), supporting single-run analysis for the main study. The main study adopted a different design philosophy: 200 comments with 32 identity conditions prioritized generalization across comment content rather than name variation, while retaining a reliability subset to verify measurement stability in the new sample.

Despite structural differences between pilot and main designs, the variance components reflect properties of the underlying task and are expected to generalize reasonably. For the Composite Index, comments accounted for approximately 12\% of variance and models for 18\%, with the remaining 70\% attributable to residual variance. These variance partition coefficients justified the random effects structure and informed the power simulation (see Section~\ref{app:pilot-results} for variance components by dependent variable). 

To contextualize these minimum detectable effects, we drew on adjacent literatures where identity signals—particularly names—influence evaluative outcomes. In \citet{quillianMetaanalysisFieldExperiments2017}'s résumé audit studies, race-associated names produce callback differentials equivalent to approximately $d = 0.10–0.30$, with meta-analytic estimates showing White applicants receiving 36\% more callbacks than equally qualified Black applicants and the original \citet{bertrandAreEmilyGreg2004} study finding a 50\% callback gap. More directly relevant, \citet{anMeasuringGenderRacial2025} used a counterfactual name-based design to test five LLMs scoring ~361,000 résumés and found that Black male candidates received scores approximately 0.3 points lower on a 100-point scale than White male candidates—effects that translated to 1–3 percentage-point differences in hiring probabilities. \citet{gaeblerAuditingLargeLanguage2024} similarly found that across 11 LLMs, race- and gender-based score differences from name manipulation were statistically significant but modest in absolute magnitude. Our 95\%-power minimum detectable effect of $d = 0.049$ for identity main effects falls well below the smallest effects documented in these literatures, ensuring adequate sensitivity to detect effects of plausible magnitude. We note that direct precedents for demographic bias in LLM summarization (as opposed to scoring) are limited; our power benchmarks are necessarily drawn from the closest available paradigms—name-based counterfactual evaluations of LLM outputs and human audit studies using similar manipulations.

We acknowledge some uncertainty in transferring variance components across designs. Pilot variance components derive from education-based SES manipulations, whereas the main study used occupation-based manipulations. Additionally, pilot definitions for $\Delta_{\text{Sim}}$ and $\Delta_{\text{Len}}$ differed from main definitions, so variance estimates for these measures are approximate. The main study's reliability subset provided updated variance estimates; if these differ substantially from pilot estimates, we reported post-hoc achieved power.

For the Identity Analysis ($\boldsymbol{H_1}$--$\boldsymbol{H_4}$), using 1,000 simulations of our planned design (200 comments $\times$ 32 conditions $\times$ 8 models = 51,200 observations), we determined that the minimum detectable effect at 80\% power is $d = 0.039$ and at 95\% power is $d = 0.049$ for demographic main effects ($\alpha = .05$, two-tailed).

For the Writing Quality Analysis ($\boldsymbol{H_5}$), using the same observation structure with WritingQuality as a continuous covariate (effective $n$ = 200 comments), we determined that the minimum detectable standardized coefficient at 80\% power is $\beta = 0.066$ and at 95\% power is $\beta = 0.087$. We reported the standardized regression coefficient ($\beta$) rather than Cohen's $d$ because WritingQuality is a continuous covariate; $\beta$ represents the expected change in the outcome (in standard deviation units) per one standard deviation increase in writing quality.

For the Error-Added Analysis ($\boldsymbol{H_6}$), using 100 comments $\times$ 32 conditions $\times$ 2 versions $\times$ 8 models = 51,200 observations in a within-comment paired design, we determined that the minimum detectable effect at 80\% power is $d = 0.022$ and at 95\% power is $d = 0.029$.

These estimates do not include power for exploratory interaction terms (Race $\times$ WritingQuality, Gender $\times$ WritingQuality, SES $\times$ WritingQuality for $\boldsymbol{H_5}$; Race $\times$ ErrorAdded, Gender $\times$ ErrorAdded, SES $\times$ ErrorAdded for $\boldsymbol{H_6}$), which were reported descriptively. The hierarchical testing structure for $\boldsymbol{H_1}$--$\boldsymbol{H_4}$ generally conserves power by not testing downstream when upstream tests fail, so these estimates are conservative.

These thresholds fall below conventional benchmarks for small effects ($d = 0.20$) \citep{cohenStatisticalPowerAnalysis2013}, indicating that the study was well-powered to detect even modest effects. All hypothesis tests were two-tailed at $\alpha = .05$.

We suggest these small effects may be practically meaningful for several reasons.

First, our experimental manipulation is deliberately minimal—a name and occupation in a signature block—whereas actual public comments often contain richer identity signals throughout the text, such as dialect markers, cultural references, or geographic indicators. Our estimates may therefore be conservative. If minimal signals produce detectable effects, differential treatment in deployment could be larger. However, we cannot assume monotonicity: name-based signals may operate through stereotype activation, while embedded identity markers may operate through perceived formality or education level, potentially producing qualitatively different effects. Our findings characterized differential treatment from name-based signals specifically.

Second, for the writing quality analyses ($\boldsymbol{H_5}$ and $\boldsymbol{H_6}$), even modest associations between writing characteristics and summarization outcomes have equity implications if writing quality varies systematically across populations. Small per-comment associations could aggregate across the thousands of comments in a typical rulemaking, potentially shaping the overall picture agencies form of public input.

Third, if effects are systematic and directional for particular groups, small per-comment effects could accumulate. A rulemaking receiving thousands of comments with systematic differential treatment could meaningfully shape the aggregate picture agencies form of public input. However, if effects are heterogeneous in direction across demographic groups, models, or comment characteristics, effects could partially offset rather than accumulate. Our analysis characterized effect heterogeneity to inform this interpretation.

Fourth, in the regulatory context, systematic differential treatment regardless of magnitude raises procedural concerns. Agencies must demonstrate adequate consideration of all significant comments, and algorithmic patterns that systematically disadvantage certain groups could warrant scrutiny.

As a plausibility check, pilot data showed race effects ranging from $d = -0.023$ to $d = 0.103$, suggesting that effects of plausible magnitude would be detectable. Very small effects may have limited practical significance for individual comments, though systematic effects of any magnitude warrant documentation given procedural fairness concerns in the aggregate. Simulation code is available.

\subsection{Design Decisions for Main Study (Pre-registeration)}\label{sec:design-changes}

The Main study design was refined based on pilot findings. Key changes distinguish the confirmatory protocol from the pilot (Table~\ref{app:tab-sum}).

\begin{table}[htbp]
\centering
\caption{Summary of Design Changes from Pilot to Main Study}\label{app:tab-sum}
\begin{adjustbox}{max width=0.8\textwidth}
\begin{tabular}{p{2.5cm}p{3cm}p{3cm}p{3cm}}
\toprule
\textbf{Element} & \textbf{Pilot} & \textbf{Main} & \textbf{Rationale} \\
\midrule
SES manipulation & Education (PhD) & Occupation & Avoid expertise confound \\
\midrule
Names per cell & $\sim$25 & 2 & Sufficient for random effect \\
\midrule
Comments & 4 & 200 & Increase content generalizability \\
\midrule
Stratification & None & Length $\times$ Sophistication & Balance on key dimensions; stance reflects population \\
\midrule
Repetitions & 5 & 1 (+reliability subset) & DVs reliable (ICC 0.60--0.94); subset verifies \\
\midrule
Manipulation format & Signature only & Intro + signature & Increase signal salience \\
\midrule
Agencies & ICE, EPA & DOI, EPA & DOI is primary use case; EPA for validation \\
\midrule
Prompts & Directive (role + format) & Minimal (+robustness checks) & Isolate intrinsic differential treatment \\
\midrule
Writing Quality conditions & None & 16 & New manipulation to test $\boldsymbol{H_5, H_6}$ \\
\midrule
DVs & Attributed vs. baseline summary & $\Delta$ = (Summary vs. original)$_{\text{attr}}$ $-$ (Summary vs. original)$_{\text{base}}$ & Directly test identity effect on fidelity to source \\
\bottomrule
\end{tabular}
\end{adjustbox}
\end{table}

\paragraph{SES Manipulation} shifted from education (PhD credential) to occupation (high- vs. low-prestige). A PhD implies policy-relevant expertise, confounding socioeconomic status with perceived credibility; occupation-based signals (e.g., "accountant" vs. "food preparation worker") provide a cleaner test of SES-based differential treatment.

\paragraph{Number of Names} per demographic cell was reduced from approximately 25 to 2. Pilot results indicated that primary patterns emerge at the demographic group level rather than from specific name choices; two names per cell suffice to model name as a random effect while substantially reducing complexity, allowing the main study to prioritize breadth across comments.

\paragraph{Number of Comments} increased from 4 to 200. The pilot's primary limitation was restricted generalizability; the main study enables generalization across comment topics, stances, lengths, and sophistication levels through stratified sampling. Comments will be stratified by length (short vs. long) and sophistication (substantive vs. non-substantive), yielding four strata. Stance (support, oppose, neutral) will not be used for stratification; instead, the sample will reflect population distributions to maintain external validity, with stance examined as a moderator in exploratory analyses.

\paragraph{Repetitions} decreased from 5 to 1, with a 10\% reliability subset (20 comments $\times$ 5 repetitions). Although raw text varied across repetitions at temperature = 0, DV measurements showed moderate to excellent reliability (ICC = 0.60--0.94), indicating single runs provide reliable estimates. The reliability subset verifies this holds in the new sample.

\paragraph{Manipulation Format} changed from a signature-only condition
\begin{quote}
    {\ttfamily

Sincerely, 
[Name], PhD
    }
\end{quote}
    
to an introduction-plus-signature condition
\begin{quote}
    {\ttfamily
    My name is [Name], and I am a [Occupation]
    ...
    Sincerely, 
    [Name]
    }
\end{quote}
Placing signals at both the beginning and end ensures salience throughout the LLM's processing.

\paragraph{Agencies} shifted from ICE and EPA to DOI and EPA. The main study will use DOI docket DOI-2025-0004 ("National Environmental Policy Act Implementing Regulations") as the primary use case and EPA docket EPA-HQ-OAR-2025-0124 ("Repeal of Greenhouse Gas Emissions Standards for Fossil Fuel-Fired Electric Generating Units") as a validation sample. The EPA docket was selected for temporal consistency (2025), environmental/climate policy alignment, and sufficient comment volume for sampling. Using different agencies from the pilot ensures independence between exploratory and main phases.

\paragraph{Prompts} changed from directive (role-based system prompt with structured output format) to minimal ("You are an assistant to summarize contents" / "Summarize this public comment"). The pilot's directive framing could confound intrinsic differential treatment with the model's interpretation of what a "policy analyst" would want. The main study uses minimal prompts for primary analysis, with directive prompts tested in robustness checks.


\paragraph{Writing Quality Conditions} were not included in the pilot. The main study introduces writing quality analyses through two complementary approaches to test $\boldsymbol{H_5}$ and $\boldsymbol{H_6}$. For $\boldsymbol{H_5}$, we will evaluate writing quality across all 200 comments using LLM-based scoring on five dimensions and examine whether quality levels predict summarization outcomes across all 32 identity conditions. To avoid conflating rating and summarization processes, the LLMs used for writing quality classification will be distinct from the models used for summarization. For $\boldsymbol{H_6}$, we will apply error injection to comments classified in the Low and Mid-Low quality quartiles, creating matched pairs of original and error-added versions across all 32 identity conditions. These manipulations are new and have no pilot precedent; accordingly, $\boldsymbol{H_5}$ and $\boldsymbol{H_6}$ represent novel tests without prior effect size estimates from this research program.

\paragraph{Dependent Variables} changed from direct attributed-vs-baseline comparisons to difference scores relative to the original comment. The pilot computed DVs by directly comparing attributed summaries to baseline summaries (e.g., $\Delta_{\text{Sim}}$ as cosine similarity between attributed and baseline embeddings; $\Delta_{\text{Len}}$ as attributed words / baseline words). The main study computes DVs relative to the original comment, then takes the difference: $\Delta$ = (Summary vs. original)$_{\text{attributed}}$ $-$ (Summary vs. original)$_{\text{baseline}}$. This enables direct measurement of how adding identity signals affects fidelity to source content, with direction preserved in model coefficients. For $\Delta_{\text{Sent}}$ and $\Delta_{\text{Read}}$, the original cancels out algebraically, making pilot and main definitions equivalent; for $\Delta_{\text{Sim}}$ and $\Delta_{\text{Len}}$, the definitions differ substantively.

\section{Design Table for Pre-registration}
\begin{table}[htbp]
\caption{\label{tab:design} Design Table}
\centering
\begin{adjustbox}{max width=\textwidth}
\begin{tabular}{p{2.5cm}p{4cm}p{3.5cm}p{3.5cm}p{4cm}}
\toprule
\textbf{Question} & \textbf{Hypothesis} & \textbf{Sampling Plan} & \textbf{Analysis Plan} & \textbf{Interpretation Given to Different Outcomes} \\
\midrule
Does commenter race affect LLM summarization? & 
$\boldsymbol{H_{1_{1}}}$: Difference scores will vary significantly across racial groups (White, Black, Hispanic, Asian). \newline $\boldsymbol{H_{1_{0}}}$: Difference scores will not vary across racial groups. & 
200 comments $\times$ 32 identity conditions $\times$ 8 models; within-subjects design & 
Hierarchical testing: (1) Omnibus gate on Composite Index, (2) Holm correction across H1--H4 using minimum $p$-value; if significant, remaining DVs tested with Holm correction. Pairwise comparisons reported descriptively. & 
\textbf{p $<$ .05}: Reject $\boldsymbol{H_{1_{0}}}$; race affects summarization; significant measures characterize how. \newline \textbf{p $\geq$ .05 and BF$_{01}$ $\geq$ 3}: Evidence for $\boldsymbol{H_{1_{0}}}$. \newline \textbf{p $\geq$ .05 and $1/3 < BF < 3$}: Inconclusive. \\
\midrule
Does commenter gender affect LLM summarization? & 
$\boldsymbol{H_{2_{1}}}$: Difference scores will vary significantly between male and female conditions. \newline $\boldsymbol{H_{2_{0}}}$: Difference scores will not vary between genders. & 
Same & 
Same hierarchical testing procedure as H1. & 
\textbf{p $<$ .05}: Reject $\boldsymbol{H_{2_{0}}}$; gender affects summarization. \newline \textbf{p $\geq$ .05 and BF$_{01}$ $\geq$ 3}: Evidence for $\boldsymbol{H_{2_{0}}}$. \newline \textbf{p $\geq$ .05 and $1/3 < BF < 3$}: Inconclusive. \\
\midrule
Does commenter SES affect LLM summarization? & 
$\boldsymbol{H_{3_{1}}}$: Difference scores will vary significantly between high-SES and low-SES conditions. \newline $\boldsymbol{H_{3_{0}}}$: Difference scores will not vary by SES. & 
Same & 
Same hierarchical testing procedure as H1. & 
\textbf{p $<$ .05}: Reject $\boldsymbol{H_{3_{0}}}$; SES affects summarization. \newline \textbf{p $\geq$ .05 and BF$_{01}$ $\geq$ 3}: Evidence for $\boldsymbol{H_{3_{0}}}$. \newline \textbf{p $\geq$ .05 and $1/3 < BF < 3$}: Inconclusive. \\
\midrule
Does the effect of race vary by gender? & 
$\boldsymbol{H_{4_{1}}}$: The effect of race on difference scores will vary significantly between male and female conditions. \newline $\boldsymbol{H_{4_{0}}}$: The effect of race will be consistent across genders. & 
Same & 
Same hierarchical testing procedure as H1. Simple effects reported descriptively. & 
\textbf{p $<$ .05}: Reject $\boldsymbol{H_{4_{0}}}$; intersectional effects present; simple effects characterize pattern. \newline \textbf{p $\geq$ .05 and BF$_{01}$ $\geq$ 3}: Evidence for $\boldsymbol{H_{4_{0}}}$. \newline \textbf{p $\geq$ .05 and $1/3 < BF < 3$}: Inconclusive. \\
\midrule
Does comment writing quality predict LLM summarization outcomes? & 
$\boldsymbol{H_{5_{1}}}$: Raw scores will covary with writing quality levels across identity conditions. \newline $\boldsymbol{H_{5_{0}}}$: Raw scores will not covary with writing quality levels. & 
200 comments $\times$ 32 identity conditions $\times$ 8 models; separate sub-model & 
WritingQuality effect tested on 4 raw DVs with Holm-Bonferroni correction. No omnibus gate. Bayes factors computed for all tests. Secondary dimension-specific analyses with Holm-Bonferroni correction across five dimensions. & 
\textbf{Any p $<$ $\alpha_{\text{Holm}}$}: Reject $\boldsymbol{H_{5_{0}}}$; writing quality predicts summarization outcomes. \newline \textbf{All p $\geq$ $\alpha_{\text{Holm}}$ and BF$_{01}$ $\geq$ 3}: Evidence for $\boldsymbol{H_{5_{0}}}$. \newline \textbf{All p $\geq$ $\alpha_{\text{Holm}}$ and $1/3 < BF < 3$}: Inconclusive. \\
\midrule
Do error-added conditions affect LLM summarization? & 
$\boldsymbol{H_{6_{1}}}$: Raw scores will differ between error-added and original conditions (same identity, low and mid-low quality only). \newline $\boldsymbol{H_{6_{0}}}$: Raw scores will not differ between error-added and original conditions. & 
100 comments $\times$ 32 identity conditions $\times$ 2 versions (original, error-added) $\times$ 8 models; separate sub-model & 
ErrorAdded effect tested on 4 raw DVs with Holm-Bonferroni correction. No omnibus gate. Bayes factors computed for all tests. & 
\textbf{Any p $<$ $\alpha_{\text{Holm}}$}: Reject $\boldsymbol{H_{6_{0}}}$; error injection affects summarization. \newline \textbf{All p $\geq$ $\alpha_{\text{Holm}}$ and BF$_{01}$ $\geq$ 3}: Evidence for $\boldsymbol{H_{6_{0}}}$. \newline \textbf{All p $\geq$ $\alpha_{\text{Holm}}$ and $1/3 < BF < 3$}: Inconclusive. \\
\bottomrule
\end{tabular}
\end{adjustbox}

\par\medskip\noindent\small
\textit{Note.} Difference scores denote the change in the dependent variable when identity is added, computed as attributed summary minus baseline summary. Primary measures include semantic similarity ($\Delta_{\text{Sim}}$), sentiment ($\Delta_{\text{Sent}}$), length ratio ($\Delta_{\text{Len}}$), and readability ($\Delta_{\text{Read}}$). The Composite Index is the mean of absolute $z$-scored difference values across the four measures. For the Identity Analysis ($\boldsymbol{H_1}$--$\boldsymbol{H_4}$), hierarchical testing first assesses whether identity signals collectively produce differential treatment, then tests each hypothesis with family-wise error rate control. For the Writing Quality Analysis ($\boldsymbol{H_5}$, $\boldsymbol{H_6}$), $\boldsymbol{H_5}$ examines whether writing quality predicts summarization outcomes and $\boldsymbol{H_6}$ compares the same low and mid-low quality comments with and without injected errors; neither analysis requires the no-identity baseline, so $\boldsymbol{H_5}$ and $\boldsymbol{H_6}$ are tested directly on the four primary measures without an omnibus gate. $\boldsymbol{H_5}$ and $\boldsymbol{H_6}$ are exploratory analyses addressing writing quality rather than identity signals. Hypotheses are non-directional due to heterogeneous effect directions reported in prior literature. If distributional assumptions are violated, sensitivity analyses using alternative modeling approaches will be conducted as specified in Methods.
\end{table}
\FloatBarrier

\section{Master Hypothesis Table}\label{app:master-table}
\input{appendix/tables/master_results}

\end{document}

%% file: appendix/tables/master_results.tex
\begin{sidewaystable}
\centering
\caption{\label{tab:master_results}Master Results Table: All Hypotheses}
\begin{adjustbox}{max width=\textwidth}
\begin{tabular}[t]{llllrrrrrrll}
\toprule
  & Hypothesis & Test & DV & $F$ & NumDF & DenDF & $p\_{raw}$ & $p\_{adj}$ & $\eta^2\_p$ & $BF\_{10}$ & Statistical Significance\\
\midrule
delta\_sim & H1\_race & race & delta\_sim & 1.8940 & 3 & 8 & 0.2091 & 0.2091 & 0.4153 & 6.39e-13 & Not significant\\
delta\_sent & H1\_race & race & delta\_sent & 5.7250 & 3 & 46331 & 0.0007 & 0.0026 & 0.0004 & 1.21e-10 & Significant\\
delta\_len & H1\_race & race & delta\_len & 5.3322 & 3 & 46331 & 0.0011 & 0.0034 & 0.0003 & 4.50e-11 & Significant\\
delta\_read & H1\_race & race & delta\_read & 3.6753 & 3 & 8 & 0.0626 & 0.1251 & 0.5795 & 1.32e-11 & Not significant\\
\addlinespace
delta\_sim1 & H2\_gender & gender & delta\_sim & 1.9003 & 1 & 8 & 0.2054 & 0.2075 & 0.1920 & 2.74e-09 & Not significant\\
delta\_sent1 & H2\_gender & gender & delta\_sent & 2.6473 & 1 & 46331 & 0.1037 & 0.2075 & 0.0001 & 3.93e-09 & Not significant\\
delta\_len1 & H2\_gender & gender & delta\_len & 0.0002 & 1 & 46331 & 0.9887 & 0.2075 & 0.0000 & 7.04e-10 & Not significant\\
delta\_read1 & H2\_gender & gender & delta\_read & 0.4498 & 1 & 8 & 0.5213 & 0.2075 & 0.0532 & 2.99e-09 & Not significant\\
delta\_sim2 & H3\_ses & ses & delta\_sim & 319.5641 & 1 & 46323 & 0.0000 & 0.0000 & 0.0069 & 6.64e+66 & Significant\\
delta\_sent2 & H3\_ses & ses & delta\_sent & 17.0214 & 1 & 46331 & 0.0000 & 0.0001 & 0.0004 & 23.0274 & Significant\\
delta\_len2 & H3\_ses & ses & delta\_len & 7.1207 & 1 & 46331 & 0.0076 & 0.0076 & 0.0002 & 0.1631 & Significant\\
delta\_read2 & H3\_ses & ses & delta\_read & 216.6116 & 1 & 46323 & 0.0000 & 0.0000 & 0.0047 & 3.93e+44 & Significant\\
\addlinespace
delta\_sim3 & H4\_interaction & race:gender & delta\_sim & 0.0791 & 3 & 8 & 0.9695 & 0.6314 & 0.0288 & 1.24e-07 & Not significant\\
delta\_sent3 & H4\_interaction & race:gender & delta\_sent & 0.5448 & 3 & 46331 & 0.6516 & 0.6314 & 0.0000 & 2.26e-07 & Not significant\\
delta\_len3 & H4\_interaction & race:gender & delta\_len & 0.2807 & 3 & 46331 & 0.8394 & 0.6314 & 0.0000 & 1.52e-07 & Not significant\\
delta\_read3 & H4\_interaction & race:gender & delta\_read & 0.6026 & 3 & 8 & 0.6314 & 0.6314 & 0.1843 & 4.60e-07 & Not significant\\
\addlinespace
writing\_quality & H5\_wq & wq\_aggregate & semantic\_similarity & 240.2970 & 1 & 180 & 0.0000 & 0.0000 & 0.5717 & 1.06e+31 & Significant\\
writing\_quality & H5\_wq & wq\_aggregate & sentiment & 0.0001 & 1 & 180 & 0.9907 & 0.9907 & 0.0000 & 0.0046 & Not significant\\
writing\_quality & H5\_wq & wq\_aggregate & length\_ratio & 85.0327 & 1 & 180 & 0.0000 & 0.0000 & 0.3208 & 7.74e+12 & Significant\\
writing\_quality & H5\_wq & wq\_aggregate & readability & 23.3783 & 1 & 180 & 0.0000 & 0.0000 & 0.1150 & 292.4042 & Significant\\
\addlinespace
error\_injection & H6\_error & error\_added & semantic\_similarity & 3.3774 & 1 & 47946 & 0.0661 & 0.1983 & 0.0001 & 0.0247 & Not significant\\
error\_injection & H6\_error & error\_added & sentiment & 0.4982 & 1 & 47946 & 0.4803 & 0.9606 & 0.0000 & 0.0059 & Not significant\\
error\_injection & H6\_error & error\_added & length\_ratio & 11.3457 & 1 & 47957 & 0.0008 & 0.0030 & 0.0002 & 1.3268 & Significant\\
error\_injection & H6\_error & error\_added & readability & 0.2726 & 1 & 47946 & 0.6016 & 0.9606 & 0.0000 & 0.0052 & Not significant\\
\bottomrule
\multicolumn{12}{l}{\rule{0pt}{1em}\textit{Note:} Partial eta-squared values should be interpreted alongside DenDF. When DenDF is small ($< 50$), eta-squared can appear inflated even for non-significant effects.}\\
\end{tabular}
\end{adjustbox}
\end{sidewaystable}

\FloatBarrier